\documentclass[a4paper,sn-mathphys,pdflatex,twoside]{sn-jnl}

\usepackage{longtable}

\makeatletter
\input{size12.clo}
\makeatother
\normalbaroutside  
\makeatletter
{\catcode`\;=\active \gdef;{\ifmmode\semicolon\else\@semicolon\fi}}
\makeatother
\theoremstyle{thmstyleone}%
\newtheorem{theorem}{Theorem}[section]% meant for sectionwise numbers
\newtheorem{proposition}[theorem]{Proposition}%
\newtheorem{corollary}[theorem]{Corollary}%
\newtheorem{lemma}[theorem]{Lemma}%
\newtheorem{definition}[theorem]{Definition}%
\newtheorem{conjecture}[theorem]{Conjecture}%

\theoremstyle{thmstyletwo}%
\newtheorem{example}[theorem]{Example}%
\newtheorem{remark}[theorem]{Remark}%

\theoremstyle{thmstylethree}%

\renewcommand\theequation{\thesection.\arabic{equation}}

\makeatletter
\@addtoreset{theorem}{section}
\@addtoreset{equation}{section}
\makeatother

\makeatletter
\def\ps@titlepage{%
      \def\@oddhead{%
      \vbox to 0pt{\vspace*{-38pt}%
    }}%
     \let\@evenhead\@oddhead%
     \def\@oddfoot{\vbox to 18pt{\vfill\reset@font\rmfamily\hfil\thepage\hfil}}
     \def\@evenfoot{}}%
\def\ps@headings{%
    \def\@oddfoot{\hfill}%
    \let\@evenfoot\@oddfoot%
      \def\@evenhead{%
      \vbox to 0pt{\vspace*{-39pt}%
         \hbox to \hsize{\hfill \hfill}}\par%%
      \hspace*{-\textwidth}\hbox to \hsize{\headerfont\thepage\qquad\rightmark\hfill}}%
      \def\@oddhead{%
      \vbox to 0pt{\vspace*{-39pt}%
         \hbox to \hsize{\hfill \hfill}}\par%%
      \hspace*{-\textwidth}\hbox to \hsize{\headerfont\hfill\leftmark\qquad\thepage}}%
      \let\@mkboth\markboth%
      }%
\makeatother

\DeclareMathOperator{\Res}{Res}
\allowdisplaybreaks[4]

\begin{document}

\title[Blobbed topological recursion of the quartic Kontsevich 
model I]{Blobbed topological recursion of 
the quartic Kontsevich model I: Loop equations and conjectures}

\author[1]{\fnm{Johannes} \sur{Branahl}}\email{j\_bran33@uni-muenster.de}
\author[2]{\fnm{Alexander} \sur{Hock}}\email{alexander.hock@maths.ox.ac.uk}
\author*[1]{\fnm{Raimar} \sur{Wulkenhaar}}\email{raimar@math.uni-muenster.de}

\affil[1]{\orgdiv{Mathematisches Institut},
  \orgname{Westf\"alische Wilhelms-Universit\"at},
  \orgaddress{\street{Einsteinstra\ss{}e 62},
  \postcode{48149} \city{M\"unster},
  \country{Germany}}}

\affil[2]{\orgdiv{Mathematical Institute},
  \orgname{University of Oxford},
  \orgaddress{\street{Andrew Wiles Building, Woodstock Road},
    \postcode{OX2 6GG} \city{Oxford},
    \country{United Kingdom}}}

\abstract{\unboldmath
  We provide strong evidence for the conjecture that the analogue of
  Kontsevich's matrix Airy function, with the cubic potential
  $\mathrm{Tr}(\Phi^3)$ replaced by a quartic term
  $\mathrm{Tr}(\Phi^4)$, obeys the blobbed topological recursion of
  Borot and Shadrin. We identify in the quartic Kontsevich model three
  families of correlation functions for which we establish interwoven
  loop equations.  One family consists of symmetric meromorphic differential
  forms $\omega_{g,n}$ labelled by genus and number of marked points
  of a complex curve. We reduce the solution of all loop equations to
  a straightforward but lengthy evaluation of residues.  In all
  evaluated cases, the $\omega_{g,n}$ consist of a part with poles at
  ramification points which satisfies the universal formula of
  topological recursion, and of a part holomorphic at ramification
  points for which we provide an explicit residue formula.
}

\keywords{Matrix models, (Blobbed) Topological recursion, 
Meromorphic forms on Riemann surfaces, Loop equations, 
Residue calculus}

\pacs[MSC Classification]{05A15, 14H70, 30F30, 32A20, 39B32}

\maketitle
\markboth{\textit{Blobbed topological recursion of the quartic Kontsevich 
model I: Loop equations and conjectures}}{\textsc{J. Branahl, A. Hock \& R. Wulkenhaar}}

\section{Introduction}

This paper achieves decisive progress in the complete solution of a
quartic analogue of the Kontsevich model. The \emph{Kontsevich model}
\cite{Kontsevich:1992ti} is 
a $N\times N$ Hermitian matrix model obtained by deforming
a Gau\ss{}ian measure $d\mu_0(\Phi)$ with covariance
\begin{align}
  \langle \Phi(e_{ij})  \Phi(e_{kl})  \rangle_c
  =\frac{\delta_{il}\delta_{jk}}{\lambda_k+\lambda_l}
  \label{gauss}
\end{align}
(where $(e_{kl})$ is the standard matrix basis and
$\lambda_1,\dots,\lambda_N$ are positive real numbers which we rename
to $E_1,\dots,E_N$ in this paper) by a cubic term
$\exp(\frac{\mathrm{i}}{6} \mathrm{Tr}(\Phi^3))$.  Under `quartic
analogue' we understand the deformation of the same Gau\ss{}ian
measure (\ref{gauss}) by a quartic term
$\exp(-\frac{\lambda}{4} \mathrm{Tr}(\Phi^4))$. The Kontsevich model
proves a conjecture by Witten \cite{Witten:1990hr} that the generating
function of intersection numbers of tautological characteristic
classes on the moduli space $\overline{\mathcal{M}}_{g,n}$ of stable
complex curves is a $\tau$-function for the KdV hierarchy. Thereby it
beautifully connects several areas of mathematics and physics such as
integrable models, matrix models, 2D quantum gravity, enumerative
geometry, complex algebraic geometry and also noncommutative geometry.

Some 15 years ago it was understood that the Kontsevich model is also
a prime example for a universal structure called \emph{topological
  recursion} \cite{Eynard:2007kz}.  It starts with the initial data
$(\Sigma,\Sigma_0,x,\omega_{0,1},B)$, called the \emph{spectral
  curve}. Here $x:\Sigma\to\Sigma_0$ is a ramified covering of Riemann
surfaces, $\omega_{0,1}$ is a meromorphic differential 1-form on
$\Sigma$ regular at the ramification points of $x$, and $B$ the
Bergman kernel, a symmetric meromorphic bidifferential form on
$\Sigma\times \Sigma$ with double pole on the diagonal and no residue.
From these initial data, topological recursion constructs a hierarchy
$\{\omega_{g,n}\}$ with $\omega_{0,2}=B$ of symmetric meromorphic
differential forms on $\Sigma^n$ and understands them as spectral
invariants of the curve. Other examples besides the Kontsevich model
(which is described e.g.\ in \cite[Sec 6]{Eynard:2016yaa}) are the
one- and two-matrix models \cite{Chekhov:2006vd}, Mirzakhani's
recursions \cite{Mirzakhani:2006fta} for the volume of moduli spaces
of hyperbolic Riemann surfaces, recursions in Hurwitz theory
\cite{Bouchard:2007hi} and in Gromov-Witten theory
\cite{Bouchard:2007ys}.

This paper provides strong evidence that our quartic analogue of the
Kontsevich model is a prime example\footnote{Up to a small detail: we
  find a blob also for cylinder topology $\omega_{0,2}=B+\phi_{0,2}$.}
for \emph{blobbed topological recursion}, an extension of topological
recursion developed by Borot and Shadrin \cite{Borot:2015hna}. In this
setting the differential forms
\begin{align*}
  \omega_{g,n}(...,z)=\mathcal{P}_z\omega_{g,n}(...,z)
  +\mathcal{H}_z\omega_{g,n}(...,z)
\end{align*}
decompose into a part $\mathcal{P}_z\omega_{g,n}$ with poles (in a
selected variable $z$) at ramification points of
$x:\Sigma\to \Sigma_0$ and a part $\mathcal{H}_z\omega_{g,n}$ with
poles somewhere else. The $\mathcal{P}_z\omega_{g,n}$ are recursively
given by the universal formula (for simple ramifications)
\begin{align}
\mathcal{P}_z\omega_{g,n+1}(I,z)
   \label{BTR-intro}
   & =\sum_{\beta_i}
   \Res\displaylimits_{q\to \beta_i}
   K_i(z,q)\bigg(
   \omega_{g-1,n+2}(I, q,\sigma_i(q))
   \\[-2ex]
   &\hspace*{3cm}
   +\hspace*{-1cm} \sum_{\substack{g_1+g_2=g\\ I_1\uplus I_2=I\\
             (g_1,I_1)\neq (0,\emptyset)\neq (g_2,I_2)}}
   \hspace*{-1.1cm} \omega_{g_1,|I_1|+1}(I_1,q)
   \omega_{g_2,|I_2|+1}(I_2,\sigma_i(q))\!\bigg)
   \nonumber
\end{align}
of topological recursion. Here $I=\{z_1,\dots,z_n\}$ collects the
other variables besides $z$, the sum is over the ramification points
$\beta_i$ of $x$ defined by $x'(\beta_i)=0$. The kernel $K_i(z,q)$ is
defined in the neighbourhood of $\beta_i$ by
$K_i(z,q)=\frac{\frac{1}{2}\int^{q}_{\sigma_i(q)}
  B(z,q')}{\omega_{0,1}(q)-\omega_{0,1}(\sigma_i(q))}$, where
$\sigma_i\neq \mathrm{id}$ is the local Galois involution
$x(q)=x(\sigma_i(q))$ near $\beta_i$.  There is no general formula for
the other part $\mathcal{H}_z\omega_{g,n}$.  The only requirement is
that
$\omega_{g,n}=\mathcal{P}_z\omega_{g,n} +\mathcal{H}_z\omega_{g,n}$
satisfy \emph{abstract loop equations} \cite{Borot:2013lpa}. The
$\omega_{g',n'}$ on the rhs of (\ref{BTR-intro}) contain both parts
$\mathcal{P}$ and $\mathcal{H}$.

This paper identifies the $\omega_{g,n}$ for the quartic analogue of
the Kontsevich model (which is probably the most innovative step) and
establishes loop equations for them and for two families of functions
interweaved with the $\omega_{g,n}$.  These loop equations are very
complicated. We succeed in solving them for
$\omega_{0,2},\omega_{0,3},\omega_{1,1}$ and $\omega_{0,4}$. The
results are remarkably simple and structured. We prove that, although
our loop equations are much more complicated than familiar equations
of topological recursion, the solutions satisfy exactly the blobbed
topological recursion (\ref{BTR-intro}) in all four cases.  This
statement boils down to equality of more than 10 rational numbers.
This is unlikely a mere coincidence so that we conjecture that the
quartic analogue of the Kontsevich model obeys exactly the structures
of blobbed topological recursion.  In a subsequent work
\cite{Hock:2021tbl} we confirm the conjecture for genus $g=0$ (i.e.\
for all $\omega_{0,n}$). The loop equations established in this paper
are shown in \cite{Hock:2021tbl} to be equivalent to equations which
express $\omega_{0,n+1}(z_1,...,z_n,-z)$ in terms of
$\omega_{0,m+1}(z_1,...,z_m,+z)$ with $m\leq n$. Their solution is a
residue formula which implements blobbed topological recursion.

All these structures make the quartic analogue of the Kontsevich model a 
member of the family of models associated with the moduli space 
$\overline{\mathcal{M}}_{g,n}$ of stable complex curves.

\medskip 

We summarise central steps which went into the result. The model under
consideration is the result of attempts to understand quantum field
theories on noncommutative geometries. These QFT models have a matrix
formulation \cite{Grosse:2003aj, Langmann:2003if}
which was a main tool in establishing perturbative
renormalisability in four dimensions \cite{Grosse:2004yu} and
vanishing of the $\beta$-function \cite{Disertori:2006nq}. 
Exact solutions of particular choices of these matrix models have been
established in \cite{Langmann:2003if} for a complex model and most
importantly in \cite{Grosse:2005ig,Grosse:2006tc} for a quantum field
theory limit of the Kontsevich model (completed much later in 
\cite{Grosse:2019nes}). 

Building on \cite{Disertori:2006nq}, one of us (RW) with H.~Grosse
proved in \cite{Grosse:2009pa} that the planar 2-point function of
the Quartic Kontsevich Model satisfies a non-linear integral equation
\begin{align}
&(\mu^2{+}x{+}y)ZG^{(0)}(x,y)
\label{GW09}
\\
&= 1-\lambda\!
\int_0^{\Lambda^2} \!\!\! dt\;\varrho_0(t)
\Big( ZG^{(0)}(x,y) \;ZG^{(0)}(x,t) 
- \frac{ZG^{(0)}(t,y) -ZG^{(0)}(x,y)}{t-x}\Big) \,.
\nonumber
\end{align}
Here $\varrho_0$ is the spectral measure of a Laplacian on the
noncommutative geometry, $\lambda$ the coupling constant and
$\mu^2(\Lambda),Z(\Lambda)$ are renormalisation parameters to achieve
existence of $\lim_{\Lambda\to \infty} G^{(0)}(x,t)$. For the purpose
of this paper it is safe to set $\mu^2=0=Z-1$. This equation is the
first instance of a \emph{Dyson-Schwinger equation} (or loop equation) in the
Quartic Kontsevich Model. In \cite{Grosse:2012uv} a fixed point
formulation of (\ref{GW09}) was found from which in the following
years some qualitative results about the solution were deduced. But
in spite of considerable efforts, a solution of (\ref{GW09}) remained
out of reach for 9 years. In 2018, one of us (RW) with E.~Panzer found
in \cite{Panzer:2018tvy} the solution of
(\ref{GW09}) for $\varrho_0(t)=1$, $\mu^2=1-2\lambda \log(1+\Lambda^2)$ and $Z=1$.
A year later, two of us (AH, RW)
with H.~Grosse extended in \cite{Grosse:2019jnv} this solution to any
H\"older-continuous $\varrho_0$ with
$\int_0^\infty \frac{dt}{(1+t)^3}\varrho_0(t)<\infty$. The limit of
(\ref{GW09}) back to a matrix measure
$\varrho_0(t)=\frac{1}{N}\sum_{k=1}^d r_k\delta(t-e_k)$, already
considered in \cite{Grosse:2019jnv}, was understood by RW with
J.~Sch\"urmann in \cite{Schurmann:2019mzu-v3} as a problem in complex
algebraic geometry. Also the next equation for the planar 2-point
function of cycle type $(2,0)$ was solved in \cite{Schurmann:2019mzu-v3}.

\medskip

The present paper is a large-scale extension of \cite{Grosse:2019jnv,
  Schurmann:2019mzu-v3} to higher topological sectors. It was already
pointed out in \cite{Grosse:2012uv,Schurmann:2019mzu-v3} that,
although all Dyson-Schwinger equations for higher topological sectors
are affine equations, no solution theory for them seemed to exist. We
succeed in finding one. In Definition~\ref{defT0Om} we identify three
families $T_{q_1,...,q_m\|pq|},T_{q_1,...,q_m\|p|q|}$ and
$\Omega_{q_1,...,q_m}$ of auxiliary functions for which we derive in
sec.~\ref{sec:dse} loop equations. These have a graphical
interpretation which we provide in Appendix~\ref{app:graphical}.
Knowing $\Omega_{\dots}$ and $T_{\dots}$ permits a straightforward
solution of all cumulants $G_{\dots}$ of the quartically deformed
measure along the lines of
\cite{Schurmann:2019mzu-v3}. Section~\ref{sec:loopeq} extends the loop
equations of sec.~\ref{sec:dse} to functions of several complex
variables. The solution for the function $\Omega^{(0)}_2(u,z)$ in
Proposition~\ref{prop:Om02} makes first contact with the Bergman
kernel of topological recursion. We describe in
sec.~\ref{sec:recursive} how all equations can be solved by evaluation
of residues. Doing this in practice can be a longer endeavour, as
demonstrated in Appendix~\ref{app:solution}. The results strongly
suggest that our auxiliary functions $\Omega_{q_1,...q_m}$ descent
from symmetric meromorphic differential forms $\omega_{g,m}$ which
satisfy the main equation (\ref{BTR-intro}) of blobbed topological
recursion. Moreover, we provide explicit residue formulae for
$\mathcal{H}_z\omega_{g,n}(...,z)$.

\subsection*{Acknowledgements}

We thank St\'ephane Dartois for bringing blobbed topological recursion
to our attention, and for valuable discussions. This paper relies
heavily on previous results obtained with Harald Grosse, Erik Panzer
and J\"org Sch\"urmann. Our work was supported\footnote{``Funded by
  the Deutsche Forschungsgemeinschaft (DFG, German Research
  Foundation) -- Project-ID 427320536 -- SFB 1442, as well as under
  Germany's Excellence Strategy EXC 2044 390685587, Mathematics
  M\"unster: Dynamics -- Geometry -- Structure."} by the Cluster of
Excellence \emph{Mathematics M\"unster} and the CRC 1442 \emph{Geometry:
  Deformations and Rigidity}. The work of AH was additionally financed
by the RTG 2149 \emph{Strong and Weak Interactions -- from Hadrons to Dark
  Matter}.

\section{The Setup}

\label{sec:setup}

\subsection{Matrix Integrals}
Our aim is the algebraic solution of the quartic analogue of 
the Kontsevich model, i.e.\ of a matrix model with the same 
weighted covariance as the Kontsevich model \cite{Kontsevich:1992ti} 
but with quartic instead 
of cubic deformation. We employ the notation developed in 
\cite{Schurmann:2019mzu-v3} where further details are given.

Let $H_N$ be the real vector space of self-adjoint $N\times
N$-matrices and $(E_1,\dots, E_{N})$ be pairwise different\footnote{This 
is important in the first sections. After extension to several 
complex variables in sec.\ \ref{sec:complex} we can admit 
multiplicities.} positive
real numbers. Let $d\mu_{E,0}(\Phi)$ be the probability measure 
on the dual space $H_N'$ uniquely defined by
\begin{align}
\exp\Big(- \frac{1}{2N} \sum_{k,l=1}^{N}
\frac{M_{kl}M_{lk}}{E_k+E_l}\Big)
= \int_{H_N'} d\mu_{E,0}(\Phi) \,e^{\mathrm{i} \Phi(M)}\;,
\label{measure0}
\end{align}
for any $M=M^*=\sum_{k,l=1}^{N} M_{kl} e_{kl}\in H_N$  
where $(e_{kl})$ is the standard matrix basis. We deform 
$d\mu_{E,0}(\Phi)$ by a quartic potential to a measure
\begin{align}
d\mu_{E,\lambda}(\Phi) &:= 
\frac{d\mu_{E,0}(\Phi)\; e^{-\frac{\lambda N}{4}\mathrm{Tr}(\Phi^4)}}{
\int_{H_N'} d\mu_{E,0}(\Phi)\; e^{-\frac{\lambda N}{4}\mathrm{Tr}(\Phi^4)}}\;,
\label{measure4}
\end{align}
where 
$\mathrm{Tr}(\Phi^4):=
\sum_{j,k,l,m=1}^{N} \Phi(e_{jk})\Phi(e_{kl})\Phi(e_{lm})\Phi(e_{mj})$
when extending the linear forms
via 
$\Phi(M_1+\mathrm{i}M_2):=\Phi(M_1)+\mathrm{i}\Phi(M_2)$ 
to complex $N\times N$-matrices.

The Fourier transform 
\begin{align}
\mathcal{Z}(M)= \int_{H'_N}\!\! d\mu_{E,\lambda}
(\Phi) \;e^{\mathrm{i}\Phi(M)}
\label{Fourier}
\end{align}
of the measure is conveniently used to organise moments
\begin{align*}
\langle e_{k_1l_1}\dots e_{k_nl_n}\rangle &:= \int_{H'_{N}} \!\! 
d\mu_{E,\lambda}(\Phi)
\;\Phi(e_{k_1l_1}) \cdots \Phi(e_{k_nl_n}) 
\\
&= \frac{1}{\mathrm{i}^{n}}
\frac{\partial^n\mathcal{Z}(M)}{\partial 
M_{k_1l_1}\cdots \partial M_{k_nl_n}} \Big|_{M=0}
\end{align*}
and cumulants 
\begin{align}
\langle e_{k_1l_1}\dots e_{k_nl_n}\rangle_c 
&= \frac{1}{\mathrm{i}^{n}}
\frac{\partial^n\log \mathcal{Z}(M)}{\partial 
M_{k_1l_1}\cdots \partial M_{k_nl_n}} \Big|_{M=0}\;.
\label{cumulants}
\end{align}
As explained in \cite{Schurmann:2019mzu-v3}, the cumulants (\ref{cumulants})
are  only
non-zero if $(l_1,\dots,l_n)=(k_{\sigma(1)},\dots,k_{\sigma(n)})$ is a
permutation of $(k_1,\dots,k_n)$, and in this case\footnote{This assumes
that the $k_i$ are pairwise different.} the cumulant only
depends on the cycle type of this permutation $\sigma$ in the
symmetric group $ \mathcal{S}_n$. A 
non-vanishing cumulant of $b$ cycles is thus of the form
$\big\langle (e_{k_1^1k_2^1} 
e_{k_2^1k_3^1} \cdots 
e_{k_{n_1}^1k_1^1}) \cdots 
(e_{k_1^bk_2^b} e_{k_2^bk_3^b} \cdots 
e_{k_{n_b}^bk_1^b}) \big\rangle_c$ and gives, after rescaling by appropriate 
powers of $N$, for pairwise different $k_i^j$ rise to 
\begin{align}
N^{n_1+\dots+n_b} 
\big\langle (e_{k_1^1k_2^1} 
e_{k_2^1k_3^1} \cdots 
e_{k_{n_1}^1k_1^1}) \cdots 
(e_{k_1^bk_2^b} e_{k_2^bk_3^b} \cdots 
e_{k_{n_b}^bk_1^b}) \big\rangle_c 
\nonumber
\\
=: 
N^{2-b} \cdot G_{|k_1^1\dots k_{n_1}^1|\dots
|k_1^b\dots k_{n_b}^b|} \;.
\label{eq:Ggbn}
\end{align}
The goal is to compute these `correlation functions' 
$G_{\dots}$ after (at this point formal) expansion
$G_{\dots}=\sum_{g=0}^\infty N^{-2g} G^{(g)}_{\dots}$,
at least in principle. This was achieved in 
\cite{Grosse:2019jnv,Schurmann:2019mzu-v3} for $G^{(0)}_{|k_1k_2|}$ and in 
\cite{Schurmann:2019mzu-v3} for $G^{(0)}_{|k^1|k^2|}$. The results 
of \cite{DeJong} extend this solution to all  $G^{(0)}_{|k_1\dots k_n|}$. 
But starting with $G^{(0)}_{|k^1_1k^1_2|k^2_1k^2_2|}$ and 
$G^{(1)}_{|k_1k_2|}$ a new structure arises which cannot be treated 
by the methods developed so far. The present paper gives the 
solution and shows that precisely these new structures provide the 
connection to the world of topological recursion 
\cite{Eynard:2007kz, Eynard:2016yaa, Borot:2015hna}.

\subsection{Dyson-Schwinger Equations}

The Fourier transform (\ref{Fourier}) satisfies the equations of motion 
\cite[Lemma 1+2]{Schurmann:2019mzu-v3}
\begin{align}
\frac{1}{\mathrm{i}}\frac{\partial \mathcal{Z}(M)}{\partial M_{pq}} 
&= \frac{\mathrm{i} M_{qp}\mathcal{Z}(M)}{N(E_p+E_q)} 
-\frac{\lambda }{\mathrm{i}^3(E_p+E_q)} 
\sum_{k,l=1}^{N} \frac{\partial^3
\mathcal{Z}(M)}{\partial M_{pk}\partial M_{kl}\partial M_{lq}}\;,
\label{eom}
\\
\frac{1}{N} \frac{\partial \mathcal{Z}(M)}{\partial E_{p}} 
&= \Big(\sum_{k=1}^N  
\frac{\partial^2}{\partial M_{pk} \partial M_{kp}}
+ \frac{1}{N}\sum_{k=1}^N  G_{|pk|}+\frac{1}{N^2} 
G_{|p|p|}\Big)\mathcal{Z}(M)\;.
\label{eom-2}
\end{align}
The first one can be converted into \cite[eq.\ (50)]{Schurmann:2019mzu-v3}
\begin{align}
-N\sum_{k=1}^{N} 
(E_p-E_q)\frac{\partial^2 
\mathcal{Z}(M)}{\partial 
M_{pk} \partial M_{kq}} 
=  \sum_{k=1}^{N} 
\Big(
M_{kp} \frac{\partial\mathcal{Z}(M)}{\partial M_{kq}} 
-M_{qk} \frac{\partial\mathcal{Z}(M)}{\partial M_{pk}} 
\Big)\;.
\label{WTI}
\end{align}
For $E_p\neq E_q$ it is safe to divide by $(E_p-E_q)$ 
to extract $\frac{\partial^2 \mathcal{Z}(M)}{\partial 
M_{pk} \partial M_{kq}}$, whereas $\frac{\partial^2 \mathcal{Z}(M)}{\partial 
M_{pk} \partial M_{kp}}$ has to be taken from (\ref{eom-2}). 
\begin{example}
\label{ex:Gpq}
For $p\neq q$ one has with $\mathcal{Z}(0)=1$
\begin{align*}
G_{|pq|}&=-N
\frac{\partial^2 \log \mathcal{Z}(M)}{\partial M_{pq} \partial M_{qp}} 
\Big|_{M=0}= 
-N \frac{\partial^2 \mathcal{Z}(M)}{\partial M_{pq} \partial M_{qp}} 
\Big|_{M=0}
\\
&= 
\frac{1}{E_p+E_q}-\frac{\lambda N}{E_p+E_q}
\sum_{k,l=1}^N 
\frac{\partial^2}{\partial M_{lq}
\partial M_{qp}}
\frac{\partial^2 \mathcal{Z}(M)}{
\partial M_{pk}
\partial M_{kl}}\Big|_{M=0}\;.
\end{align*}
The second line results when differentiating \eqref{eom} with respect to 
$M_{qp}$. We split the $l$-sum into $l=p$ where \eqref{eom-2} is used and 
$l\neq p$ where \eqref{WTI} for $q\mapsto l$ is inserted:
\begin{align*}
(E_p+E_q)G_{|pq|}
&= 1
-\lambda
\frac{\partial^2}{\partial M_{pq}
\partial M_{qp}}
\Big[
\frac{\partial \mathcal{Z}(M)}{\partial E_p}
-\mathcal{Z}(M)\Big(\sum_{k=1}^N G_{|pk|}+\frac{1}{N}G_{|p|p|}
\Big)\Big]\Big|_{M=0}
\\
&
-\lambda\sum_{k=1}^N 
\sum_{\substack{l=1\\l\neq p}}^N 
\frac{\partial^2}{\partial M_{lq}
\partial M_{qp}}\Big[
\frac{
M_{kp} \frac{\partial\mathcal{Z}(M)}{\partial M_{kl}} 
-M_{lk} \frac{\partial\mathcal{Z}(M)}{\partial M_{pk}} 
}{E_l-E_p}\Big]\Big|_{M=0}\;.
\end{align*}
Inserting $\mathcal{Z}(M)= 
1- \frac{1}{N^2} \sum_{j,k=1}^N \big(\frac{N}{2}
G_{|jk|} M_{jk}M_{kj}
+ \frac{1}{2} G_{|j|k|} M_{jj}M_{kk}\big)+ \mathcal{O}(M^4)$, 
the differentiation yields
the \emph{Dyson-Schwinger equation} (DSE) for the 2-point function
\begin{align}
(E_p+E_q)G_{|pq|}
&= 1+\frac{\lambda}{N} \frac{\partial G_{|pq|}}{\partial E_p}
-\lambda G_{|pq|}\Big(\frac{1}{N}\sum_{k=1}^N G_{|pk|}+\frac{1}{N^2}G_{|p|p|}
\Big)
\label{eq:Gpq}
\\*
&+\frac{\lambda}{N} \sum_{\substack{l=1\\l\neq p}}^N 
\frac{G_{|lq|}-G_{|pq|}}{E_l-E_p}
+\frac{\lambda}{N^2} \frac{G_{|q|q|}-G_{|p|q|}}{E_q-E_p}\;.
\nonumber
\end{align}
\end{example}
\begin{example}
\label{ex:G1plus1}
For $p\neq q$ one has with $\mathcal{Z}(0)=1$
\begin{align*}
G_{|p|q|}&=-N^2
\frac{\partial^2 \log \mathcal{Z}(M)}{\partial M_{pp} \partial M_{qq}} 
\Big|_{M=0}= 
-N^2 \frac{\partial^2 \mathcal{Z}(M)}{\partial M_{pp} \partial M_{qq}} 
\Big|_{M=0}
\\
&= 
-\frac{\lambda N^2}{E_p+E_p}
\sum_{k,l=1}^N 
\frac{\partial^2}{\partial M_{lp}
\partial M_{qq}}
\frac{\partial^2 \mathcal{Z}(M)}{
\partial M_{pk}
\partial M_{kl}}\Big|_{M=0}\;.
\end{align*}
The second line results when differentiating \eqref{eom} taken at
$q\mapsto p$ with respect to 
$M_{qq}$. We split the $l$-sum into $l=p$ where \eqref{eom-2} is used and 
$l\neq p$ where \eqref{WTI} for $q\mapsto l$ is inserted:
\begin{align*}
(E_p+E_p)G_{|p|q|}
&= 
-\lambda
\frac{N\partial^2}{\partial M_{pp}
\partial M_{qq}}
\Big[
\frac{\partial \mathcal{Z}(M)}{\partial E_p}
-\mathcal{Z}(M)\Big(\sum_{k=1}^N G_{|pk|}+\frac{1}{N}G_{|p|p|}
\Big)\Big]\Big|_{M=0}
\\
&
-\lambda\sum_{k=1}^N 
\sum_{\substack{l=1\\l\neq p}}^N 
\frac{N \partial^2}{\partial M_{lp}
\partial M_{qq}}\Big[
\frac{
M_{kp} \frac{\partial\mathcal{Z}(M)}{\partial M_{kl}} 
-M_{lk} \frac{\partial\mathcal{Z}(M)}{\partial M_{pk}} 
}{E_l-E_p}\Big]\Big|_{M=0}\;.
\end{align*}
Inserting $\mathcal{Z}(M)= 
1- \frac{1}{N^2} \sum_{j,k=1}^N \big(\frac{N}{2}
G_{|jk|} M_{jk}M_{kj}
+ \frac{1}{2} G_{|j|k|} M_{jj}M_{kk}\big)+ \mathcal{O}(M^4)$, 
the differentiation yields
the Dyson-Schwinger equation for the $(1{+}1)$-point function
\begin{align}
(E_p+E_p)G_{|p|q|}
&= \frac{\lambda}{N} \frac{\partial G_{|p|q|}}{\partial E_p}
-\lambda G_{|p|q|}\Big(\frac{1}{N}\sum_{k=1}^N G_{|pk|}+\frac{1}{N^2}G_{|p|p|}
\Big)
\label{eq:G1plus1}
\\*
&+\frac{\lambda}{N} \sum_{\substack{l=1\\l\neq p}}^N 
\frac{G_{|l|q|}-G_{|p|q|}}{E_l-E_p}
+\lambda \frac{G_{|qq|}-G_{|pq|}}{E_q-E_p}\;.
\nonumber
\end{align}
\end{example}

\noindent
The arising repeated differentiations with respect to the $E_q$ 
suggest to introduce:
\begin{definition}\label{defT0Om}
Generalised correlation functions are defined for pairwise 
different $q_j, p^s_i$ by 
\begin{align*}
T_{q_1,...,q_m\|p_1^1...p_{n_1}^1|p_1^2...p_{n_2}^2|...|p_1^b...p_{n_b}^b|}
:=\frac{(-N)^m\partial^m}{\partial E_{q_1}...\partial 
E_{q_m}}G_{|p_1^1...p_{n_1}^1|p_1^2...p_{n_2}^2|...|p_1^b...p_{n_b}^b|}\;.
\end{align*}
Moreover, assume that for 
\begin{align*}
\Omega_{q} &:= \frac{1}{N}\sum_{k=1}^N G_{|qk|}+\frac{1}{N^2}G_{|q|q|}
\end{align*}
there is a function $\mathcal{F}(E,\lambda)$ with 
$\Omega_{q}=-N\frac{\partial \mathcal{F}}{\partial E_q}$. Then  
functions $\Omega_{q_1,...,q_m}$, symmetric in their indices, 
are defined by 
\begin{align*}
\Omega_{q_1,...,q_m} &:= 
\frac{(-N)^{m-1}\partial^{m-1}\Omega_{q_1}}{
\partial E_{q_2}...\partial 
E_{q_{m}}} +\frac{\delta_{m,2}}{(E_{q_1}-E_{q_2})^2}\;,\qquad
m\geq 2\;.
\end{align*}
\end{definition}
For the aim of the paper it will be sufficient to focus on the
following special cases, the generalised 2-point and $1+1$-point
functions:
\begin{align*}
  T_{q_1,...,q_m\|pq|}
  &:=\frac{(-N)^m\partial^m}{\partial E_{q_1}...\partial 
    E_{q_m}}G_{|pq|}\;,
  &
T_{q_1,...,q_m\|p|q|}
&:=\frac{(-N)^m\partial^m}{\partial E_{q_1}...\partial 
E_{q_m}}G_{|p|q|}\;.
\end{align*}
These functions will be considered as formal power series 
in $\frac{1}{N^2}$, 
\begin{align}
	G=\sum_{g=0}^\infty N^{-2g} G^{(g)}\;,\quad
	T=\sum_{g=0}^\infty N^{-2g} T^{(g)}\;,\quad
	\Omega=\sum_{g=0}^\infty N^{-2g} \Omega^{(g)}\;,
	\label{genus-expansion}
\end{align}
where $g$ plays the r\^ole of a genus.

\begin{figure}[h]
\includegraphics[width= 1.0\textwidth]{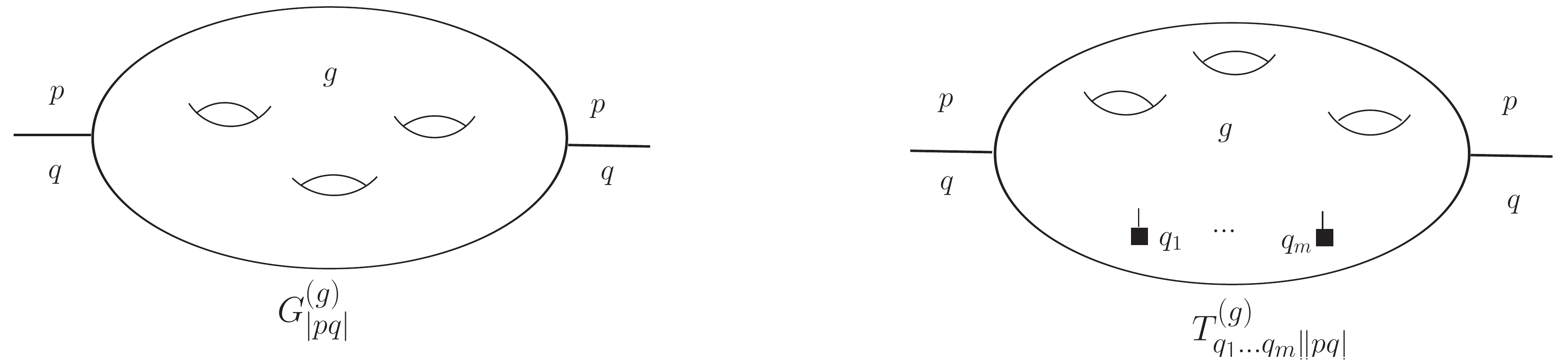}
\caption{The action of $\frac{\partial}{\partial E_{q_i}}$
  on $G^{(g)}_{|pq|}$ (shown on the left, with $b=1$ boundary
  and no marked point) 
  turns an inner face into a marked face with a square-vertex.
  The result of $m$ such operations is depicted on the right; 
  it corresponds to a genus $g$-Riemann surface with one boundary
  (carrying two 1-valent vertices) and $m$ marked points.
  The operation is described in \cite{Branahl:2020uxs} in details.
     \label{fig:pert0}}
\end{figure}

It is well-known that the 2-point function $G^{(g)}_{|pq|}$ and the
$1+1$-point function $G^{(g)}_{|p|q|}$ have an expansion into ribbon
graphs, see \cite{Branahl:2020uxs} for the definition of those ribbon
graphs associated to the quartic Kontsevich model. The generalised
correlation functions of Def. \ref{defT0Om} have a topological meaning
as well, since the derivative $\frac{\partial}{\partial E_{q_i}}$ acts
on an internal edge and fixes it to $E_{q_i}$. This operation is
extensively discussed in \cite{Branahl:2020uxs} and depends
topologically on the labelling of the two faces adjacent to the edge. Let
us briefly sum up the major ideas on which the detailed graphical
discussion in App.~\ref{app:graphical} will build:

We consider genus-$g$ Riemann surfaces $X$ with $b$ boundaries and $m$ marked
points. We let $\chi=2-2g-m-b$ be the Euler characteristic of $X$.
\begin{enumerate}
\item  The $\Omega^{(g)}_{q_1,...,q_m}$ correspond to no boundary ($b=0$)
  and $m$ marked points;

\item  The $T^{(g)}_{q_1,...,q_m\|pq|}$ correspond to $b=1$ boundary and 
  $m$ marked points;  

\item  The $T^{(g)}_{q_1,...,q_m\|p|q|}$ correspond to $b=2$ boundaries and 
  $m$ marked points.
\end{enumerate}  
The ribbon graphs, consisting of vertices, edges and faces,
are embedded into these Riemann surfaces. The faces carry a label from
1 to $N$, and we distinguish three types:
\begin{enumerate}
\item internal faces whose labellings have to be summed over;
\item external faces in which at least one edge is part of the
  boundary (with fixed label);
\item marked faces (with fixed label, too), where no edge is
  part of a boundary. 
\end{enumerate}
For the vertices, also three types show up:
\begin{enumerate}
\item 4-valent vertices in the interior, carrying the weight $(-\lambda)$;
\item 1-valent vertices on the boundaries belonging to one (in $G_{|p|q|}$)
  or two (in $G_{|pq|}$) external face(s);
\item one square-vertex within every marked face (see Fig.\ \ref{fig:pert0}).
\end{enumerate}

Let us return to the actual aim of the paper: We will show in this
paper that a complexification of the $\Omega^{(g)}_{\dots}$ gives rise to
meromorphic differentials $\omega_{g,n}$ conjectured to obey blobbed
topological recursion \cite{Borot:2015hna}. This conjecture has been
proved for $g=0$ in \cite{Hock:2021tbl}. In a supplement
\cite{Branahl:2020uxs} we express the functions $\Omega_{q_1,...,q_m}$
as distinguished polynomials in the correlation functions
$G_{|\dots|}$, which themselves are quickly growing series in Feynman
ribbon graphs. Both results together imply that intersection numbers
\cite{Borot:2015hna} are accessible by a particular combination of
graphs.

\subsection{Dyson-Schwinger Equations for 
Generalised Correlation Functions}

\label{sec:dse}

\begin{definition}
For a set $I=\{q_1,\dots,q_m\}$ we introduce:
\begin{itemize}
\item $|I|=m$, $|\emptyset|=0$.

\item $I,q:=I\cup \{q\}=\{q_1,\dots,q_m,q\}$. 

\item $I{\setminus} q:=I\setminus \{q\}$, with
$I{\setminus} q=\emptyset$ if $q\notin I$. 
\end{itemize}
We let $\sum_{I_1\uplus I_2=I}$ be the sum over all disjoint partitions
of $I$ into two possibly empty subsets $I_1,I_2$.
\end{definition}
\noindent
Equations for $T_{\dots}$ result by application of
$D_I:=\frac{(-N)^m \partial^m}{\partial E_{q_1}\cdots \partial
  E_{q_m}}$ to (\ref{eq:Gpq}) and (\ref{eq:G1plus1}) when taking
Definition~\ref{defT0Om} into account. We will give three important
types of Dyson-Schwinger equations that will determine the recursive
structure of our model:

\begin{proposition}
\label{prop:TOm}
Let $I=\{q_1,\dots,q_m\}$. For pairwise different 
$q_j,p,q$, the generalised correlation functions $T_{\dots}$ 
and $\Omega_{\dots}$ satisfy the following three types of
Dyson-Schwinger equations: 
\begin{itemize}[1)]
\item[1)] The DSE determining the generalised 2-point function:
\begin{align}
&\Big(E_q+E_p+\frac{\lambda}{N}
\sum_{\substack{l=1\\l\neq p}}^N  \frac{1}{E_l-E_p}\Big)T_{I\|pq|}
- \frac{\lambda}{N}\sum_{\substack{l=1 \\ l\notin I,p}}^N
 \frac{T_{I\|lq|}}{E_l-E_p}
\label{DSE-T2}
\\
&= \delta_{0,|I|}
-
\lambda\bigg\{
\sum_{I'\uplus I''=I}
\Omega_{I',p} T_{I''\|pq|}
-\frac{1}{N}\frac{\partial T_{I\|pq|}}{\partial E_p}
\nonumber
\\
&+ 
\sum_{j=1}^m \frac{\partial}{\partial E_{q_j}}
\Big(\frac{T_{I\setminus q_j\|q_jq|}}{E_{q_j}-E_p}\Big)
- \frac{1}{N^2}\frac{T_{I\|q|q|}-T_{I\|p|q|}}{E_q-E_p}
\bigg\}\;.
\nonumber
\end{align}
\item[2)] The DSE determining the generalised 1+1-point function:
\begin{align}
&\Big(E_p+E_p+\frac{\lambda}{N}
\sum_{\substack{l=1\\l \neq p}}^N  \frac{1}{E_l-E_p}\Big)
T_{I\|p|q|}
-
\frac{\lambda}{N}\sum_{\substack{l=1\\l\notin I,p}}^N  
\frac{T_{I\|l|q|}}{E_l-E_p}
\label{DSE-T1}
\\
&=(-\lambda)
\bigg\{
\sum_{I'\uplus I''=I}
\Omega_{I',p}T_{I''\|p|q|} 
-\frac{1}{N}\frac{\partial T_{I\|p|q|}}{\partial E_p}
\nonumber
\\
&+\sum_{j=1}^m \frac{\partial}{\partial E_{q_j}}
\Big(\frac{T_{I\setminus q_j\|q_j|q|}}{E_{q_j}-E_p}\Big)
-\frac{T_{I\|qq|}-T_{I\|pq|}}{E_{q}-E_p}
\bigg\}\;.
\nonumber
\end{align}
\item[3)] The DSE for $\Omega_{I,q}$ in terms of $T$:
\begin{align}
&\Omega_{I,q}
= \frac{\delta_{|I|,1}}{
(E_{q_1}-E_q)^2}
+
\frac{1}{N}\sum_{\substack{l=1\\l\notin I}}^NT_{I\|ql|}
- \sum_{j=1}^m 
\frac{\partial T_{I\setminus q_j\|qq_j|}}{\partial E_{q_j}} 
+\frac{1}{N^2} T_{I\|q|q|}\;,
\label{DSE-Om}
\end{align}
where $T_{\emptyset\|qq_j|}:=G_{|qq_j|}$ and
$T_{\emptyset\|q_j|q|}:=G_{|q_j|q|}$. 
\end{itemize}
\begin{proof}
We apply the operator $D_I:=\frac{(-N)^m \partial^m}{\partial E_{q_1}\cdots 
  \partial E_{q_m}}$ to (\ref{eq:Gpq}) and take Definition~\ref{defT0Om}
into account. Here we have
\begin{align*}
D_I\Big(  
G_{|pq|}\Big(\frac{1}{N}\sum_{k=1}^N G_{|pk|}+\frac{1}{N^2}G_{|p|p|}
\Big)\Big)= \!\!
\sum_{I_1\uplus I_2=I} \!\!
 \Big(\Omega_{I_1,p}-\frac{\delta_{|I_1|,1}}{(E_{I_1}-E_p)^2}\Big) T_{I_2\|pq|}
 \tag{*}
\end{align*}
and, separating the cases $l\in I$ and $l\notin I$,
\begin{align*}
D_I\Big(  \sum_{\substack{l=1\\l\neq p}}^N 
\frac{G_{|lq|}-G_{|pq|}}{E_l-E_p}\Big)
&=\sum_{\substack{l=1\\l\notin I,p}}^N 
\frac{T_{I\|lq|}}{E_l-E_p}-\sum_{\substack{l=1\\l\neq p}}^N 
\frac{T_{I\|pq|}}{E_l-E_p}
\\
&-N \sum_{j=1}^m \Big\{
\frac{\partial}{\partial E_{q_j}}
\Big(\frac{T_{I\setminus q_j\|q_jq|}}{E_{q_j}-E_p}\Big)
+\frac{T_{I\setminus q_j\|pq|}}{(E_{q_j}-E_p)^2}\Big\}\;.
\tag{**}
\end{align*}
The last term in (**) cancels the contribution with $\delta_{|I_1|,1}$ in (*).
All other derivatives are clear and arrange into (\ref{DSE-T2}).

The same considerations give (\ref{DSE-T1}) when applying $D_I$
to (\ref{eq:G1plus1}). Equation (\ref{DSE-Om}) is clear.
\end{proof}
\end{proposition}

\begin{remark}
  In a similar manner one can derive Dyson-Schwinger equations for
  $G_{|k_1^1\dots k_{n_1}^1|\dots |k_1^b\dots k_{n_b}^b|}$. In fact
  \cite{Grosse:2012uv} they can be reduced to polynomials in
  correlation functions with $n_s\in \{1,2\}$, in $\lambda$ and in
  $\frac{1}{E_{k_i}-E_{k_j}}$. We refer to \cite{Hock:2020rje} for the
  precise form of the recursion.  Applying $D_I$ then gives rise to
  Dyson-Schwinger equations for generalised correlation functions
  $T_{q_1,...,q_m\|k_1^1\dots k_{n_1}^1|\dots|k_1^b\dots
    k_{n_b}^b|}$. We have provided them in prior versions of this
  paper (e.g.\ \url{https://arxiv.org/abs/2008.12201v3}). For the sake
  of readability we decided to suppress them here.
\end{remark}

The Dyson-Schwinger equations of Proposition \ref{prop:TOm} are in
one-to-one correspondence with a graphical representation via a 
perturbative expansion into ribbon graphs. We derive them once again
in App.~\ref{app:graphical} for illustrative purposes.

\section{Loop Equations in Several 
Complex Variables}

\label{sec:loopeq}

\subsection{Complexification}
\label{sec:complex}

The equations in Proposition~\ref{prop:TOm} are not sufficient to
determine the functions $G,T,\Omega$ because there is no equation for
derivatives with respect to matrix indices (e.g.\ in
$\frac{\partial T_{I\|pq|}}{\partial E_p} $ ) or functions
with coincident matrix indices (e.g.\ $G_{|qq|}$, $G_{|q|q|}$ or $T_{p\|pq|}$),
which however are needed.  Our strategy is
therefore to meromorphically extend these equations, where the
extension is not necessarily unique, but unique at the points $E_p$.
\begin{definition}
\label{def:complexification}
Proposition \ref{prop:TOm} suggests the following extension:
\begin{itemize}[(a)]

\item[\textup{(a)}] Introduce holomorphic functions 
$G,T,\tilde{\Omega}$ in several complex variables, 
defined on Cartesian products of a neighbourhood 
$\mathcal{V}$ of $\{E_1,...,E_N\}$ in $\mathbb{C}$, 
which at $E_1,\dots,E_N$ agree with the previous correlation functions:
\begin{align*}
  G(E_p,E_q)&\equiv G_{|pq|}\;,\qquad
  G(E_p|E_q)\equiv G_{|p|q|}\;,
\\
T(E_{q_1},...,E_{q_m}\| E_p,E_q|)
&\equiv T_{q_1,\dots,q_m\|pq|}\;,\qquad
\\
T(E_{q_1},...,E_{q_m}\| E_p|E_q|)
&\equiv T_{q_1,\dots,q_m\|p|q|}\;,
\\
\tilde{\Omega}(E_{q_1},...,E_{q_m})
&\equiv \Omega_{q_1,\dots,q_m}\;.
\end{align*}

\item[\textup{(b)}] Write the equations in Proposition \ref{prop:TOm}
  in terms of $G,T,\tilde{\Omega}$ and postulate that they extend to
  pairwise different points
  $\{E_p\mapsto\zeta,E_q\mapsto \eta,E_{q_j} \mapsto \eta_j\}$ of
  $\mathcal{V}$.

\item[\textup{(c)}] Complexify the derivative by 
\begin{align*}
  \frac{\partial }{\partial E_q} f(E_q)\mapsto
  \frac{f(\eta)-f(E_q)}{\eta-E_q}
  +\frac{\partial }{\partial E_q}\Big\vert_{E_q\mapsto \eta} f(\eta)
\end{align*}
such that the
$\frac{\partial }{\partial E_q}\big\vert_{E_q\mapsto \eta}$-derivative
acts in the sense of Definition \ref{defT0Om} with extension to
$E_q\mapsto \eta$, and a difference quotient which tends for
$\eta\to E_q$ to the derivative on the argument of $f$.

\item[\textup{(d)}] Keep the $E_l$ in summations over $l \in \{1,\dots,N\}$ and
  complete the $l$-summation with the difference quotient term 
of \textup{(c)}.
  Consider the equations for
  $\zeta,\eta,\zeta^s_i,\eta_j \in \mathcal{V}\setminus
  \{E_1,\dots,E_N\}$.

\item[\textup{(e)}] Define the values of $G,T,\tilde{\Omega}$ at 
$\zeta=E_p,\eta=E_q,\zeta^s_i =E_{p^s_i},\eta_j=E_{q_j}$ and 
at coinciding points by a limit procedure.
\end{itemize}
\end{definition}

\begin{remark}
\label{rem:complex}
The complexification of the derivative defined in \textup{(c)}
distinguishes
between Definition \ref{defT0Om} and a derivative acting on the
argument of the function. The derivative on the argument is split
into a difference quotient to generate all missing terms in the
$l$-summation, e.g. for $l=q$ by
\begin{align}\label{del}
	\lim_{\eta\to E_q}\frac{f(\eta)-f(E_q)}{\eta-E_q}
=\lim_{E_l\to E_q} \frac{f(E_l)-f(E_q)}{E_l-E_q}=:\frac{\partial^{ext}}{\partial E_q} f(E_q),
\end{align} 
where the analyticity property at $E_q$ holds by \textup{(b)}.
Consequently, the summation over $l$ gets unrestricted in the
extension to $\mathcal{V}$ and coincides with
Proposition~\ref{prop:TOm} on the points $E_p$. Notice that the
extension of Definition \ref{def:complexification} is a meaningful
extension but cannot be unique.
\end{remark}

\noindent
The complexification procedure allows to relax the condition that 
all $E_1,\dots,E_N$ are pairwise different. From now on we agree that 
$(E_1,\dots,E_N)$ is made of pairwise different 
$e_1,\dots,e_d$ which arise with multiplicities 
$r_1,\dots,r_d$ in the tuple, with $r_1+...+r_d=N$.

We search for a solution of the equations after expansion
(\ref{genus-expansion}) of all arising functions as formal 
power series in $\frac{1}{N^2}$.
It will become clear later on that $g$ plays the r\^ole of the genus of 
a Riemann surface so that we call (\ref{genus-expansion}) the 
\emph{genus expansion}. When splitting the equations into homogeneous 
powers of $N^{-2}$ we agree that 
$\frac{1}{N}\sum_{k=1}^d r_k G^{(g)}(e_k,\dots)$ 
contributes to order $g$. Similarly for $T^{(g)}$ and $\tilde{\Omega}^{(g)}$.

\begin{example}
The complexification procedure of Definition~\ref{def:complexification}
turns \eqref{eq:Gpq} in Example~\ref{ex:Gpq} in presence of multiplicities 
of the $e_i$ into 
\begin{align}
(\zeta+\eta)G(\zeta,\eta)
&= 1 -\frac{\lambda}{N^2} T(\zeta\|\zeta,\eta)
-\lambda G(\zeta,\eta) \tilde{\Omega}(\zeta)
\label{eq:Gzetaeta0}
\\*
&+\frac{\lambda}{N} \sum_{k=1}^d r_k
\frac{G(e_k,\eta)-G(\zeta,\eta)}{e_k-\zeta}
+\frac{\lambda}{N^2} \frac{G(\eta|\eta)-G(\zeta|\eta)}{\eta-\zeta}\;.
\nonumber
\end{align}
For instance, we have $\frac{\partial G_{|pq|}}{\partial E_p}
\mapsto -\frac{1}{N} T(\zeta\|\zeta,\eta)+ 
r_p \frac{G(\zeta,\eta)-G(e_p,\eta)}{\zeta-E_p}$. The last term extends
$\sum_{l=1, l\neq p}^N \frac{G_{|lq|}-G_{|pq|}}{E_l-E_p}
\mapsto 
\sum_{l=1, l\neq p}^d r_l \frac{G(e_l,\eta)-G(\zeta,\eta)}{e_l-\zeta}$
to the unrestricted sum in (\ref{eq:Gzetaeta0}).
After genus expansion and with $\tilde{\Omega}^{(0)}(\zeta)
= \frac{1}{N}\sum_{k=1}^d r_k G^{(0)}(\zeta,e_k)$ from 
Definition~\ref{defT0Om} we have
\begin{align}
&\Big(\zeta+\eta + 
\frac{\lambda}{N}\sum_{k=1}^d r_k G^{(0)}(e_k,\zeta)
+\frac{\lambda}{N} \sum_{k=1}^d \frac{r_k}{e_k-\zeta}
\Big) G^{(g)}(\zeta,\eta)
-\frac{\lambda}{N}
\sum_{k=1}^d \frac{r_k G^{(g)}(e_k,\eta)}{
e_k-\zeta} 
\nonumber
\\
&= \delta_{g,0} -\lambda \sum_{g'=1}^g 
G^{(g-g')}(\zeta,\eta) \tilde{\Omega}^{(g')}(\zeta)
\nonumber
\\
&-\lambda T^{(g-1)}(\zeta\|\zeta,\eta)
+\lambda \frac{G^{(g-1)}(\eta|\eta)-G^{(g-1)}(\zeta|\eta)}{\eta-\zeta}\;.
\label{eq:Gzetaeta}
\end{align}
Note that the sum over $g'$ restricts to $g'\geq 1$ because the case 
$g'=0$ is explicitly included in the lhs.
\end{example}

For $g=0$ this becomes a non-linear equation for 
the function $G^{(0)}(\zeta,\eta)$ of $\zeta,\eta\in\mathcal{V}$. It has
been recently solved:
\begin{theorem}[\cite{Schurmann:2019mzu-v3}, building heavily
on \cite{Grosse:2019jnv}] 
\label{throm1}
Let $\lambda,e_k>0$. 
Assume that there is a rational function 
$R:\hat{\mathbb{C}}\to \hat{\mathbb{C}}$ with
\begin{enumerate}
\item $R$ has degree $d+1$, is normalised to 
  $R(\infty)=\infty$ and biholomorphically maps a domain $\mathcal{U}
  \subset \mathbb{C}$ to a neighbourhood 
(which can be assumed to be $\mathcal{V}$)
  in $\mathbb{C}$ of a real interval that contains $e_1,\dots,e_d$.

\item In terms of $G^{(0)}(R(z),R(w))=:\mathcal{G}^{(0)}(z,w)$ and 
$e_k=:R(\varepsilon_k)$ for $z,w,\varepsilon_k\in \mathcal{U}$ one has
\begin{align}
-R(-z)= R(z)+\frac{\lambda}{N}\sum_{k=1}^d \frac{r_k}{R(\varepsilon_k)
-R(z)}
+\frac{\lambda}{N} \sum_{k=1}^d  r_k \mathcal{G}^{(0)}(z,\varepsilon_k)\;.
\label{ansatz}
\end{align}
\end{enumerate}
Then $R$ and $\mathcal{G}^{(0)}$ are uniquely determined by the case $g=0$ 
of \eqref{eq:Gzetaeta} to
\begin{align}
R(z)&=z-\frac{\lambda}{N} \sum_{k=1}^d \frac{\varrho_k}{\varepsilon_k+z}\;,\qquad
R(\varepsilon_k)=e_k\;,\quad
\varrho_k R'(\varepsilon_k)=r_k\;,
\nonumber
\\
\mathcal{G}^{(0)}(z,w)&=\frac{\displaystyle 
1 -\frac{\lambda}{N} \sum_{k=1}^d \frac{r_k}{
(R(z)-R(\varepsilon_k))(R(\varepsilon_k)-R({-}w))}
\prod_{j=1}^d \frac{
R(w){-}R({-}\widehat{\varepsilon_k}^j)}{ R(w)-R(\varepsilon_j)}
}{R(w)-R(-z)}\;.
\label{Gzw-final}
\end{align}
The ansatz \eqref{ansatz} is identically fulfilled by
\eqref{Gzw-final}. In these equations, the solutions of $R(v)=R(z)$ 
are denoted by $v\in\{z,\hat{z}^1,\dots,\hat{z}^d\}$ with $z\in \mathcal{U}$
when considering $R:\hat{\mathbb{C}}\to \hat{\mathbb{C}}$. 
The function $\mathcal{G}^{(0)}(z,w)$ is symmetric. Its poles are located 
at $z+w=0$ and $z,w\in \{\widehat{\varepsilon_k}^j\}$ for $k,j\in\{1,...,d\}$.
\end{theorem}

\begin{figure}[h]
    \includegraphics[width= 0.99\textwidth]{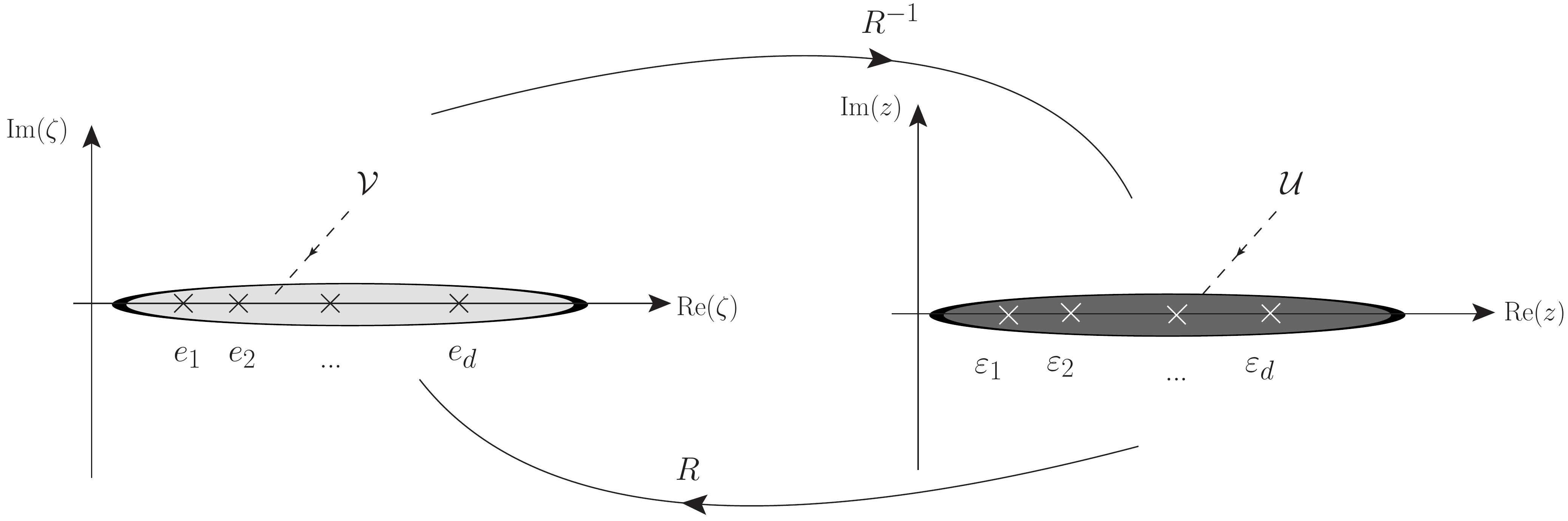}
    \caption{Illustration of the complexification procedure: The
      biholomorphic map $R:\mathcal{U}\to \mathcal{V}$ with
      $R(\varepsilon_k)=e_k$ will later be enlarged
      to a ramified cover $R:\hat{\mathbb{C}}\to \hat{\mathbb{C}}$.
      Functions on $\mathcal{U}$ will meromorphically continue
      to the Riemann sphere $\hat{\mathbb{C}}=\mathbb{C}\cup\{\infty\}$.
      \label{fig:complexification}}
\end{figure}

Figure~\ref{fig:complexification} sketches 
the map $R$.
The rational function $R$ introduces another change of variables.
\begin{definition}\label{defcompl}
Let $G^{(g)},T^{(g)},\tilde{\Omega}^{(g)}$ be the functions in several 
complex variables obtained by the complexification of
Definition~\ref{def:complexification}, genus expansion and by 
admitting multiplicities of the $e_k$. Then functions 
$\mathcal{G}^{(g)},\mathcal{T}^{(g)},\Omega^{(g)}_m$ of several 
complex variables are introduced by
\begin{align*}
  \mathcal{G}^{(g)}(z,w)&:=G^{(g)}(R(z),R(w))\;,\qquad
  \mathcal{G}^{(g)}(z|w):=G^{(g)}(R(z)|R(w))\;,
\\
\mathcal{T}^{(g)}(u_1,...,u_m\|z,w|)
&:=
T^{(g)}(R(u_1),...,R(u_m)\|R(z),R(w)|)\;,
\\
\mathcal{T}^{(g)}(u_1,...,u_m\|z|w|)
&:=
T^{(g)}(R(u_1),...,R(u_m)\|R(z)|R(w)|)\;,
\\
\Omega^{(g)}_{m}(u_1,....,u_m)&:=
\tilde{\Omega}^{(g)}(R(u_1),...,R(u_m)) \;.
\end{align*}
We let $\mathcal{T}^{(g)}(\emptyset\|z,w|):=
\mathcal{G}^{(g)}(z,w)$ and
$\mathcal{T}^{(g)}(\emptyset\|z|w|):=
\mathcal{G}^{(g)}(z|w)$.
\end{definition}
\noindent
Originally defined as holomorphic functions on Cartesian products of
$\mathcal{U}$, we assume (and will show) that 
$\mathcal{G}^{(g)},\mathcal{T}^{(g)},\Omega^{(g)}_m$ 
extend to meromorphic functions on $\hat{\mathbb{C}}
=\mathbb{C}\cup\{\infty\}$.

\subsection{Complexified Dyson-Schwinger Equations}

We now combine the complexification according to 
Definition~\ref{def:complexification} with the change of variables 
of Definition~\ref{defcompl}:
\begin{corollary}
\label{cor:DSEcomplex2}
Let $I=\{u_1,...,u_m\}$. The complexification of
Definition~\ref{def:complexification} 
turns after genus expansion \eqref{genus-expansion}, inclusion 
of multiplicities of $e_i$ and the change of variables given 
in Definition~\ref{defcompl}, which involves 
the rational function $R$ of Theorem~\ref{throm1},
the Dyson-Schwinger equations \eqref{DSE-T2} of 
Proposition \ref{prop:TOm} into equations for meromorphic functions in
several complex variables:
\begin{align}
&(R(w)-R(-z))\mathcal{T}^{(g)}(I\|z,w|)
-\frac{\lambda}{N}\sum_{k=1}^d \frac{r_k
\mathcal{T}^{(g)}(I\|\varepsilon_k,w|)}{R(\varepsilon_k)-R(z)}
\label{DSE-cT2}
\\
&=\delta_{0,m}\delta_{g,0}-\lambda\bigg\{
\sum_{\substack{ I_1\uplus I_2=I,\;g_1+g_2=g \\
    (g_1,I_1)\neq(0,\emptyset)}} \hspace*{-1em}
\Omega^{(g_1)}_{|I_1|+1}(I_1,z)
\mathcal{T}^{(g_2)}(I_2\|z,w|) 
+ \mathcal{T}^{(g-1)}(I,z\|z,w|)
\nonumber
\\
&+
  \sum_{i=1}^m\frac{\partial}{\partial R(u_i)}
\frac{\mathcal{T}^{(g)}(I\setminus u_i\|u_i,w|)}{
R(u_i)-R(z)}
+\frac{\mathcal{T}^{(g-1)}(I\|z|w|)
  -\mathcal{T}^{(g-1)}(I\|w|w|)}{R(w)-R(z)} 
\bigg\}\,.
\nonumber
\end{align}
\end{corollary}
Equation (\ref{DSE-cT2}) reduces for $(g,m)=(0,0)$ to
\begin{align}
\mathcal{G}^{(0)}(z,w)=\frac{1+\frac{\lambda}{N}\sum_{k=1}^d 
\frac{r_k \mathcal{G}^{(0)}(\varepsilon_k,w)}{
R(\varepsilon_k)-R(z)}}{R(w)-R(-z)}\;.
\label{eq:Gzw0}
\end{align}

\begin{corollary}
\label{cor:DSEcomplex1}
Let $I=\{u_1,...,u_m\}$. The complexification of
Definition~\ref{def:complexification} 
turns after genus expansion \eqref{genus-expansion}, inclusion 
of multiplicities of $e_i$ and the change of variables given 
in Definition~\ref{defcompl}, which involves 
the rational function $R$ of Theorem~\ref{throm1},
the Dyson-Schwinger equations \eqref{DSE-T1} of 
Proposition \ref{prop:TOm} into equations for meromorphic functions in
several complex variables:
\begin{align}
&(R(z)-R(-z))\mathcal{T}^{(g)}(I\|z|w|)
-\frac{\lambda}{N}\sum_{k=1}^d r_k\frac{
\mathcal{T}^{(g)}(I\|\varepsilon_k|w|)}{R(\varepsilon_k)-R(z)}
\label{DSE-cT1}
\\
&=-\lambda\bigg\{
\sum_{\substack{ I_1\uplus I_2=I,\;  g_1+g_2=g \\
    (I_1,g_1)\neq (\emptyset,0)}} \hspace*{-1em}
\Omega^{(g_1)}_{|I_1|+1}(I_1,z)
 \mathcal{T}^{(g_2)}(I_2\|z|w|) 
+\mathcal{T}^{(g-1)}(I,z\|z|w|)
 \nonumber
 \\
&+\sum_{i=1}^m
\frac{\partial}{\partial R(u_i)}\frac{
\mathcal{T}^{(g)}(I\setminus u_i\|u_i|w|) }{R(u_i)-R(z)}
+\frac{\mathcal{T}^{(g)}(I\|z,w|)
 -\mathcal{T}^{(g)}(I\|w,w|)}{R(w)-R(z)} 
\bigg\}\;.
\nonumber
\end{align}
\end{corollary}

\begin{remark}\label{rem:2P}
  The DSE of Corollary \ref{cor:DSEcomplex2} has a very intricate
  form. Actually, this structure of DSEs is well-known from the
  2-matrix model. We refer to \cite[eq. (2-19)]{Eynard:2007nq} with
  the correspondence $\Omega \mapsto W$, $\mathcal{T}\mapsto H$ and
  $\sum_k\frac{\mathcal{T}(I\|\varepsilon_k,w|)}{R(\varepsilon_k)-R(z)}
  \mapsto  \tilde{U}$. However, the last term of (\ref{DSE-cT2})
  does not have a corresponding counterpart.
  Thus, the quartic Kontsevich model is a priori similar to the
  2-matrix model, but in some sense richer in its structure. This difference
  could explain why it is (conjecturally) governed by the extension to
  Blobbed Topological Recursion \cite{Borot:2015hna}. The
  fundamental building blocks in the 2-matrix model are the
  $W^{(g)}_{m,0}$ which were proved \cite{Chekhov:2006vd} to satisfy
  topological recursion. For this reason, the
  main interest lies on computation and structure of
  $\Omega^{(g)}_m$.
\end{remark}

\subsection{The DSE for $\Omega^{(g)}_m(u_1,...,u_m)$}

To solve the system of equations (\ref{DSE-cT2}) and
(\ref{DSE-cT1}) we have to 
establish another DSE for $\Omega^{(g)}_m$. The same steps as in
Corollaries~\ref{cor:DSEcomplex2} and
\ref{cor:DSEcomplex1} turn (\ref{DSE-Om}) into
\begin{align}
&\Omega^{(g)}_{|I|+1}(I,z)
= \frac{\delta_{g,0}\delta_{|I|,1}}{(R(u_1)-R(z))^2}
+ \frac{1}{N}\sum_{k=1}^d r_k \mathcal{T}^{(g)}(I\|z,\varepsilon_k|)
 \label{DSE-cOm}
\\
&\hspace*{3.4cm}
- \sum_{i=1}^m \frac{\partial \mathcal{T}^{(g)}(I{\setminus} u_i\|z,u_i|)}{
\partial R(u_i)} 
+ \mathcal{T}^{(g-1)}(I\|z|z|)\,.
\nonumber
\end{align}
We will prove another relation:
\begin{proposition}
\label{prop:Omega0}
Let $I=\{u_1,....,u_m\}$.
The meromorphic functions $\Omega^{(g)}_{m+1}$ satisfy for $(g,m)\neq (0,0)$
the DSE
\begin{align}
&R'(z)\mathfrak{G}_0(z)\Omega^{(g)}_{|I|+1}(I,z)
-\frac{\lambda}{N^2}\sum_{n,k=1}^d r_nr_k
\frac{\mathcal{T}^{(g)}(I\|\varepsilon_k,\varepsilon_n|)}{
(R(\varepsilon_k)-R(z))(R(\varepsilon_n)-R(-z))}
\nonumber
\\
&=\frac{\delta_{g,0}\delta_{|I|,1}}{(R(z)-R(u_1))^2}
-\sum_{\substack{I_1\uplus I_2=I,\; g_1+g_2=g\\
(I_1,g_1)\neq (\emptyset,0)\neq (I_2,g_2)
}} \hspace*{-2em}
\Omega^{(g_1)}_{|I_1|+1}(I_1,z)
\frac{\lambda}{N}\sum_{n=1}^d r_n 
\frac{ \mathcal{T}^{(g_2)}(I_2\|z,\varepsilon_n|)}{
R(\varepsilon_n)-R(-z)} 
\nonumber
\\
&-\sum_{j=1}^m\frac{\partial}{\partial R(u_j)}
\frac{\frac{\lambda}{N}\!\sum_{n=1}^d r_n
\frac{\mathcal{T}^{(g)}(I\setminus u_j\|u_j,\varepsilon_n|) \!\!}{
R(\varepsilon_n)-R(-z)} }{R(u_j)-R(z)}
-\frac{\lambda}{N}\sum_{n=1}^d r_n 
\frac{\mathcal{T}^{(g-1)}(I,z\|z,\varepsilon_n|)}{
R(\varepsilon_n)-R(-z)}
\nonumber
\\
&-\frac{\lambda}{N}\sum_{n=1}^d r_n
\frac{ \mathcal{T}^{(g-1)}(I\|z|\varepsilon_n|)
  - \mathcal{T}^{(g-1)}(I\|\varepsilon_n|\varepsilon_n|)}{
(R(\varepsilon_n)-R(z))(R(\varepsilon_n)-R(-z))}
\nonumber
\\
&  -\sum_{j=1}^m\frac{\partial}{\partial R(u_j)} 
\mathcal{T}^{(g)}(I\setminus u_j\|u_j,z|)
+\mathcal{T}^{(g-1)}(I\|z|z|)\;,
\label{DSE-Omega-complex}
\end{align}
where $\mathfrak{G}_0(z):=\Res_{v\to -z} \mathcal{G}^{(0)}(z,v)dv$.
\begin{proof}
  Take (\ref{DSE-cT2}), multiply it by
  $\frac{r_n}{N (R(w)-R(-z))}$, set $w=\varepsilon_n$ and sum over
  $n$.  The lhs has the term
  $\frac{1}{N}\sum_{n=1}^d r_n
  \mathcal{T}^{(g)}(I\|z,\varepsilon_n|)$, which by \eqref{DSE-cOm}
  equals $\Omega_{m+1}^{(g)}(I,z)$ plus other terms.
  Another $\Omega_{m+1}^{(g)}(I,z)$ arises with a prefactor
  $\frac{\lambda}{N}\sum_n \frac{r_n
    \mathcal{G}^{(0)}(\varepsilon_n,z)}{R(\varepsilon_n)-R(-z)}$ from the case
  $(I_2,g_2)=(\emptyset,0)$ in the second line of  (\ref{DSE-cT2}). Moving it to
  the lhs, we reconstruct a common prefactor $R'(z)\mathfrak{G}_0(z)$
  via \eqref{eq:Gzw0}.
\end{proof}
\end{proposition}

\begin{remark}\label{Rem:2MM}
  The DSE of $\Omega^{(g)}_m$ gives the possibility for a comparison
  with the 2-matrix model as well. The DSE of Prop.~\ref{prop:Omega0}
  coincides (up to two additional terms) with
  \cite[eq. (2-20)]{Eynard:2007nq} after setting $q=p$, where those
  functions are related by $\Omega\mapsto W$, $\mathcal{T}\mapsto H$,
  $\sum_k\frac{\mathcal{T}(I\|\varepsilon_k,w|)}{
    R(\varepsilon_k)-R(z)}\mapsto  \tilde{U}$ and
  $\sum_{k,n}\frac{\mathcal{T}(I\|\varepsilon_k,\varepsilon_n|)}{
    (R(\varepsilon_k)-R(z))(R(\varepsilon_n)-R(-z))}\mapsto E$.
  The two terms violating the exact coincidence are the last term
  and third last term of (\ref{DSE-Omega-complex}).
\end{remark}

\subsection{Solution for $\Omega^{(0)}_2(u,z)$}

For $(g,m)=(0,1)$ the equation in Proposition~\ref{prop:Omega0} reduces to
\begin{align}
&R'(z)\mathfrak{G}_0(z)\Omega^{(0)}_{2}(u,z)
-\frac{\lambda}{N^2}\sum_{n,k=1}^d r_nr_k
\frac{\mathcal{T}^{(0)}(u\|\varepsilon_k,\varepsilon_n|)}{
(R(\varepsilon_k)-R(z))(R(\varepsilon_n)-R(-z))}
\nonumber
\\
&=
-\frac{\partial}{\partial R(u)}
\frac{1+\frac{\lambda}{N}\sum_{n=1}^d r_n
\frac{\mathcal{G}^{(0)}(u,\varepsilon_n)}{
R(\varepsilon_n)-R(-z)} }{R(u)-R(z)}
-\frac{\partial}{\partial R(u)} 
\mathcal{G}^{(0)}(u,z)
\nonumber
\\
&=-\frac{\partial}{\partial R(u)}\big(
\mathcal{G}^{(0)}(u,-z)+
\mathcal{G}^{(0)}(u,z)\big)\;.
\label{eq:Omega02}
\end{align}
The last line follows from (\ref{eq:Gzw0}).
In \cite{Grosse:2019jnv} the following representation was proved:
\begin{align}
\mathcal{G}^{(0)}(z,u)
= \frac{1}{z+u} \Big(1+
\frac{\lambda^2}{N^2} \sum_{k,l,m,n=1}^d \frac{C_{k,l}^{m,n}}{
(z-\widehat{\varepsilon_k}^m)(u-\widehat{\varepsilon_l}^n)}
\Big)\;,
\label{G2p}
\end{align}
where $C_{k,l}^{m,n}=\frac{(\widehat{\varepsilon_k}^m +\widehat{\varepsilon_l}^n) 
r_k r_l \mathcal{G}^{(0)}(\varepsilon_k,\varepsilon_l)}{
R'(\widehat{\varepsilon_k}^m)R'(\widehat{\varepsilon_l}^n)
(R(\varepsilon_l)-R(-\widehat{\varepsilon_k}^m))
(R(\varepsilon_k)-R(-\widehat{\varepsilon_l}^n))}$. On one hand this shows
\begin{align}
\mathfrak{G}_0(z)= 1-
\frac{\lambda^2}{N^2} \sum_{k,l,m,n=1}^d \frac{C_{k,l}^{m,n}}{
(z-\widehat{\varepsilon_k}^m)(z+\widehat{\varepsilon_l}^n)}\;,
\label{frakG0}
\end{align}
on the other hand we have the partial fraction decomposition
\begin{align}
\mathcal{G}^{(0)}(z,u)
= \frac{\mathfrak{G}_0(z)}{z+u} 
+
\frac{\lambda^2}{N^2} \sum_{k,l,m,n=1}^d \frac{C_{k,l}^{m,n}}{
(z+\widehat{\varepsilon_l}^n)(z-\widehat{\varepsilon_k}^m)
(u-\widehat{\varepsilon_l}^n)}\;.
\label{G2p-pfd}
\end{align}
Both together imply:
\begin{proposition}
\label{prop:Om02}
Assume that (for generic $u$) the function 
$\Omega^{(0)}_2(u,z)$ is regular at any zero $z$ of $\mathfrak{G}_0$. Then
\[
\Omega^{(0)}_2(u,z)=\frac{1}{R'(z)R'(u)}\Big(
\frac{1}{(u-z)^2}+\frac{1}{(u+z)^2}\Big)\;.
\]
\begin{proof}
Inserting (\ref{G2p}) and (\ref{G2p-pfd}) into 
(\ref{eq:Omega02}) gives
\begin{align*}
&R'(z)\Omega^{(0)}_2(u,z) -\frac{1}{R'(u)(u+z)^2}-\frac{1}{R'(u)(u-z)^2}
\\
&=
\frac{1}{\mathfrak{G}_0(z)}\bigg[
\frac{\lambda}{N^2} \sum_{n,k=1}^d \frac{r_kr_n\mathcal{T}^{(0)}
(u\|\varepsilon_k,\varepsilon_n|)}{
(R(\varepsilon_k)-R(z))(R(\varepsilon_n)-R(-z))}
\\
&\qquad 
+\frac{\lambda^2}{N^2} \sum_{k,l,m,n=1}^d \frac{C_{k,l}^{m,n}
\big(\frac{1}{R'(u)(u-\widehat{\varepsilon_l}^n)^2}
+\frac{1}{R'(u)(u-\widehat{\varepsilon_k}^m)^2}\big)
}{
(z+\widehat{\varepsilon_l}^n)(z-\widehat{\varepsilon_k}^m)}
\bigg]\;.
\nonumber
\end{align*}
Since $\mathfrak{G}_0(z)$ has poles at every 
$z=\pm \widehat{\varepsilon_n}^j$, 
the rhs of the above equation has poles at most at the zeros of 
$\mathfrak{G}_0$. By assumption, the lhs is regular there. Thus, 
both sides must be constant
by Liouville's theorem and then, when considering 
$z\to \infty$, identically zero. 
\end{proof}
\end{proposition}
\begin{remark}
  \label{rem:Om02}
  Proposition \ref{prop:Om02} indicates that we are on the right
  track.  The solution $\Omega^{(0)}_2(u,z)$ is symmetric, and its
  part $\frac{1}{(u-z)^2}$ is closely related to the Bergman kernel
  $B(u,z)=\frac{du dz}{(u-z)^2}$ of topological recursion. Looking
  back into Remark~\ref{rem:complex} we can be confident that the
  complexification procedure of Definition~\ref{def:complexification}
  is reasonable.
	
  Comparing again with the 2-matrix model, our DSE \eqref{eq:Omega02}
  of $\Omega^{(0)}_{2}$ differs slightly. The last term
  $\mathcal{G}^{(0)}(u,z)$ in \eqref{eq:Omega02}, which generates the
  pole on the antidiagonal, is not present in the 2-matrix model. We
  refer for instance to \cite[eq. (2-20)]{Eynard:2007nq} with the same
  identifications as in Remark \ref{Rem:2MM} and with $g=0$, $q=p$,
  $\mathbf{p_K}=\emptyset$ and $|\mathbf{q_L}|=1$. The last term $H$
  in \cite[eq. (2-20)]{Eynard:2007nq} corresponds to our
  $\mathcal{G}^{(0)}(u,z)$, which is not present for
  $\mathbf{p_K}=\emptyset$. The distinction between the two different
  sets $\mathbf{p_K}$ and $\mathbf{q_L}$ is significant for the
  2-matrix model.
\end{remark}

\section{Recursive Solution}

\label{sec:recursive}

\begin{figure}[ht]
	\includegraphics[width= 1.0\textwidth]{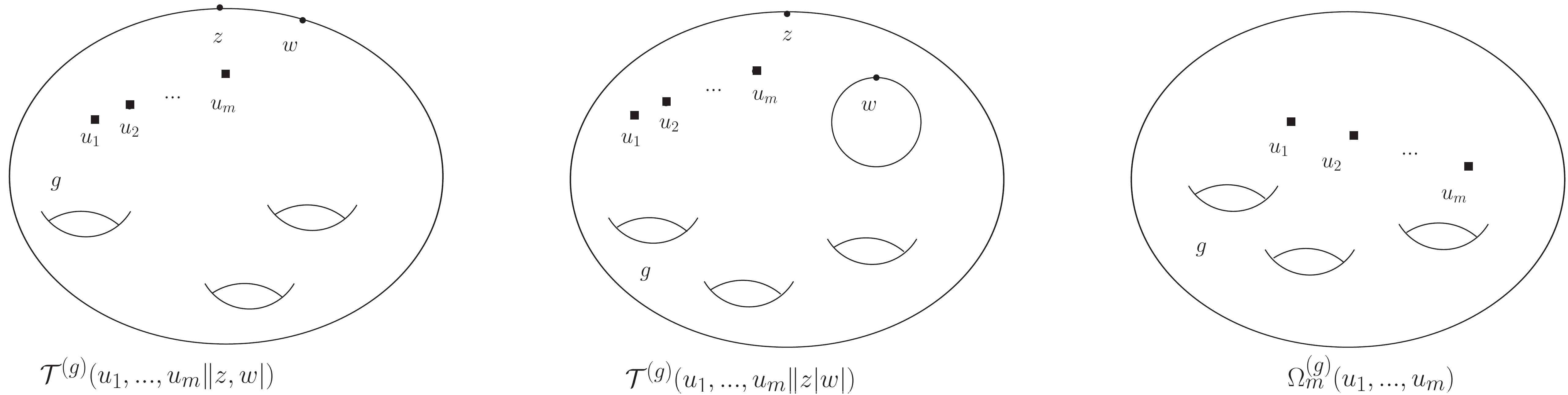}
\caption{The correlation functions relate to topological data of
  Riemann surfaces $X$. The Riemann surface corresponding to
  $\mathcal{T}^{(g)}(u_1,...,u_m\|z,w|)$
 has $g$ handles, $m$ marked points labelled
 $u_1,..,u_m$ and one boundary component with two defects
 (the 1-valent vertices) labelled $z$ and $w$. The Riemann surface
 corresponding to $\mathcal{T}^{(g)}(u_1,...,u_m\|z|w|)$ has $g$ handles,
 $m$ marked points labelled $u_1,...,u_m$ and two boundary components
 each with one defect labelled $z$ or $w$, respectively.
 The Riemann surface corresponding to $\Omega^{(g)}(u_1,...,u_m)$ has
 $g$ handles, $m$ marked points labelled $u_1,...,u_m$ and no
 boundary.
          \label{fig:gtomega}}
\end{figure}

In previous sections we have introduced and studied certain families 
of functions
$\Omega^{(g)}_m(u_1,..,u_m)$,
$\mathcal{T}^{(g)}(u_1,...,u_m\|z,w|)$, $\mathcal{T}^{(g)}(u_1,...,u_m\|z|w|)$
and $\mathcal{G}^{(g)}(z,w)$, $\mathcal{G}^{(g)}(z|w)$. 
As already outlined after Def.~\ref{defT0Om},
the integers $(g,m,b)$ are interpreted as topological data of a
Riemann surface $X$ (see Fig.~\ref{fig:gtomega}):  
\begin{itemize}
\item $g$ is the \emph{genus} of $X$,
\item $m$ is the number of \emph{marked points} on $X$, 
\item $b$ is the number of \emph{boundary components} of $X$; more
  precisely $b=1$ for $\mathcal{T}^{(g)}(u_1,...,u_m\|z,w|)$ and $b=2$
  for $\mathcal{T}^{(g)}(u_1,...,u_m\|z|w|)$.
\end{itemize}
We let $\chi=2-2g-m-b$ be the Euler characteristic of $X$. 

\begin{figure}[h]
  \includegraphics[width= 1.0\textwidth]{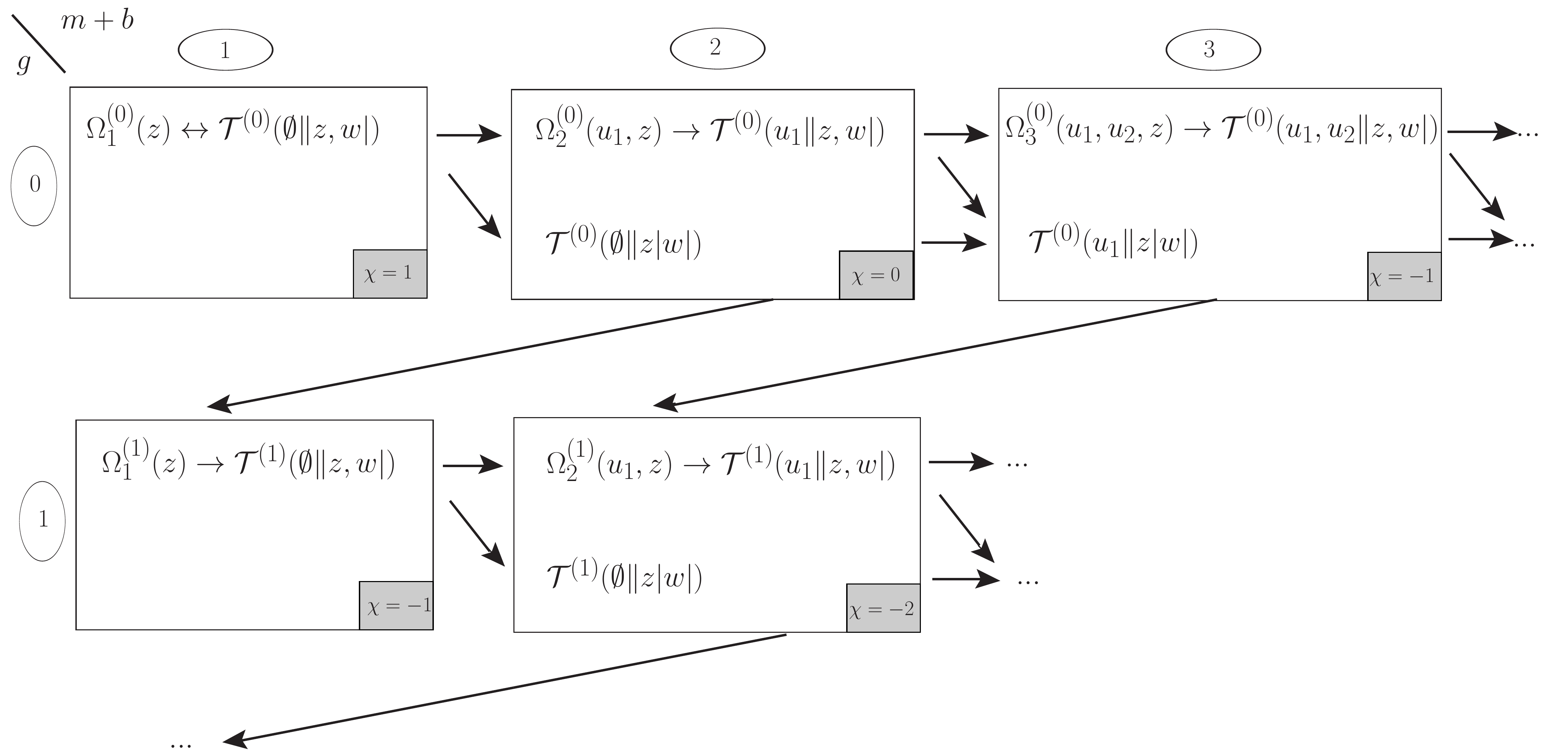}
\caption{The three types of Dyson-Schwinger equations
(\ref{DSE-cT2}), (\ref{DSE-cT1}) and (\ref{DSE-Omega-complex})
are recursively built up and ordered in the
negative Euler characteristic $-\chi$.
This illustration shows for each
$-\chi=2g+m+b-2$, where $b=0$ for $\Omega^{(g)}_m(u_1,...,u_{m})$, $b=1$
for $\mathcal{T}^{(g)}(u_1,...,u_{m1}\|z,w|)$
and $b=2$ for $\mathcal{T}^{(g)}(u_1,...,u_{m}\|z|w|)$, the order
in which to compute previous functions via solution of
(\ref{DSE-cT2}), (\ref{DSE-cT1}) and (\ref{DSE-Omega-complex}), respectively.
The initial case, $\chi=1$ is solved simultaneously by
combining two equations. 
		\label{fig:euler1}}
\end{figure}

The Dyson-Schwinger equations for the generalised correlation
functions $\Omega^{(g)}_m(u_1,..,u_m)$,
$\mathcal{T}^{(g)}(u_1,...,u_m\|z,w|)$ and
$\mathcal{T}^{(g)}(u_1,...,u_m\|z|w|)$ follow a recursion in the Euler
characteristic: To compute a generalised correlation function of
negative Euler characteristic $-\chi=2g+(m+b)-2$ we need generalised
correlation functions of negative Euler characteristic
$-\chi'< -\chi$.  In case of equality one builds
$\mathcal{T}^{(g)}(u_1,...,u_{m-1}\|z,w|)$ from
$\Omega^{(g)}_m(u_1,...,u_{m-1},z)$.
Figure \ref{fig:euler1} shows the recursive structure in
solving the correlation function for small $-\chi$.
This structure
extends in obvious manner to higher topologies $(g,m+b)$.

We will prove that the solution of $\mathcal{T}^{(g)}(u_1,...,u_m\|z,w|)$ and
$\mathcal{T}^{(g)}(u_1,...,u_m\|z|w|)$ are obtained by a simple evaluation of
residues. For that the following analyticity property is necessary:
\begin{lemma}\label{lem:analy}
 Let $I=\{u_1,...,u_m\}$.  The generalised correlation functions
 $\mathcal{T}^{(g)}(I\|w,z)|$, $\mathcal{T}^{(g)}(I\|w|z|)$,
and
  $\Omega^{(g)}_m(I)-\frac{\delta_{g,0}\delta_{m,2}}{(R(u_1)-R(u_2))^2}$
are analytic at any two coinciding variables in the domain $\mathcal{U}$
which includes all $\varepsilon_k$ but excludes $0$.
\begin{proof}
  The analyticity property is proved by induction in the Euler
  characteristic. It is true for
  $\mathcal{G}^{(0)}(z,w)=\mathcal{T}^{(0)}(\emptyset\|z,w|)$ by the
  explicit form \eqref{Gzw-final} and
  $\mathcal{G}^{(0)}(z|w)=\mathcal{T}^{(0)}(\emptyset\|z|w|)$ by the
  explicit form given in \cite{Schurmann:2019mzu-v3}. From Proposition
  \ref{prop:Om02}, we have the limit
  $\lim_{u\to z} \big(\Omega^{(0)}_2(u,z)-\frac{1}{(R(u)-R(z))^2}
  \big) =\frac{1}{4z^2(R'(z))^2 }-\frac{R'''(z)}{6 (R'(z))^3}
  +\frac{(R''(z))^2}{4 (R'(z))^4}$.
		
  Then by induction all terms $\mathcal{T}^{(g)}$ and $\Omega^{(g_1)}_{|I_1|+1}$
  with $2g_1+|I_1|\geq 2$ are analytic at coinciding arguments.
The only critical terms for $z\to u_i$ arise in combination
\begin{align*}
&\Omega^{(0)}_2(u_i,z)\mathcal{T}^{(g)}(I\backslash
  u_i\|z,w|)
+\frac{\partial}{\partial R(u_i)}\Big(\frac{\mathcal{T}^{(g)}(I\backslash
  u_i\|u_i,w|)}{R(u_i)-R(z)}\Big)
\\
&= \Big(\Omega^{(0)}_2(u_i,z)-\frac{1}{(R(u_i)-R(z))^2}\Big)
\mathcal{T}^{(g)}(I\backslash
  u_i\|z,w|)
\\
&+\frac{\partial}{\partial R(u_i)}\Big(\frac{\mathcal{T}^{(g)}(I\backslash
  u_i\|u_i,w|)-\mathcal{T}^{(g)}(I\backslash
  u_i\|z,w|)}{R(u_i)-R(z)}\Big)\;,
\end{align*}
which is analytic for $z\to u_i$. Regularity 
for $w\to u_i$ is obvious, and regularity for $z\to w$ holds by 
induction. 
Thus, $\mathcal{T}^{(g)}(I\|z,w|)$ is 
regular for any $z,w\to u_i$ and $z\to w$. 
Similarly for \eqref{DSE-cT1}.
The same argument in the rhs of 
Proposition~\ref{prop:Omega0} shows analyticity of
$\Omega^{(g)}_{|I|+1}$ with $2g+I\geq 2$ at any $u_i\to u_j$.
\end{proof}
\end{lemma}

\subsection{Recursive Solution of 
\texorpdfstring{$\mathcal{T}^{(g)}(u_1,...,u_m\|z,w|)$
}{T\textasciicircum\{(g)\}(u\_1,...,u\_m||z,w|)} and \texorpdfstring{$\mathcal{T}^{(g)}(u_1,...,u_m\|z|w|)$
}{T\textasciicircum\{(g)\}(u\_1,...,u\_m||z|w|)}}\label{Sec:RecursT}

The main observation when solving the DSEs (\ref{DSE-cT2}) (or
(\ref{DSE-cT1})) is the rationality of the second term of the lhs in
$R(z)$. After multiplication with
$\prod_{k=1}^d(R(\varepsilon_k)-R(z))$, the resulting second term
becomes a polynomial in $R(z)$ of degree $d-1$. This observation
suggests an application of the interpolation formula (see
Lemma~\ref{lem:interpol}), where the $d$ distinct numbers are chosen
as $x_j=R(-\hat{w}^j)$ (or $x_j=R(\alpha_j)$) in order to let the
first term of the lhs vanish at $z=-\hat{w}^j$ (or at
$z=\alpha_j$). The analyticity is easily shown by induction and
similar to Lemma \ref{lem:analy}.  For the sake of readability we have
outsourced Propositions~\ref{prop2} and \ref{prop1+1} and their proofs
to the Appendix \ref{App:RecT}. Here we only give their corollaries:
\begin{corollary}\label{cor2+} 	Let $I=\{u_1,...,u_m\}$.
The generalised 2-point function is given by
\begin{align}
\mathcal{T}^{(g)}(I\|z,w|)
&=\lambda\mathcal{G}^{(0)}(z,w)\Res\displaylimits_{t\to z,-\hat{w}^j,u_i,w}
\frac{R'(t)\, dt}{(R(z)-R(t))(R(w)-R(-t))\mathcal{G}^{(0)}(t,w)} 
\nonumber
\\
&\times\bigg[\sum_{\substack{I_1\uplus I_2=I \\g_1+g_2=g\\
(I_1,g_1)\neq (\emptyset,0)}} 
\Omega^{(g_1)}_{|I_1|+1}(I_1,t)\mathcal{T}^{(g_2)}(I_2\|t,w|)
\nonumber
\\[-4ex]
&\qquad\qquad\qquad\qquad+ \mathcal{T}^{(g-1)}(I,t\|t,w|)
+\frac{\mathcal{T}^{(g-1)}(I\|t|w|)}{R(w)-R(t)}
		\bigg].
\nonumber
\end{align}
\end{corollary}
\begin{figure}[h]
	\includegraphics[width= 1.0\textwidth]{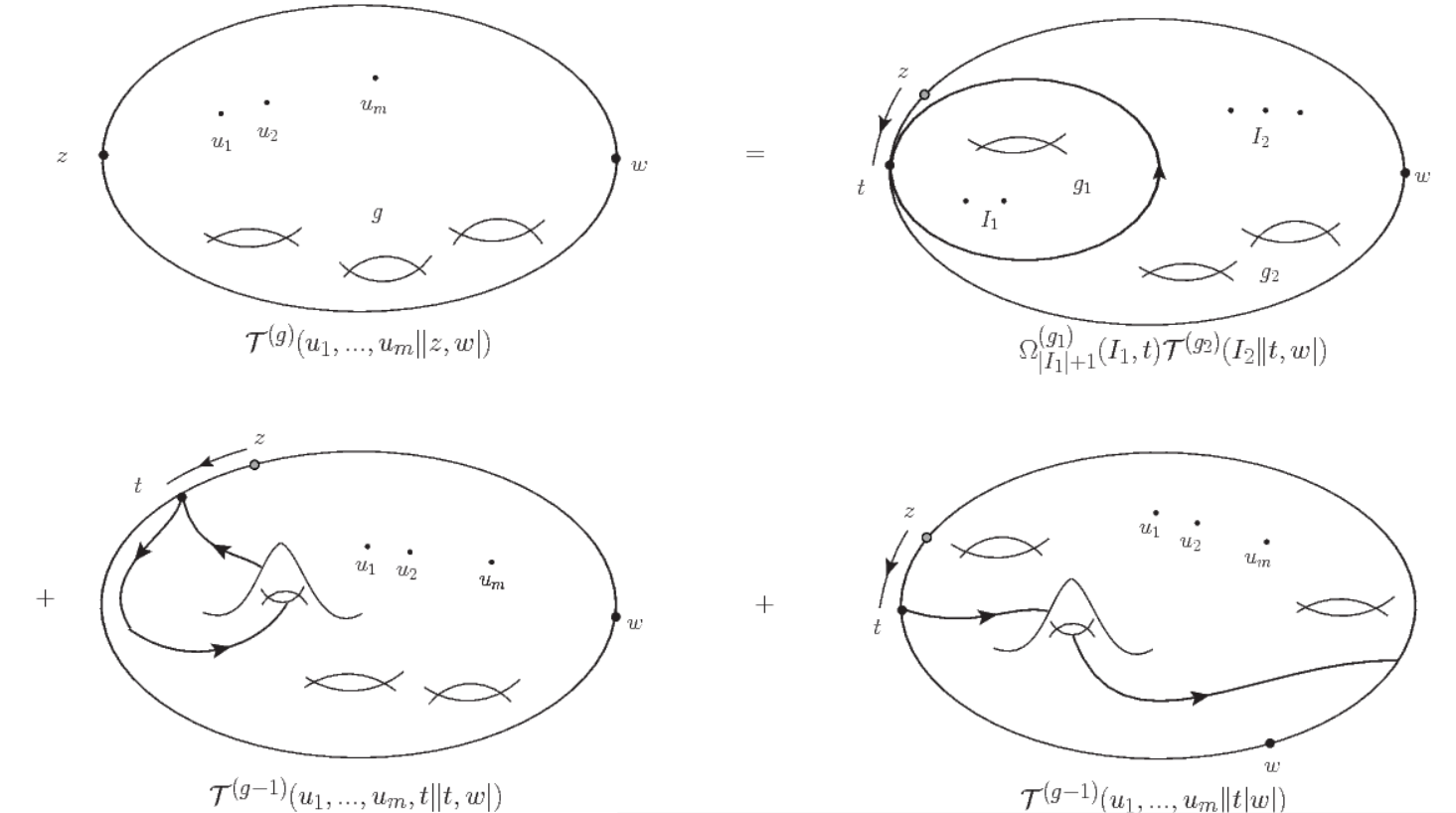}
	\caption{Graphical illustration of Corollary \ref{cor2+}.
		\label{fig:rec1}}
\end{figure}
Instead of providing the technical proof, we have decided to give a
graphical interpretation of Corollary \ref{cor2+} by cutting the
Riemann surface corresponding to the generalised 2-point
function. The cut operation itself, as shown in Fig.~\ref{fig:rec1},
generates for the generalised 2-point function the factor
\begin{align*}
	\lambda\mathcal{G}^{(0)}(z,w)
	\frac{R'(t)\, dt}{(R(z)-R(t))(R(w)-R(-t))\mathcal{G}^{(0)}(t,w)} 
\end{align*}
together with a residue operation of $t$ at $z,-\hat{w}^j, u_i$ and
$w$. Now, the generalised 2-point function can be cut in three
topologically different ways:

\begin{enumerate}
\item The cut starts from $t$ and ends at $t$ by encircling
  the set $I_1\subset I$ of marked points and $g_1$ handles.
  This separates $\mathcal{T}^{(g)}(I\|z,w|)$ into
  $\Omega^{(g_1)}_{|I_1|+1}(I_1,t)$ and $\mathcal{T}^{(g_2)}(I_2\|t,w|)$.
  Take the sum over all possibilities with
  Euler characteristic $\chi\leq 0$. 

\item The cut starts at $t$, paces through a handle and ends
  again at $t$. This removes the handle (reduces the genus by $1$) at
  expense of an additional marked point labelled $t$.

\item The cut starts at $t$, paces through a handle and ends at
  the boundary next to $w$ (not at $t$). This reduces the genus by
  one and generates the factor $\frac{1}{R(w)-R(t)}$ and two
  separated boundaries with one defect on each. 
\end{enumerate}
It seems that another possible case would be the variant of 3.
where the cut does
not pace through a handle but ends next to $w$. This would generate
two separated correlation functions of the form
$\mathcal{T}^{(g')}(I'\|t|)$, but these do no exist since the
quartic Kontsevich model has no 1-point function (and
therefore no generalised 1-point function). 

\begin{remark}
  \label{rem:cut}
  This graphical description was already invented for the 2-matrix
  model to understand graphically any correlation function as a
  recursion depending on correlation functions of lower topology
  \cite{Eynard:2007gw}. However, in the 2-matrix model two different
  sets of marked points exists, whereas the quartic Kontsevich model
  has a mixing of those sectors. In general, a proof of graphical
  rules is achieved by direct derivation via DSEs.
\end{remark}

\begin{corollary}\label{cor1+}
Let $I=\{u_1,...,u_m\}$.
Proposition \ref{prop1+1} is equivalent to
 \begin{align*}
 &\mathcal{T}^{(g)}(I\|z|w|)
\\
& =\frac{\lambda\prod_{j=1}^d\frac{R(z)-R(\alpha_j)}{
R(z)-R(\varepsilon_j)}}{(R(z)-R(-z))}
\Res\displaylimits_{t\to z ,\alpha_j,u_i,w}
\frac{R'(t)\, dt\prod_{k=1}^d(R(t)-R(\varepsilon_k))}{(R(z)-R(t)) 
\prod_{k=1}^d(R(t)-R(\alpha_k))}  
\\
&\times\bigg[\sum_{\substack{I_1\uplus I_2=I\\g_1+g_2=g\\ 
		(I_1,g_1)\neq (\emptyset,0)
}}\!\!\!\!\! \Omega^{(g_1)}_{|I_1|+1}(I_1,t) \mathcal{T}^{(g_2)}(I_2\|t|w|) 
+ \mathcal{T}^{(g-1)}(I,t\|t|w|)
+\frac{\mathcal{T}^{(g)}(I\|t,w|)}{R(w)-R(t)}\bigg]\:.
 \end{align*}
\end{corollary}
\begin{figure}[h]
	\includegraphics[width= 1.0\textwidth]{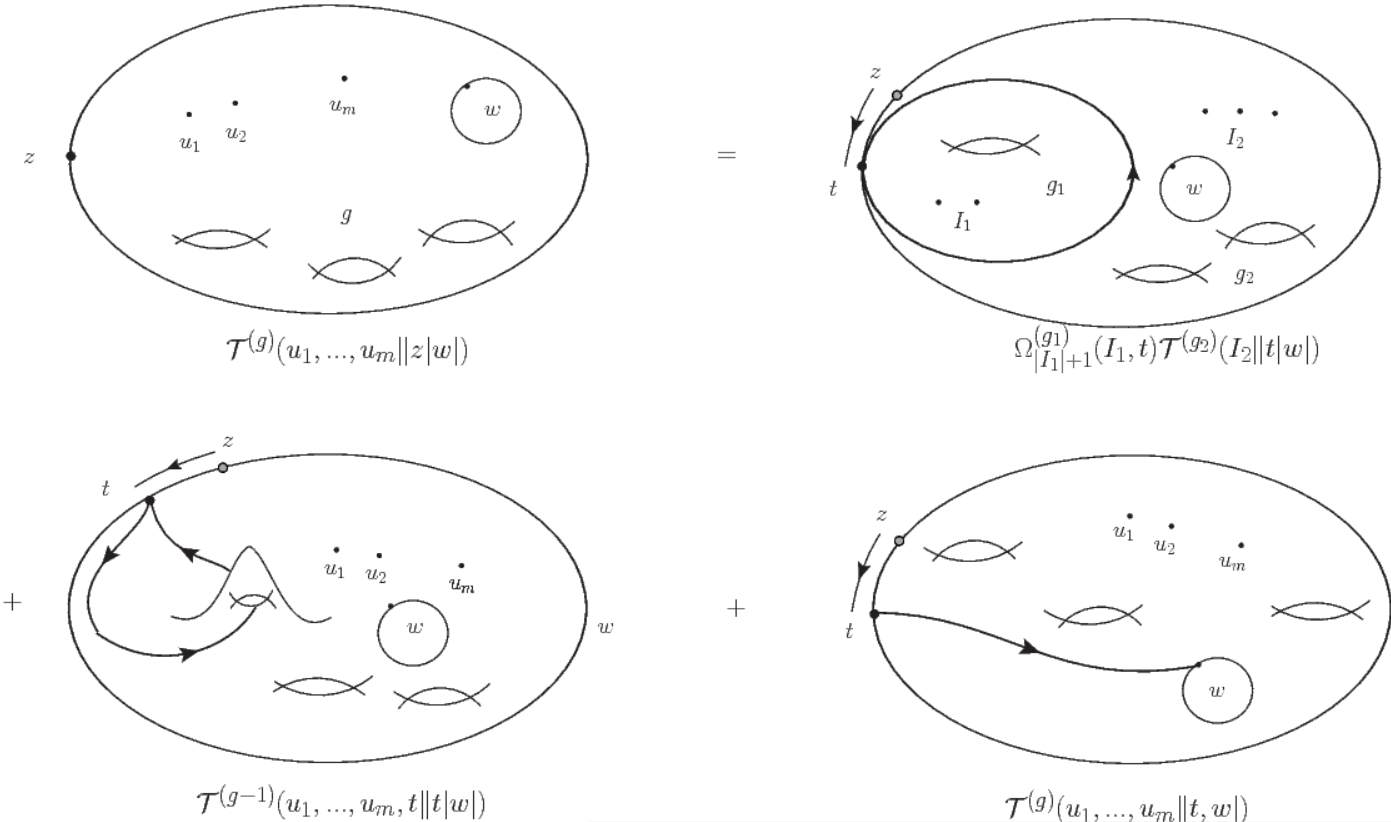}
	\caption{Graphical illustration of Corollary \ref{cor1+}.
		\label{fig:rec2}}
\end{figure}
The graphical intepretation of Corollary \ref{cor1+} differs slightly
from Corollary \ref{cor2+} since the initial topological data of
$\mathcal{T}^{(g)}(I\|z|w|)$ differs from
$\mathcal{T}^{(g)}(I\|z,w|)$. The cutting operation itself generates
the factor
\begin{align*}
	\frac{\lambda\prod_{j=1}^d\frac{R(z)-R(\alpha_j)}{
			R(z)-R(\varepsilon_j)}}{(R(z)-R(-z))}
	\frac{R'(t)\, dt\prod_{k=1}^d(R(t)-R(\varepsilon_k))}{(R(z)-R(t)) 
		\prod_{k=1}^d(R(t)-R(\alpha_k))},
\end{align*}
where residues of $t$ are taken at $z ,\alpha_j,u_i,w$. This is due to
the fact that the starting boundary has only one defect. The
first two cases are essentially the same as in Corollary
\ref{cor2+}. However, the third case differs since two boundaries each
with one defect are present. The third case has a cut starting
at $t$ and merging into the second boundary next to $w$. Both
boundaries are merged to a single boundary with two defects.
Furthermore, the cut not ending at its starting point
generates again a factor $\frac{1}{R(w)-R(t)}$, similar to the third
case of Corollary \ref{cor2+}.

\begin{remark}
  We have focused the computation in this paper to the generalised
  2-point and $1+1$-point function. 
  In a previous version of this paper (e.g.\
  \url{https://arxiv.org/abs/2008.12201v3})
 we have also defined more general
  correlation functions. These can be solved exactly with the same
  graphical rules, but with more possibilities in cutting the Riemann
  surfaces into different topologies.
\end{remark}

\subsection{Recursive Solution for 
\texorpdfstring{$\Omega^{(g)}_m$}{Omega\textasciicircum\{(g)\}\_m} under 
 Assumptions on its Poles}

The solution of the DSE for $\Omega^{(g)}_m(u_1,...,u_m)$ in Proposition
\ref{prop:Omega0} to low $2g+m$ (see Appendix~\ref{app:solution}) suggests
the following:
\begin{conjecture}
\label{conj:Omega-poles}
The function $\Omega^{(g)}_{m+1}(u_1,...,u_m,z)$ is holomorphic in every 
$z\in \{\pm \widehat{u_l}^j,\pm \widehat{\varepsilon_k}^j,
\pm \varepsilon_k,\pm \alpha_k\}$, where 
$k,j=1,...,d$ and $l=1,...,m$.
\end{conjecture}
We prove this conjecture in Appendix~\ref{app:poles-g0} 
for the planar sector $g=0$. Conjecture
\ref{conj:Omega-poles} and 
Lemma~\ref{lem:analy} imply that 
$\Omega^{(g)}_m(u_1,...,u_m,z)$ can only have poles at 
$z=\{0,-u_1,...,-u_m,\beta_1,...,\beta_{2d}\}$, where the $\beta_i$ are
the ramification points of $R$ given by $R'(\beta_i)=0$. 
Being by an easy induction argument a rational function, 
$\Omega^{(g)}_m(u_1,...,u_m,z)$ must coincide with the partial
fraction
decomposition about its set of poles. This partial fraction
decomposition can be written as a residue 
which applied to Proposition~\ref{prop:Omega0} gives:
\begin{corollary}
\label{corr:Omega-residue}
Assume Conjecture \ref{conj:Omega-poles} is true for all $(g,m)$.
Then 
for $(g,m)\neq (0,0)$ and $(g,m)\neq (0,1)$ one has
\begin{align*}
&R'(z)\Omega^{(g)}_{m+1}(u_1,...,u_m,z)
\\
&=\Res\displaylimits_{q\to 0,-u_l,\beta_i}
\frac{dq}{(q-z)\mathfrak{G}_0(q)}
\bigg[\sum_{\substack{I_1\uplus I_2=\{u_1,...,u_m\}\\ g_1+g_2=g\\
(I_1,g_1)\neq (\emptyset,0)\neq (I_2,g_2)}} 
\hspace*{-1cm}
\Omega^{(g_1)}_{|I_1|+1}(I_1,q)\frac{\lambda}{N}\sum_{n=1}^d 
\frac{ r_n\mathcal{T}^{(g_2)}(
I_2\|q,\varepsilon_n|)}{R(\varepsilon_n)-R(-q)}
\\
&
+\sum_{j=1}^m \frac{\partial}{\partial R(u_j)}
\mathcal{T}^{(g)}(u_1,..\check{u_j}..,u_m\|u_j,q|)
+\frac{\lambda}{N}\sum_{n=1}^d 
\frac{r_n\mathcal{T}^{(g-1)}(u_1,...,u_m,q\|q,\varepsilon_n|)}{
R(\varepsilon_n)-R(-q)}
\\
&
+\frac{\lambda}{N}\sum_{n=1}^d \frac{r_n 
\mathcal{T}^{(g-1)}(u_1,...,u_m\|q|\varepsilon_n|)}{
(R(\varepsilon_n)-R(q))(R(\varepsilon_n)-R(-q))}
-\mathcal{T}^{(g-1)}(u_1,...,u_m\|q|q|)\bigg]\,.
\end{align*}
\end{corollary}

We evaluate in Appendix~\ref{app:solution} the residues of
Corollary~\ref{corr:Omega-residue} for $\Omega^{(0)}_{3}(u_1,u_2,z)$,
$\Omega^{(0)}_{4}(u_1,u_2,u_3,z)$ and $\Omega^{(1)}_{1}(z)$. For
convenience we collect these results in
Subsection~\ref{sec:omega-low}.  The outcome suggests that the
$\Omega^{(g)}_m$ are closely related to structures in
blobbed topological recursion (BTR) \cite{Borot:2015hna}. We review in
Appendix~\ref{recbtr} central aspects of the BTR.  To make contact
with BTR we reformulate the solution formulae of our loop equations in
terms of \emph{meromorphic differential forms}:
\begin{definition}
  For integers $g\geq 0$ and $m\geq 1$ we introduce meromorphic
  differentials $\omega_{g,m}$ on $\hat{\mathbb{C}}^m$ by
\begin{align}
\omega_{0,1}(z) &:= -R(-z)R'(z)dz \;,
\\
  \nonumber
  \omega_{g,m}(z_1,\dots,z_m) 
&:= \lambda^{2-2g-m}
\Omega^{(g)}_m(z_1,\dots,z_m) \prod_{j=1}^m R'(z_j) dz_j\quad
\text{for $2g+m\geq 2$}\;.
\end{align}
\end{definition}

Corollary \ref{corr:Omega-residue} takes the following form:
\begin{align}
&\omega_{g,m+1}(u_1,...,u_m,z)
\label{res-omgm}
=\Res\displaylimits_{q\to 0,-u_l,\beta_i}
\frac{dz}{(q-z)\mathfrak{G}_0(q)}
\bigg[
\\
&\sum_{\substack{I_1\uplus I_2=\{u_1,...,u_m\}\\ g_1+g_2=g\\
(I_1,g_1)\neq (\emptyset,0)\neq (I_2,g_2)}} 
\hspace*{-0.5cm}
\omega_{g_1,|I_1|+1}(I_1,q)
\frac{\lambda}{N}\sum_{n=1}^d 
\frac{ r_n\mathfrak{t}_{g_2,|I_2|}(
I_2\|q,\varepsilon_n|)}{R'(q)(R(\varepsilon_n){-}R({-}q))}
\nonumber
\\
&
+\sum_{j=1}^m 
d_{u_j}[\mathfrak{t}_{g,m-1}(u_1,..\check{u_j}..,u_m\|u_j,q|)]dq
\nonumber
\\
&+\frac{\lambda}{N}\sum_{n=1}^d 
\frac{r_n\mathfrak{t}_{g-1,m+1}(u_1,...,u_m,q\|q,\varepsilon_n|)}{
R'(q)(R(\varepsilon_n)-R(-q))}
\nonumber
\\
&
+\frac{\lambda}{N}\sum_{n=1}^d \frac{r_n 
\mathfrak{t}_{g-1.m}(u_1,...,u_m\|q|\varepsilon_n|)}{
(R(\varepsilon_n)-R(q))(R(\varepsilon_n)-R(-q))}dq
-\mathfrak{t}_{g-1,m}(u_1,...,u_m\|q|q|)dq\bigg]\,,
\nonumber
\end{align}
where $d_{u_j}$ is the exterior differential in $u_j$ 
and
$\mathfrak{t}_{g,m}(u_1,...,u_m\|z,w|)
:= \lambda^{2g-m}
\mathcal{T}^{(g)}(z_1,...,z_m\|z,w|) \prod_{j=1}^m R'(z_j) dz_j$
as well as $\mathfrak{t}_{g,m}(u_1,...,u_m\|z|w|)
:= \lambda^{-1-2g-m-}
\mathcal{T}^{(g)}(z_1,...,z_m\|z|w|) \prod_{j=1}^m R'(z_j) dz_j$.

The residue in (\ref{res-omgm}) provides a natural decomposition 
\begin{align}
\label{decom}
\omega_{g,m+1}(u_1,...,u_m,z)=
\mathcal{P}_{z}\omega_{g,m+1}(u_1,...,u_m,z)
+\mathcal{H}_{z}\omega_{g,m+1}(u_1,...,u_m,z)
\end{align}
into a part
\[
\mathcal{P}_{z}\omega_{g,m+1}(u_1,...,u_m,z)
:=\sum_{i=1}^{2d}\Res\displaylimits_{q\to\beta_i}
\frac{dz}{(q-z)\mathfrak{G}_0(q)}\Big[\dots\Big]
\]
whose poles in $z$ are located only at the ramification points 
$\beta_i$ of $R$ and a part
\[
\mathcal{H}_{z}\omega_{g,m+1}(u_1,...,u_m,z)
:=\Big(\Res\displaylimits_{q\to 0}+\sum_{l=1}^m 
\Res\displaylimits_{q\to -u_l}\Big)
\frac{dz}{(q-z)\mathfrak{G}_0(q)}
\Big[\dots\Big]
\]
which is holomorphic in $z$ at the ramification points. We will
discuss these projectors in the context of blobbed topological
recursion in Appendix \ref{recbtr}, too.

\subsection{Solution of 
\texorpdfstring{$\omega_{g,m}$}{omega\_\{g,m\}} to Low Degree}

\label{sec:omega-low}

This subsection lists the results for 
$\omega_{0,3}$, $\omega_{0,4}$ and $\omega_{1,1}$ obtained by
evaluating the residues in the system 
(\ref{res-omgm}), (\ref{res-tgm2}) and (\ref{res-tgm1}). 
Appendix \ref{app:solution} gives details about 
the procedure and provides a few intermediate results. 

We let $\sigma_i$ be the local Galois involution near the ramification
point $\beta_i$, i.e.\ $R(z)=R(\sigma_i(z))$,
$\lim_{z\to \beta_i} \sigma_i(z)=\beta_i$ and
$\sigma_i\neq \mathrm{id}$. We let $B(z,w)=\frac{dz\,dw}{(z-w)^2}$ be
the Bergman kernel and define $x(z)=R(z)$ and
$y(z)=-R(-z)$. Moreover, we introduce two kernel forms
\begin{align}
  K_i(z,q)&=\frac{\frac{1}{2}\int_{q'=\sigma_i(q)}^{q'=q}B(z,q') }{
    (y(q)-y(\sigma_i(q))dx(\sigma_i(q))}\;, &
  K_u(z,q)&=\frac{\int_{q'=-q}^{q'=-u} B(z,q')}{y(-q)-y(-u)}\;.
\end{align}
The evaluation of $\omega_{0,3}$ and $\omega_{0,4}$ in
Appendix~\ref{app:solution} confirms\footnote{Conjecture
  \ref{conj-omplanar} is now a Theorem proved in
  \cite{Hock:2021tbl}. Several equivalent expressions for the residues at
  $z=u_k$ are given there.}
for $m\in \{2,3\}$:
\begin{conjecture}
\label{conj-omplanar}
  For any $I=\{u_1,..,u_m\}$ with $m\geq 2$ one has
\begin{align}
&\omega_{0,m+1}(I,z)
\\
&=
\sum_{i=1}^{2d}\Res\displaylimits_{q\to \beta_i} K_i(z,q)
\sum_{\substack{I_1\uplus I_2=I\\  I_1,I_2\neq \emptyset}}
\omega_{0,|I_1|+1}(I_1,q) \omega_{0,|I_2|+1}(I_2,\sigma_i(q))
\nonumber
\\
&+
\sum_{k=1}^{m} d_{u_k}
\Big(\Res\displaylimits_{q\to u_k} K_{u_k}(z,q) \!\!\!\!
\sum_{\substack{I_1\uplus ... \uplus I_s=I{\setminus} u_k\\ I_1,...,I_s \neq  \emptyset}}
\!\!\!\!
\frac{\omega_{0,|I_1|+1}(I_1,-q)}{y(q)-y(u_k)}
\prod_{r=2}^s \frac{\omega_{0,|I_r|+1}(I_r,u_k)}{(y(q)-y(u_k))dx(u_k)}
  \Big).
\nonumber
\end{align}
\end{conjecture}
We remark that $\omega_{0,m}$ given by Conjecture~\ref{conj-omplanar}
automatically satisfy the linear and quadratic loop equations given later in
Definition~\ref{def:absloopeq} or inside Conjecture~\ref{mconj}.

The residues at $q=\beta_i$ can be evaluated with the formulae given in
Appendix~\ref{app:Galois};
the residues at $q=u_k$ are straightforward. In terms of
\begin{align}
x_{n,i}:=\frac{R^{(n+2)}(\beta_i)}{R''(\beta_i)}\;,\qquad
y_{n,i}:=(-1)^n \frac{R^{(n+1)}(-\beta_i)}{R'(-\beta_i)}\;.
\end{align}
and with $Q(u;z):= \frac{1}{u-z}+\frac{1}{u+z}$ arising in 
$\omega_{0,2}(u,z)=-d_u[Q(u;z)]dz$, we find
\begin{align}
\mathcal{P}_z\omega_{0,3}(u_1,u_2,z)&=
-d_{u_1}d_{u_2}\Big[\sum_{i=1}^d 
\frac{Q(u_1;\beta_i)Q(u_2;\beta_i)}{
R'(-\beta_i)R''(\beta_i) (z-\beta_i)^2}\Big]dz\;,
\label{Pom03}
\\
\mathcal{H}_z\omega_{0,3}(u_1,u_2,z)
&= d_{u_1} \Big[\frac{\omega_{0,2}(u_2,u_1)}{(dR)(u_1)
  R'(-u_1)(z+u_1)^2}\Big]  
  +[u_1\leftrightarrow u_2]
  \label{Hom03}
\end{align}
and 
\begin{align}
&\mathcal{P}_z\omega_{0,4}(u_1,u_2,u_3,z)
\label{Pom04}
\\
  &=d_{u_1}d_{u_2}d_{u_3}\Big[
  \sum_{i=1}^{2d} \frac{Q(u_1;\beta_i)Q(u_2;\beta_i)}{
(R''(\beta_i))^2 (R'(-\beta_i))^2} 
\Big\{-
\frac{Q(u_3;\beta_i)}{(z-\beta_i)^4} 
+\frac{Q(u_3;\beta_i)}{(z-\beta_i)^3}
\frac{x_{1,i}}{3}
\nonumber
\\
& ~
+\frac{Q'(u_3;\beta_i)}{(z-\beta_i)^2}\frac{x_{1,i}}{2}
- \frac{Q''(u_3;\beta_i)}{2(z-\beta_i)^2}
+\frac{Q(u_3;\beta_i)}{(z-\beta_i)^2}
\Big(\frac{x_{2,i}}{6}
-\frac{x_{1,i}^2}{4}
-\frac{y_{1,i}x_{1,i}}{6}
+\frac{y_{2,i}}{6}
\Big)
\Big\}
\nonumber
\\
&- \frac{Q(u_3;\beta_i)}{R'(-\beta_i)R''(\beta_i)
(z-\beta_i)^2}
\Big\{
\frac{Q(u_2;u_1)}{R'(u_1)R'(-u_1)(u_1+\beta_i)^2}
\nonumber
\\
& \quad
+\frac{Q(u_1;u_2)}{R'(u_2)R'(-u_2)(u_2+\beta_i)^2}
+ \sum_{n\neq i} \frac{Q(u_1;\beta_n)Q(u_2;\beta_n)}{
R'(-\beta_n)R''(\beta_n)(\beta_i-\beta_n)^2}\Big\}
\nonumber
\\[-1ex]
&+[u_3\leftrightarrow u_1]+[u_3\leftrightarrow u_2]
     \Big]dz\;,
     \nonumber
\\[1ex plus 2pt]
&\mathcal{H}_z\omega_{0,4}(u_1,u_2,u_3,z)  
\label{Hom04}
\\
&=d_{u_3}\Big[
\frac{2\omega_{0,2}(u_1,u_3)\omega_{0,2}(u_2,u_3)}{(dR(u_3))^2 (R'(-u_3))^2}
\Big(-\frac{1}{(z+u_3)^3}+\frac{R''(-u_3)}{2R'(-u_3)(z+u_3)^2}\Big)
\nonumber
\\
&\qquad\qquad+\frac{\omega_{0,3}(u_1,u_2,u_3)}{dR(u_3) R'(-u_3) (z+u_3)^2}
\Big]dz
+[u_3\leftrightarrow u_1]+[u_3\leftrightarrow u_2]\;,
\nonumber
\end{align}
where in $Q'(u;z),Q''(u;z)$ the derivative is with respect to the
second argument. We have simplified (\ref{Hom04}) using the reflection
(\ref{flip-W03}).

We also have a result for $g=1$:
\begin{proposition}
\label{prop:om11}
The solution of \eqref{res-omgm} for $m=0$ and $g=1$ is
\begin{align*}
\mathcal{P}_z\omega_{1,1}(z)
&=\sum_{i=1}^{2d} \frac{dz}{R'(-\beta_i) R''(\beta_i)}
\Big\{
-\frac{1}{8  (z-\beta_i)^4} 
+ \frac{x_{1,i}}{24 (z-\beta_i)^3}
\nonumber
  \\
  &+\frac{1}{(z-\beta_i)^2}\Big(
\frac{x_{2,i}}{48}
- \frac{x_{1,i}^2}{48}
-\frac{x_{1,i}y_{1,i}}{48}
+
\frac{y_{2,i}}{48}
-  \frac{1}{8 \beta_i^2}\Big)
    \Big\}
\;,
  \\
 \mathcal{H}_z\omega_{1,1}(z)&=
-\frac{dz}{8 (R'(0))^2 z^3}
+\frac{R''(0)dz}{16(R'(0))^3z^2}\;.
\end{align*}
The differential form $\omega_{1,1}$
satisfies for $z$ near $\beta_i$ the loop equations
given in Definition~\ref{def:absloopeq}.
\end{proposition}

\section{Main Conjecture}
\label{mainconj}

We established with the proof of Conjecture~\ref{conj-omplanar}
for $m=2$ and $m=3$
as well as with Proposition
\ref{prop:om11} the unexpected result that all 
$\omega_{g,m}$ evaluated so far satisfy the linear and quadratic loop
equations. This is very unlikely a mere coincidence, which suggests: 
\begin{conjecture}
\label{mconj}
Let 
$R:\hat{\mathbb{C}}\to \hat{\mathbb{C}}$ be 
the ramified covering defined in 
\eqref{Gzw-final}. Let  
$\beta_1,...,\beta_{2d}$ be its ramification points and 
$\sigma_i$ the corresponding local Galois
involution in the vicinity of $\beta_i$.
For all $g\geq 0$ and $m\geq 1$, the meromorphic 
differentials $\omega_{g,m}$ given by 
$\omega_{0,1}(z)=-R(-z)R'(z)dz$, 
$\omega_{0,2}(u_1,z)=\frac{du_1\,dz}{(u_1-z)^2}
+\frac{du_1\,dz}{(u_1+z)^2}$ 
and for $2-2g-m< 0$ by evaluation of the system 
\eqref{res-omgm}, \eqref{res-tgm2} and \eqref{res-tgm1} are
symmetric and satisfy the linear loop equation
\[
\omega_{g,m}(u_1,...,u_{m-1},z)+
\omega_{g,m}(u_1,...,u_{m-1},\sigma_i(z))=
\mathcal{O}(z-\beta_i)dz
\]
and the quadratic loop equation
\begin{align*}
&\omega_{g-1,m+1}(u_1,...,u_{m-1},z,\sigma_i(z))
+  \sum_{\substack{I_1\uplus I_2=\{u_1,...,u_{m-1}\}\\g_1+g_2=g}}
\hspace*{-1cm}
   \omega_{g_1,|I_1|+1}(I_1,z)
    \omega_{g_2,|I_2|+1}(I_2,\sigma_i(z))
\\[-2ex]
&=\mathcal{O}((z-\beta_i)^2)(dz)^2\;.
\end{align*}
\end{conjecture}
If the conjecture is true\footnote{As shown in \cite{Hock:2021tbl},
  Conjecture \ref{mconj} is true at least for $g=0$.}, it is a general
fact established in \emph{blobbed topological recursion}
\cite{Borot:2015hna} (and recalled in Appendix~\ref{recbtr}) that the
projection to the polar part is given by the universal formula of
topological recursion:
\begin{align}
\label{toprec}
&\mathcal{P}_z \omega_{g,m}(u_1,...,u_{m-1},z)
\\
&=\sum_{i=1}^{2d}\Res\displaylimits_{q\to \beta_i}
\frac{\frac{1}{2}\int_{q'=\sigma(q)}^{q'=q} B(z,q')}{
\omega_{0,1}(q)-\omega_{0,1}(\sigma_i(q))}
\Bigg(
\omega_{g-1,m+1}(u_1,...,u_{m-1},q,\sigma_i(q))
\nonumber
\\[-2ex]
&\hspace*{4cm}+  \sum_{\substack{I_1\uplus I_2=\{u_1,...,u_{m-1}\}\\g_1+g_2=g
\\(I_1,g_1)\neq (\emptyset,0)\neq 
(I_2,g_2)}}\hspace*{-1cm}
   \omega_{g_1,|I_1|+1}(I_1,q)
    \omega_{g_2,|I_2|+1}(I_2,\sigma_i(q))\Bigg)\;,
\nonumber
\end{align}
where $B(u,z)=\frac{du\,dz}{(u-z)^2}$ is the Bergman kernel.

\section{Conclusion and Outlook}
\label{Sec:Conclusion}

This paper makes the Quartic Kontsevich Model a member of a rich
family of models affiliated with the moduli space
$\overline{\mathcal{M}}_{g,n}$ of stable complex curves. Common to all
these models is the possibility to construct all functions of interest
(cumulants of a measure, correlation functions, generating functions
of something) recursively in decreasing Euler characteristic
$\chi=2-2g-n$. The quartic analogue of the Kontsevich model originates
from attempts to put the $\lambda\varphi^4$-quantum field theory model
on a noncommutative geometry. It is a Hermitian matrix model in which
a Gau\ss{}ian measure with non-trivial covariance (\ref{measure0}) is
deformed by a quartic potential, see (\ref{measure4}).  This paper
shows that the loop equation for the planar 2-point function of the
Quartic Kontsevich model, found in \cite{Grosse:2009pa} and eventually
solved in \cite{Grosse:2019jnv}, is indeed the initial datum for a
novel structure affiliated with $\overline{\mathcal{M}}_{g,n}$.

We find that the primary structure of the Quartic Kontsevich Model is
not the entirety of cumulants of the quartically deformed measure (as
thought before) but a family of auxiliary functions
$\Omega^{(g)}_{q_1,...,q_m}$ introduced in Definition
\ref{defT0Om}. They are particular polynomials of cumulants
\cite{Branahl:2020uxs}.  
The $\Omega^{(g)}_{q_1,...,q_m}$ are extended first to meromorphic functions 
$\Omega^{(g)}_m$ and then to meromorphic forms 
$\omega_{g,m}$ on $\hat{\mathbb{C}}^m$. It is convenient to view 
$\hat{\mathbb{C}}^m$ as the space of (complex, compactified) lines through
the $m$ marked points of a Riemann surface of genus $g$, see 
Fig.~\ref{diagrams}. The 
$\Omega^{(g)}_m$ do not exist alone; there are other families of
functions $T^{(g)}_{\dots}$ which interpolate between cumulants and
$\Omega$'s. These  $T^{(g)}_{\dots}$ extend to meromorphic functions
$\mathcal{T}^{(g)}(u_1,...,u_m\|z,w|)$
and $\mathcal{T}^{(g)}(u_1,...,u_m\|z|w|)$
on the space of lines through 
\begin{enumerate}
\item the $m$ marked points of a bordered Riemann surface of genus $g$
  with $b=1$ or $b=2$ boundary components,
\item defects on the boundary component(s); it is enough to consider
  two defects for $b=1$ and one defect on each boundary for $b=2$.
\end{enumerate}

This distinction is nothing new for matrix models. It already appeared
for the Hermitian 2-matrix model (2MM) \cite{Staudacher:1993xy} which
has mixed-coloured and non-mixed coloured boundaries.  The underlying
structure of monochromatic boundary correlation functions of the 2MM
was proved to follow a topological recursion \cite{Chekhov:2006vd}.
However, to compute non-mixed coloured boundary correlation functions
the knowledge of mixed-coloured boundary correlation functions is
inevitable \cite{Eynard:2007nq}.

The Quartic Kontsevich Model, discussed here, almost shares its
structure with the 2MM (cf. \eqref{DSE-cT2} with \cite[eq.\
(1-3)]{Eynard:2007gw}), even though it is by definition a completely
different model. We have shown that the resulting Dyson-Schwinger
equations are structurally almost of the same form.  We have found an
algorithm consisting of three steps (see Figure \ref{fig:euler1}) to
compute a given correlation function of Euler characteristic $\chi-1$
from correlation functions of Euler characteristic $\geq \chi$. We
showed that this calculation reduces to an evaluation of residues.

A look upon the explicitly given results for small $(-\chi)$ suggests
that the quartic analogue of the Kontsevich model is governed by
\emph{blobbed topological recursion} \cite{Borot:2015hna}.  This is an
extension of topological recursion by an infinite family of initial
data $\phi_{g,m}$.  For convenience we provide in Appendix
\ref{recbtr} some background information about the BTR. 

The final proof of our Main Conjecture \ref{mconj} is on the way.
The proof for $g=0$ is accomplished in \cite{Hock:2021tbl};
there remains little doubt that the result holds in general.
The geometric structure is apparent:
The spectral curve (of genus zero) is identified and parametrised by
\begin{align*}
x(z)&=R(z)\\
y(z)&=-R(-z),\qquad \text{where}\quad
R(z)=z-\frac{\lambda}{N}\sum_{n=1}^d \frac{r_n}{R'(\varepsilon_n)(\varepsilon_n+z)}.
\end{align*}
The numbers $\varepsilon_n$ are related by $e_p=R(\varepsilon_p)$ 
to the distinct values $e_p$ occurring with multiplicity $r_p$ in the 
parameters $E_1,...,E_N$ of the Gau\ss{}ian measure (\ref{measure0}).

Our blobbed topological recursion is defined by:
\begin{enumerate}
\item the covering
$x=R:\hat{\mathbb{C}}\to \hat{\mathbb{C}}$ of the Riemann sphere ramified at 
$\{\beta_1,\dots,\beta_{2d}\}$;

\item two meromorphic differentials
\begin{align}\label{speccurve}
\omega_{0,1}(z)&=y(z)dx(z) \qquad \text{ on } \hat{\mathbb{C}}\;, \\\nonumber
\omega_{0,2}(z,u)&=B(z,u)+\phi_{0,2}(z,u) \qquad \text{ on } \hat{\mathbb{C}}^2 \;,
\end{align}
both regular at the ramification points, where
$B(z,u)=\frac{dz\,du}{(z-u)^2}$ is the usual Bergman kernel 
and $\phi_{0,2}(z,u)=\frac{dz\,du}{(z+u)^2}$
a symmetric 2-form blob with a double pole on the antidiagonal;

\item the recursion kernel  $K_i(z,q)
  =\frac{\frac{1}{2}\int_{q'=\sigma_i(q)}^{q'=q} B(z,q')}{
\omega_{0,1}(q)-\omega_{0,1}(\sigma_i(q))}$ constructed
with the usual Bergman kernel and the local Galois involution $\sigma_i$
near $\beta_i$. 
  
\end{enumerate}
The presence of a blob $\phi_{0,2}(z,u)$ is an important difference to
the standard approach \cite{Borot:2015hna}. Moreover, we recall that
for the proof of BTR \cite{Borot:2015hna} it was sufficient to assume
$\omega_{g,m}$ to be defined on disjoint unions $\cup_{i}U_i$ about
the ramification points. In contrast, our differential forms
$\omega_{g,m}$ are globally defined meromorphic forms on
$\hat{\mathbb{C}}^m$.

We noticed an intriguing r\^ole of the global involution
$z\mapsto -z$ on $\hat{\mathbb{C}}$.
This involution is of central importance in \cite{Hock:2021tbl}
for proving Conjecture~\ref{mconj} for genus $g=0$. 
The blobs of higher genus have poles
at the fixed point $z=0$ of this involution; also the other poles
at $z_i=-z_j$ are related in this way. Since $z\to -z$ is a very
natural structure, we expect that the corresponding intersection
numbers have a topological significance. It seems
worthwhile to work out details and to compute these numbers.
Moreover, comparing our spectral curve \eqref{speccurve} to \cite{Borot:2009ix},
we already realised that a subset of the normalised part generates
\textit{simple Hurwitz numbers}. Our partition function is, however,
considerably easier and more natural than that of \cite{Borot:2009ix}.

One can take the point of view that the linear and quadratic loop
equations \cite{Borot:2013lpa} are the heart of TR.  Their general
solution is blobbed topological recursion \cite{Borot:2015hna};
further conditions are necessary to reduce it to pure TR. This raises
the question why the original Kontsecich model
\cite{Kontsevich:1992ti} and a large class of generalisations
\cite{Belliard:2021jtj} satisfy these further conditions, whereas the
quartic analogue of the Kontsevich model does not.  At the moment we
do not have a good intuitive explanation, but on a technical level
there are several reasons. In Remarks \ref{rem:2P}, \ref{Rem:2MM},
\ref{rem:Om02} and \ref{rem:cut} we have indicated similarities and
decisive differences to the Hermitean 2-matrix model.  Precisely the
additional terms compared with the 2-matrix model are responsible for
the poles of $\omega_{g,n}$ away from ramification points of $x$ (and the
diagonal in case of $\omega_{0,2}$). The Laurent series about these
additional poles is completely fixed by our global (on $\hat{\mathbb{C}}^n$)
  loop equations; there is
absolutely no freedom in choosing the blobs. This is a clear
difference with the original formulation of blobbed topological
recursion \cite{Borot:2015hna} in which the abstract loop equations
are only considered \emph{locally in a neighbourhood}
of the ramification points (so
that the blobs can be chosen freely within the constraints of the loop
equations).

Another technical reason for BTR is
that the $\omega_{g,n}$ in matrix models
are typically related to correlations of diagonal matrix elements
$\Phi_{aa}$ (such as in the (generalised) Kontsevich model
\cite{Belliard:2021jtj}) or correlations of
resolvents $\mathrm{Tr}((z-\Phi)^{-1})$ (such as in the
2-matrix model \cite{Chekhov:2006vd}). 
Because of the invariance of the quartic Kontsevich model
under $\Phi \mapsto -\Phi$, these special correlations only give rise to
even Euler characteristics. In particular, the initial
$\Omega^{(0)}_1$ cannot be obtained in this way. We have shown in
this paper that $\Omega^{(0)}_1$ in the quartic Kontsevich model
is built from the two-point function $\mathcal{G}^{(0)}(z,w)$,
which has a pole at $z+w=0$. This pole at opposite diagonals 
proliferates into the $\omega_{g,n}$ for all $2g+n\geq 2$ and induces poles
at $z_i=0$ for $g\geq 1$.

Private discussions with B.~Eynard and E.~Garcia-Failde also suggest
that there is hope to formulate the current version of blobbed
topological recursion in terms of pure TR by increasing the genus of
the spectral curve by 1. The appearance of the same phenomenon in the
$O(n)$ model \cite{Borot:2009ia} and the remarkable structural analogies
of holomorphic and polar parts in the quartic Kontsevich model 
make this hope a justified research goal for the future, among other
stimulating questions arising from this model.

\appendix

\section{Notations and Relations}

For the sake of readability, and because we partly deviate from
conventions in the literature, we list in the table below a few
important notations and symbols used in this paper.

\begin{longtable}[h]{|c|p{0.7\textwidth}|}
\hline
\textbf{Symbol} &\textbf{ Explanation} \\
\hline\endhead
  $E_1,...,E_N$; $\lambda$ & Parameters of Gau\ss{}ian measure and its
                             quartic deformation
\\\hline
  $e_k$, $r_k$, $d$ & Distinct values in $(E_l)$, their multiplicities,
                      their number
\\\hline
  $R(z)$ & Implicitly defined by $ R(z) = z- \frac{\lambda}{N}
           \sum_{k=1}^d \frac{r_k}{R'(\varepsilon_k)(z+\varepsilon_k)}$,\\
         &  $e_k=R(\varepsilon_k)$ \\
\hline
$\varepsilon_k$& Unique solutions in neighbourhood of $\lambda=0$ 
of $e_k=R(\varepsilon_k)$ 
\\\hline
$\hat z^j$ & $d$ preimages with $R(z)=R(\hat z^j)$ and $z\neq \hat{z}^j$ \\
\hline
$\{0,\pm \alpha_i\}$ & $2d+1$ solutions of $R(z)-R(-z)=0$\\
\hline
$\beta_i$ & $2d$ ramification points, solutions of $R'(z)=0$\\
\hline
  $\sigma_i(z)$ & Local Galois involution in the vicinity of $\beta_i$,
                  with \\
                & $R(z)=R(\sigma_i(z))$, $\lim_{z\to \beta_i}\sigma_i(z)=\beta_i$
                  and $\sigma_i\neq \mathrm{id}$
\\
 \hline
$G_{|\dots|}$ & Correlation functions/cumulants of the deformed measure  \\
\hline
$\mathcal{G}^{(g)}(...) $& Complexification and transformation via $R$
of $G_{|\dots|}$, plus genus expansion. Satisfies
 $\mathcal{G}(\varepsilon_k,...)=G_{|k...|}$
\\ \hline
$\mathcal{G}^{(0)}(z,w)$ & Given in Thm.~\ref{throm1} as solution of a
non-linear equation
\\
\hline
  $T_{...\|\dots|}$&Generalised correlation functions: $E_q$-derivatives of
  $G_{|\dots|}$ given in Def.~\ref{defT0Om}
  \\
\hline
  $\mathcal{T}^{(g)}(...\|...|)$& 
Complexification, transformation via $R$ and  genus expansion of $T_{...\|...|}$ 
  \\ \hline
 $\Omega_{q_1...q_m}$ & Derivative of
 $\frac{1}{N}\sum_p G_{|q_1p|}+\frac{1}{N^2}G_{|q_1|q_1|}$ with respect to \\ &
                                 $E_{q_2},...,E_{q_{m}}$ 
(see Def.~\ref{defT0Om})
\\ \hline
 $\Omega^{(g)}_m(z_1,...,z_m)$ & Complexification, transformation via
 $R$ and  
genus expansion of $\Omega_{q_1...q_m}$ 
\\\hline
$\omega_{g,m}(z_1,\dots,z_m) $& \parbox[t]{0.7\textwidth}{Meromorphic
  differential \\ $=\lambda^{2-2g-m}
\Omega^{(g)}_m(z_1,\dots,z_m) \prod_{j=1}^m R'(z_j) dz_j$} 
\\\hline
$L(x),L_w(x)$ & Lagrange interpolation polynomials \\
\hline
  $\chi$ & Euler characteristic $\chi=2-2g-m-b$; the $\Omega^{(g)}_m$ have $b=0$
  \\ \hline
  $\mathfrak{G}_0(z)$& Auxiliary function
  $\mathfrak{G}_0(z)=\Res\displaylimits_{w\to -z}
  \mathcal{G}^{(0)}(z,w)dw $ \\
\hline
$C_{k,l}^{m,n}$ & Partial fraction coefficients of $\mathcal{G}^{(0)}(z,w)$\\
\hline
  $Q(w;z)$ &Auxiliary function $Q(w;z):=\frac{1}{w+z}+\frac{1}{w-z}$,
             derivatives $Q'$, $Q''$ etc. with respect to the second argument
  \\ \hline
$B(z,u)$ & Bergman kernel $B(z,u)=\frac{dz\, du}{(z-u)^2}$\\
\hline
$\phi_{0,2}$ & A blob given by $\frac{dz\, du}{(z+u)^2}$ \\

\hline
$K_i(z,q)$ & Recursion kernel  \\
\hline
  $\mathcal{H}_z, \mathcal{P}_z$ & Projections to 
                                   holomorphic and polar parts
                                   (near ramification points)
of meromorphic $m$-forms \\
\hline
\end{longtable}

\section{Recap of Blobbed Topological Recursion}
\label{recbtr}

The outstanding applicability of topological recursion (TR) to a great
bandwidth of mathematical phenomena is clearly undoubted. However,
there exist models showing a certain recursive behaviour regarding
their solutions of loop equations, but not perfectly fitting into the
recursion of ordinary TR, for instance in the Hermitian 1-matrix model
extended by multi-trace contributions \cite{Borot:2013fla} or in the
quartic melonic tensor model \cite{Bonzom:2016kqf}. The appearance of
poles at $z\in \{0,-u_l\}$ in Corollary \ref{corr:Omega-residue} gave
a first hint\footnote{We thank St\'ephane Dartois for pointing out
  this extension of topological recursion.} to focus on a framework
that even enlarges the mentioned bandwidth.  Discovered in 2015, it
extends the usual TR by additional topological quantities baptised
\textit{blobs} to \textit{blobbed topological recursion}
\cite{Borot:2015hna}.

It was observed that the loop equations of several (matrix) models can be
reduced to a system of linear and quadratic loop equations:
\begin{definition}
\label{def:absloopeq}
Let $x:\Sigma\to \Sigma_0$ be a ramified covering with simple
ramification points $\beta_i$ and $\sigma_i$ be the local
Galois involution around $\beta_i$, i.e.\ 
$x(z)=x(\sigma_i(z))$, $\lim_{z\to \beta_i} \sigma_i(z)=\beta_i$ and 
$\sigma_i\neq \mathrm{id}$. A
family of meromorphic differential forms $\omega_{g,m}$ on $\Sigma^m$,
with $g \geq 0$ and $m >0$, fulfils the \textbf{linear loop equation}
if
\begin{align}
\omega_{g,m+1}(u_1,...,u_m,z)+
\omega_{g,m+1}(u_1,...,u_m,\sigma_i(z))=
\mathcal{O}(z-\beta_i)dz
\end{align}
is a holomorphic linear form for $z \to \beta_i$ with (at least) a simple zero
at $\beta_i$. The family of $\omega_{g,m}$ fulfils the
\textbf{quadratic loop equation} if
\begin{align}
Q^i_{g,m+1}&:=\omega_{g-1,m+2}(u_1,...,u_m,z,\sigma_i(z))
\nonumber
\\
&+ \hspace*{-0.5cm}
\sum_{\substack{g_1+g_2=g \\ I_1\uplus I_2=\{u_1,...,u_m\}}}
\hspace*{-0.8cm}
   \omega_{g_1,|I_1|+1}(I_1,z)
    \omega_{g_2,|I_2|+1}(I_2,\sigma_i(z))
    \nonumber
 \\
&=\mathcal{O}((z-\beta_i)^2)(dz)^2
\label{qle}
\end{align}
is a holomorphic quadratic form with at least a double zero at $z \to \beta_i$.
\end{definition}
An important subclass of solutions is given by differentials governed by TR
\cite{Borot:2013lpa}. The entirety of solutions, instead, is provided
by BTR. According to Subsection
\ref{sec:omega-low}, the solutions $\omega_{0,2}$,
$\omega_{0,3}$, $\omega_{0,4}$  and $\omega_{1,1}$ of the loop equations
of the  Quartic Kon\-tse\-vich Model fulfil the
linear and quadratic loop equations. We hope to provide in
near future the proof of the natural Main Conjecture~\ref{mconj}
that all $\omega_{g,m}$ obey these loop equations.

\begin{figure}[h!t]
  \centering
    \includegraphics[width= 0.99\textwidth]{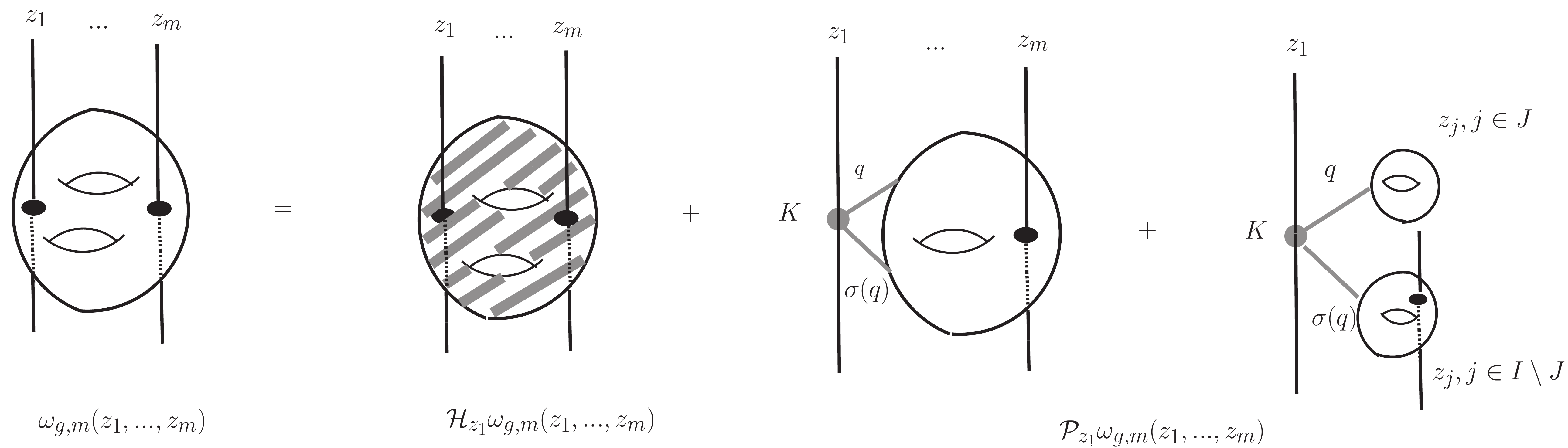}
    \caption{Diagrammatic representation of blobbed topological
      recursion: $\omega_{g,m}$ is a meromorphic form on a product of 
      $\hat{\mathbb{C}}$ (each shown as a line) each attached to a marked point
      (shown as black dot) on a genus-$g$ Riemann surface (here $g=2$). It is
      recursively generated. The second and third graph on the rhs  
     are copies of ordinary
      topological recursion; these provide the part of $\omega_{g,m}$
      with poles in $z_1$ at ramification points of $x=R$. The first graph  
      on the rhs, however, depicts the holomorphic part as an additional input of each recursion step. 
      Its poles in $z_1$ are located
      outside the ramification points of $x=R$.  
\label{diagrams}}
\end{figure}

The suggestive notation in \eqref{decom} was inspired by
\cite{Borot:2015hna} and shall be explained now. In the framework of
BTR, one defines projectors $\mathcal{H}_z$ and $\mathcal{P}_z$ acting on
\[
\omega_{g,m}(...,z)
= \mathcal{H}_z\omega_{g,m}(...,z)+ \mathcal{P}_z\omega_{g,m}(...,z).
\]
It is shown in \cite{Borot:2015hna} that the part
$\mathcal{P}_z\omega_{g,m}(z_1...,z_{m-1},z)$ is produced by the
universal formula of topological recursion from $\omega_{g',m'}$ with
$2g'+m'-2<2g+m-2$.  The mechanism of BTR can be depicted as in
Fig.~\ref{diagrams}.
Applying these projections in every variable decomposes $\omega_{g,m}$
into $2^m$ pieces, among them the purely holomorphic part (for
$2g+m-2>0$)
$\phi_{g,m}(z_1,...,z_{m-1},z)=
\mathcal{H}_{z_1}...
\mathcal{H}_{z_{m-1}}\mathcal{H}_z\omega_{g,m}(z_1...,z_{m-1},z)$,
called the \textit{blob}, and the purely polar part
$\mathcal{P}_{z_1}...\mathcal{P}_{z_{m-1}}\mathcal{P}_z
\omega_{g,m}(z_1...,z_{m-1},z)$.  In the special case where
$
\mathcal{P}_{z_1}...\mathcal{P}_{z_{m-1}}\mathcal{P}_z\omega_{g,m}(z_1...,z_{m-1},z)
=\omega_{g,m}(z_1...,z_{m-1},z)$,
the solution of abstract loop equations shall be called a
\textit{normalised} one, denoted by $\omega^o_{g,m}$. In
\cite{Borot:2015hna} there was developed a diagrammatic representation
of (products of) projectors $\mathcal{H}$ and $\mathcal{P}$ acting on
$\omega_{g,m}$.

We will slightly deviate from the above conventions by choosing the
unstable blobs $\phi_{0,1},\phi_{0,2}$ differently and by adopting a
global formulation. First, we set $\phi_{0,1}=0$ 
and  $\phi_{0,2}(z,u)=\frac{dzdu}{(z+u)^2}$ with $\omega_{0,1}(z)=y(z)dx(z)$
as usual and $\omega_{0,2}(z,u)=B(z,u)+\phi_{0,2}(z,u)$, see (\ref{speccurve}).
In the original formulation \cite{Borot:2015hna},
the Riemann surface $\mathcal{C}$ is a disjoint union $\cup_{i}U_i$ of
sufficiently small neighbourhoods of the ramification points
$\beta_i$. Then $\mathcal{H}_z\omega_{g,m}$ is indeed holomorphic in
every $z\in \mathcal{C}$.  In contrast, our Quartic Kontsevich Model
is defined globally on $\hat{\mathbb{C}}$ so that the term
\textit{holomorphic part} should be treated with more caution. It is
rather a relic of previous namings and means holomorphic in
ramification points, but with poles somewhere else on
$\hat{\mathbb{C}}$.

The global formulation also suggests a more natural
definition of the projection $\mathcal{P}_z$, namely
\begin{equation}
\label{Pzop}
\mathcal{P}_z\omega(z)=\sum_{i=1}^{2d} 
\mathcal{P}^i_z\omega(z)\;,\qquad 
\mathcal{P}^i_z\omega(z):=
\Res\displaylimits_{q\to \beta_i}\Big[\omega(q) \int_{\infty}^q B(z,.)\Big]
\end{equation}
for a 1-form $\omega$ in a selected variable
(in case there are $2d$ ramification points). Here
$B(z,z')$ is the Bergman kernel; whereas
\cite{Borot:2015hna} defines $\mathcal{P}_z$ with the given
bidifferential $\omega_{0,2}(z,z')$.
The global formulation allows us to start the contour integral at
the special point $\infty \in \hat{\mathbb{C}}$ instead of $\beta_i$
chosen in \cite{Borot:2015hna}.
In particular, our projector (\ref{Pzop}) sees the residue
and thus gives the whole principal part of the Laurent series about $\beta_i$.

A main achievement in \cite{Borot:2015hna} is a simple 
proof (which adapts arguments of \cite[Prop.~2.7]{Borot:2013lpa}) 
that meromorphic $m$-forms $\omega_{g,m}$ which satisfy
the abstract loop equations of Definition \ref{def:absloopeq}
have a polar part given by the universal TR-formula. The essence of the proof
remains unchanged when defining the polar part via (\ref{Pzop}).
We find it convenient to sketch the arguments.
For $z$ near $\beta_i$ and $I=\{z_1,...,z_{m-1}\}$ define 
\begin{align*}
  \mathcal{S}^i_z \omega_{g,m}(I,z)
  &= \omega_{g,m}(I,z)+ \omega_{g,m}(I,\sigma_i(z))\;,
 \\
  \Delta^i_z \omega_{g,m}(I,z) &= \omega_{g,m}(I,z)- \omega_{g,m}(I,\sigma_i(z))\;.
\end{align*}
The quadratic loop equation (\ref{qle})  can be written as
$\mathcal{P}^i_z \big[\frac{Q^i_{g,m}(I,z) }{\Delta^i_z \omega_{0,1}(z)}
\big] = 0$.  Indeed, $\Delta^i_z \omega_{0,1}(z)$ has a double zero at
every $z=\beta_i$ so that
$\frac{Q^i_{g,m}(I,z) }{\Delta^i_z \omega_{0,1}(z)}$ is holomorphic in
$z=\beta_i$.  Write $Q^i_{g,m}(I,z)=\omega_{0,1}(z) S^i_z\omega_{g,m}(I,z) 
-\omega_{g,m}(I,z)\Delta^i_z\omega_{0,1}(z)
+\tilde{Q}^i_{g,m}(I,z)$ where $\tilde{Q}^i_{g,m}(I,z)$ excludes both terms with
$\omega_{0,1}$ in $Q^i_{g,m}$.
Both $\omega_{0,1}(z)$ and (by the linear loop equation) $S^i_z\omega_{g,m}(I,z)$ 
have a simple zero at $z=\beta_i$ so that we arrive at
\[
\mathcal{P}^i_z \omega_{g,m}(I,z)=
\mathcal{P}^i_z \Big[\frac{\tilde{Q}^i_{g,m}(I,z)}{\Delta^i_z\omega_{0,1}(z)}\Big]
=\frac{1}{2}\mathcal{P}^i_z \Big[\frac{\tilde{Q}^i_{g,m}(I,z)}{
\Delta^i_z\omega_{0,1}(z)}\Big]
-\frac{1}{2}\mathcal{P}^i_{\sigma_i(z)}
\Big[\frac{\tilde{Q}^i_{g,m}(I,z)}{\Delta^i_z\omega_{0,1}(z)}\Big]\;.
\]
The second equality follows from the antisymmetry of
$\frac{\tilde{Q}^i_{g,m}(I,z)}{\Delta^i_z\omega_{0,1}(z)}$
under the involution $z\leftrightarrow \sigma_i(z)$. Inserting this result into
(\ref{Pzop}) establishes 
\begin{align*}
  \mathcal{P}_z \omega_{g,m}(I,z)
  &= \sum_{i=1}^{2d} \Res\displaylimits_{q \to \beta_i} K_i(z,q)
  \tilde{Q}^i_{g,m}(I,q)  
  \\
  \text{ with }K_i(z,q)
  &=\frac{\frac{1}{2}\int_{q'=\sigma_i(q)}^{q'=q} B(z,q')}{
\omega_{0,1}(q)-\omega_{0,1}(\sigma_i(q))}\;.
\end{align*}
It writes out as in (\ref{toprec}).

\section{Local Galois Involution and Recursion Kernel}

\label{app:Galois}

Let $x:\hat{\mathbb{C}}\to \hat{\mathbb{C}}$ be a ramified
covering of the Riemann sphere with simple ramification points,
$\omega_{0,1}(z)=y(z)dx(z)$ a meromorphic 1-form which is holomorphic
in the ramification points of $x$, and 
$B(z_1,z_2)=\frac{dz_1dz_2}{(z_1-z_2)^2}$ be 
the standard Bergman kernel on $\hat{\mathbb{C}}^2$.
For a ramification point $\beta_i$ of $x$, 
determined by $x'(\beta_i)=0$, let $\sigma_i$ be the local Galois 
involution in a neighbourhood $\mathcal{U}_i$ of $\beta_i$, determined by 
$x(\sigma_i(z))=x(z)$,
$\lim_{z\to \beta_i}\sigma_i(z)=\beta_i$ and 
$\sigma_i\neq \mathrm{id}$. Let
\begin{align}
x_{n,i}:=\frac{x^{(n+2)}(\beta_i)}{x''(\beta_i)}\;,\qquad
y_{n,i}:=\frac{y^{(n+1)}(\beta_i)}{y'(\beta_i)}\;.
\end{align}
\begin{lemma}
\label{lemma:Galois}
The local Galois involution
$\sigma_i$ in $\mathcal{U}_i$ has a formal power series expansion
$\sigma_i(q)=\beta_i+\sum_{n=0}^\infty c_{n,i}(q-\beta_i)^{n+1}$ whose
coefficients are recursively given by $c_{0,i}=-1$ and for $n \geq 1$ by
\begin{align*}
  c_{n,i}&=\frac{(-1)^n-1}{(n+2)!}x_{n,i}
  +\frac{1}{2}\sum_{k=1}^{n-1}c_{k,i}c_{n-k,i}+\frac{1}{(n+2)!}\sum_{k=3}^{n+1}x_{k-2,i}b_{n+2,k,i}(x)\;,
  \\
&\text{where }  b_{n,k,i}(x):=
B_{n,k}(1!c_{0,i},2!c_{1,i},
...,(n{-}k{+}1)!c_{n-k,i})\;.
\end{align*}
Here $B_{n,k}$ are the Bell polynomials. The first examples are
\begin{align*}
c_{1,i}&=-\frac{x_{1,i}}{3}\;, &
c_{2,i}&=-\frac{x_{1,i}^2}{9}\;,\quad
\\
c_{3,i} &=-\frac{2 x_{1,i}^3}{27}
+\frac{x_{1,i}x_{2,i}}{18}-\frac{x_{3,i}}{60}\;,\quad &
c_{4,i}&=
-\frac{4x_{1,i}^4}{81}+\frac{x_{1,i}^2x_{2,i}}{18}
-\frac{x_{1,i}x_{3,i}}{60}\;.
\end{align*}
\begin{proof}
Insert the power series ansatz into the identity
$0=x(\sigma_i(q))-x(q)$ for $q$ in a neighbourhood of $\beta_i$. 
Then all derivatives with respect to $q$ vanish at $q=\beta_i$ so that 
we have from Fa\`a di Bruno's formula and with $x'(\beta_i)=0$
\begin{align*}
x^{(n)}(\beta_i) 
= \sum_{k=2}^n x^{(k)}(\beta_i)\cdot 
b_{n,k,i}(x)\;.
\end{align*} 
This gives $c_{0,i}^2=1$ for $n=2$. The solution $c_{0,i}=1$ selects
the primary branch $c_{n,i}=0$ for all $n\geq 1$. For the local Galois
involution we thus have $c_{0,i}=-1$. Solving the resulting equations
by using the definition for the following Bell polynomials
$b_{n,n,i}(x)=(-1)^n$ and
\begin{align*}
b_{n,2,i}(x)= n! \Big( -c_{n-2,i}+\frac{1}{2}\sum_{k=1}^{n-3}c_{k,i}c_{n-2-k,i}\Big)
\end{align*}
gives after shifting $n\to n+2$ and dividing by $x''(\beta_i)$ the
desired recursion.
\end{proof}
\end{lemma}

The recursion kernel near a ramification point $\beta_i$ specifies to
\begin{align}
K_i(z,q) = 
\frac{\big(\frac{1}{z-q}-\frac{1}{z-\sigma_i(q)}\big)dz}{
2 (y(q)-y(\sigma_i(q))) x'(\sigma_i(q)) d\sigma_i(q)}\;.
\end{align}
In terms of $b_{n,k,i}(x)$,
the terms in the recursion kernel expand into
\begin{align*}
\frac{1}{z-q}-\frac{1}{z-\sigma_i(q)}
&=\sum_{n=1}^\infty\frac{(q-\beta_i)^n}{n!} \Big(
\frac{n!}{(z-\beta_i)^{n+1}}
-\sum_{k=1}^n 
\frac{k!}{(z-\beta_i)^{k+1}}b_{n,k,i}(x)\Big)\;,
\\
y(q)-y(\sigma_i(q))
&= y'(\beta_i)\sum_{n=1}^\infty \frac{(q-\beta_i)^n}{n!}\Big( y_{n-1,i}
- \sum_{k=1}^n  y_{k-1,i} b_{n,k,i}(x)\Big)\;,
\\
x'(\sigma_i(q))
&= x''(\beta_i)\sum_{n=1}^\infty \frac{(q-\beta_i)^n}{n!}
\sum_{k=1}^n x_{k-1,i} b_{n,k,i}(x)\;.
\end{align*}
Up to order $\mathcal{O}((q-\beta_i)^3)$ we thus get
\begin{align*}
&K_i(z,q) 
\\
&=\frac{dz}{x''(\beta_i)y'(\beta_i) d\sigma_i(q)}\Big\{
-\frac{1}{2(z-\beta_i)^2(q-\beta_i)}
-\frac{x_{1,i}}{12(z-\beta_i)^2}
\\
&+\Big[
\Big(- \frac{x_{1,i}^2 }{8}
-\frac{x_{1,i} y_{1,i}}{12}  
+ \frac{x_{2,i} }{12} 
+ \frac{y_{2,i} }{12} 
\Big)
\frac{(q-\beta_i)}{(z-\beta_i)^2}
\\
&+\frac{x_{1,i}}{6}\frac{(q-\beta_i)}{(z-\beta_i)^3}
-\frac{1}{2}\frac{(q-\beta_i)}{(z-\beta_i)^4}\Big]
\\
&+\Big[\Big(
- \frac{37 x_{1,i}^3}{432}
-\frac{x_{1,i}^2 y_{1,i}}{24} 
+ \frac{x_{1,i} x_{2,i}}{12}
+ \frac{x_{1,i}y_{2,i} }{24} 
- \frac{x_{3,i}}{80}
\Big)
\frac{(q-\beta_i)^2}{(z-\beta_i)^2}
\\
&+ \frac{x_{1,i}^2}{12}\frac{(q-\beta_i)^2}{(z-\beta_i)^3}
- \frac{x_{1,i}}{4}
\frac{(q-\beta_i)^2}{(z-\beta_i)^4}\Big]
\\
&+\Big[\Big(
- \frac{209 x_{1,i}^4}{2592}
- \frac{19 x_{1,i}^3 y_{1,i} }{432}
-\frac{x_{1,i}^2 y_{1,i}^2 }{72} 
+ \frac{17 x_{1,i}^2 x_{2,i}}{144}
+ \frac{x_{1,i} x_{2,i} y_{1,i}}{36}
\\
& 
+ \frac{19 x_{1,i}^2 y_{2,i}}{432}
+ \frac{x_{1,i} y_{1,i} y_{2,i}}{36} 
- \frac{x_{2,i}^2}{72}
- \frac{23 x_{1,i} x_{3,i}}{720} 
- \frac{x_{3,i} y_{1,i}}{240}
- \frac{x_{2,i} y_{2,i}}{72} 
\\
& 
- \frac{y_{2,i}^2}{72} 
- \frac{x_{1,i} y_{3,i}}{72} 
+ \frac{x_{4,i}}{240}
+ \frac{y_{4,i}}{240}
\Big)
\frac{(q-\beta_i)^3}{(z-\beta_i)^2}
\\
&
+ \Big(
\frac{19 x_{1,i}^3}{216}  
+\frac{x_{1,i}^2 y_{1,i}}{36}
- \frac{x_{1,i} x_{2,i}}{18}
- \frac{x_{1,i} y_{2,i}}{36}
+ \frac{x_{3,i}}{120}
\Big)
\frac{(q-\beta_i)^3}{(z-\beta_i)^3}
\\
&+\Big(
- \frac{19 x_{1,i}^2}{72}
- \frac{x_{1,i} y_{1,i}}{12}
+ \frac{x_{2,i}}{12}
+ \frac{y_{2,i}}{12}
\Big)
\frac{(q-\beta_i)^3}{(z-\beta_i)^4}
\\
&
+ \frac{x_{1,i}}{3} \frac{(q-\beta_i)^3}{(z-\beta_i)^5} 
- \frac{1}{2}\frac{(q-\beta_i)^3}{(z-\beta_i)^6}\Big]
\\
&+\mathcal{O}((q-\beta_i)^4)\Big\}\;.
\end{align*}

\section{Graphical Derivation of the
  Dyson-Schwinger Equations}

\label{app:graphical}

Since all the correlation functions have a combinatorial
interpretation, also the DSE's of Prop. \ref{prop:TOm} can be
described combinatorially in terms of ribbon graphs. The correlation
function are generating series of those ribbon graphs; we refer to
\cite{Branahl:2020uxs} for more details and precise definitions.

Before, we need two further relations between generalised correlation functions achieved by applying \eqref{WTI}.
\subsection{Ward-Takahashi Identity for Generalised Correlation Functions}
We will represent higher correlation function through lower ones by
summing over one of the indices:
\begin{lemma}\label{lem:WTI4P}
Let $I=\{q_1,...,q_m\}$. Then we have the identity
\begin{align*}
  -\frac{T^{(g)}_{I\|nq|}-T^{(g)}_{I\|pq|}}{E_n-E_p}
  &=\frac{1}{N}\sum_{\substack{k=1\\ k\notin I}}^N
  T^{(g)}_{I\|pknq|}+\sum_{\substack{g_1+g_2=g\\ I_1 \uplus I_2=I}}
	T^{(g_1)}_{I_1\|pq|}T^{(g_2)}_{I_2\|qn|}
\\ \nonumber
&
-\sum_{q_i\in I}\frac{\partial}{\partial E_{q_i}}
T^{(g)}_{I\setminus q_i\|pq_inq|}
+T^{(g-1)}_{I\|p|pnq|}+T^{(g-1)}_{I\|n|pnq|}+T^{(g-1)}_{I\|pq|nq|}.
\end{align*}
\begin{proof}
The Ward-Takahashi identity for the
$4$-point function was already derived
in \cite[eq.~(B.5)]{Schurmann:2019mzu-v3}, that is
\begin{align*}
-\frac{G^{(g)}_{|qn|}-G^{(g)}_{|pq|}}{E_n-E_p}
&=\frac{1}{N}\sum_{k=1}^NG^{(g)}_{|pknq|} +\sum_{g_1+g_2=g}
G^{(g_1)}_{|pq|}G^{(g_2)}_{|qn|}\\
&+G^{(g-1)}_{|p|pnq|}+G^{(g-1)}_{|n|pnq|}
+G^{(g-1)}_{|pq|nq|}.
\end{align*}
Applying the operator
$D_I:=\frac{(-N)^m \partial^m}{\partial E_{q_1}\cdots \partial
  E_{q_m}}$ to the identity and making use of Definition \ref{defT0Om}
yields the assertion.
\end{proof}
\end{lemma}

\begin{lemma}\label{lem:WTI1+3P}
Let $I=\{q_1,...,q_m\}$. Then we have the identity
\begin{align*}
  -\frac{T^{(g)}_{I\|p|q|}-T^{(g)}_{I\|n|q|}}{E_p-E_n}
  &=\frac{1}{N}\sum_{\substack{k=1\\ k\notin I}}^NT^{(g)}_{I\|pkn|q|}
-\sum_{q_i\in I}\frac{\partial}{\partial E_{q_i}}
T^{(g)}_{I\setminus q_i\|pq_in|q|}+T^{(g)}_{I\|pqqn|}
\\\nonumber
&+ \!\! \sum_{\substack{g_1+g_2=g\\ I_1 \uplus I_2=I}}
T^{(g_1)}_{I_1\|pn|}\bigg(T^{(g_2)}_{I_2\|q|n|}+T^{(g_2)}_{I_2\|q|p|}\bigg)
+T^{(g-1)}_{I\|q|n|pn|}+T^{(g-1)}_{I\|q|p|pn|}.
\end{align*}
\begin{proof}
The Ward-Takahashi identity for the
$3+1$-point function was already derived in
\cite[eq.~(B.6)]{Schurmann:2019mzu-v3}, that is
\begin{align*}
  -\frac{G^{(g)}_{|n|q|}-G^{(g)}_{|p|q|}}{E_n-E_p}&=
  \frac{1}{N}\sum_{k=1}^NG^{(g)}_{|pkn|q|}
	+G^{(g)}_{|pqqn|}
	+ \!\!\! \sum_{g_1+g_2=g} \!\!\!
	G^{(g_1)}_{|pn|}\Big(G^{(g_2)}_{|q|n|}+G^{(g_2)}_{|q|p|}\Big)\\
	&+G^{(g-1)}_{|q|n|pn|}+G^{(g-1)}_{|q|p|pn|}.
\end{align*}
Applying the operator
$D_I:=\frac{(-N)^m \partial^m}{\partial E_{q_1}\cdots \partial
  E_{q_m}}$ to the identity and making use of Definition \ref{defT0Om}
yields the assertion.
\end{proof}
\end{lemma}

\subsection{Graphical Derivation of
  \eqref{DSE-Om}}\label{Sec.App.Om}

We will start with the DSE 3) of Prop.~\ref{prop:TOm}, which is
achieved combinatorially by an action of the operation
$-N\frac{\partial}{\partial E_q}$ on $\Omega_{I}$ with $q\notin I$
(this was discussed in \cite{Branahl:2020uxs} in greater detail). The
derivative acts on an edge adjacent to an internal face, splitting the
edge and fixing the label of the internal face to $q$. There are three
different cases depending on the other adjacent face to this edge:
\begin{itemize}[a)]
\item[a)] If the other adjacent face is also an internal face labelled
  by $l$, a generalised 2-point function appears with a sum over this
  external face $l$, which is $\frac{1}{N}\sum_{l=1}^NT_{I\|ql|}$.

\item[b)] If the other adjacent face is labelled by $q_i\in I$, a
  generalised 2-point function appears with external faces $q$ and
  $q_i$. A derivative of the form
  $-N\frac{\partial}{ \partial E_{q_i}}$ has to be taken only acting
  on the face labelled by $E_{q_i}$ since this face has one additional
  edge already split, that is
  $-\frac{\partial^{ext} T_{I\setminus q_i\|qq_i|}}{\partial E_{q_i}}$
  as defined in \eqref{del}. A sum over all possible $q_i$ has to be
  taken.

\item[c)] If the other adjacent face is the same face, a generalised
  $1+1$-point function appears with one genus less and two external
  faces labelled by $q$ each of length one, that is  
  $\frac{1}{N^2}T_{I\|q|q|}$.
	\end{itemize}
Combining case a) and b) via 
\begin{align}\label{delext}
  -N\frac{\partial }{\partial E_{q_i}}T_{I\setminus q_i\|qq_i|}
  =T_{I\|qq_i|}-N\frac{\partial^{ext} }{\partial E_{q_i}}T_{I\setminus q_i\|qq_i|},
\end{align}
where we have used \eqref{del}, we obtain the DSE \eqref{DSE-Om}.

\subsection{Graphical Derivation of \eqref{DSE-T2}}

\begin{figure}[h]
 	\includegraphics[width= 1.0\textwidth]{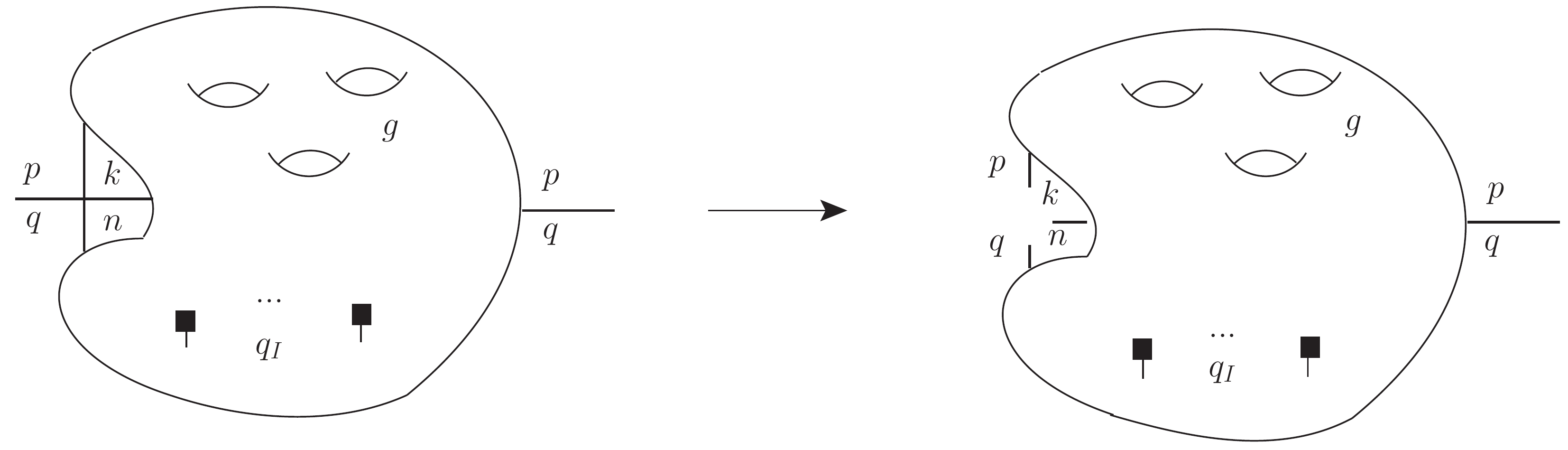}
 	\caption{Generic deletion of the first vertex for the generalised 2-point function, which separates the free propagator from the rest. This operation generates the factor $\frac{-\lambda }{E_p+E_q}$.
 		\label{fig:pert1}}
 \end{figure}

The DSE 1) of Prop.~\ref{prop:TOm} is achieved by a bijection between
$T^{(g)}_{I\|pq|}$, a generating series of ribbon graphs with a
certain structure, and other generating series via an operation we
call vertex deletion (of the first vertex). The generic deletion is
drawn in Fig.~\ref{fig:pert1}.
 We will distinguish between six topologically different cases depending
 on the choice of $n$ in Fig.~\ref{fig:pert1}. For each choice of $n$,
 we get several subcases according to the choice of $k$, which we
 do not draw in
 general. However, it is evident by the distinctions in $n$ how these
 different topologies for $k$ should look like.

\begin{itemize}[a)]
\item[a)] Generic $n$: Here, $k$ can be an internal face label running from
  1 to $N$ with the prefactor $\frac{1}{N}$; the correlation function
  on the rhs of Fig.~\ref{fig:pert1} therefore is
	\begin{align*}
		\frac{1}{N}\sum_{k=1}^NT^{(g)}_{I\|pknq|}.
	\end{align*} 
However, five further subcases occur, depending on the choice of $k$:
\begin{itemize}
\item For $k=q$: The rhs can be split into two components with the
  structure $T^{(g_1)}_{I_1\|pq|}T^{(g_2)}_{I_2\|qn|}$ with
  $g_1+g_2=g$ and $I_1\uplus I_2=I$.
\item For $k=q_i\in I$: The rhs has an extra derivative on the face
  labelled by $q_i$ coming from the split edge of the square-vertex,
  that is
  $-\frac{\partial^{ext}}{\partial E_{q_i}}T^{(g)}_{I\setminus
    q_i\|pq_inq|}$.
\item For $k=p$: A reduction of the genus by one can be generated with
  two boundary components one of length 1 and the other of length 3,
  so we obtain $T^{(g-1)}_{I\|p|pnq|}$.
\item For $k=n$: A reduction of the genus by one can be generated with
  two boundary components one of length 1 and the other of length 3,
  so we obtain $T^{(g-1)}_{I\|n|pnq|}$.
\item For $k=q$: A reduction of the genus by one can be generated with
  two boundary components of length 2, so we obtain
  $T^{(g-1)}_{I\|pq|nq|}$.
\end{itemize}
These six terms can be combined and surprisingly the identity
of Lemma \ref{lem:WTI4P} can be applied exactly. Consequently, including the
deleted free propagator and the vertex, we obtain after summation
over $n$
\begin{align}\nonumber
  \frac{-\lambda}{E_p+E_q}\frac{1}{N}\sum_{n=1}^N  \bigg[&
  \frac{1}{N}\sum_{k=1}^NT^{(g)}_{I\|pknq|}
  +\sum_{\substack{g_1+g_2=g\\ I_1 \uplus I_2=I}}
  T^{(g_1)}_{I_1\|pq|}T^{(g_2)}_{I_2\|qn|}
  \\ \nonumber
  & \!\!\!\!\!
  -\sum_{q_i\in I}\frac{\partial^{ext}}{\partial E_{q_i}}
  T^{(g)}_{I\setminus q_i\|pq_inq|}
  +T^{(g-1)}_{I\|p|pnq|}+T^{(g-1)}_{I\|n|pnq|}+T^{(g-1)}_{I\|pq|nq|}\bigg]
 \\\label{pert1}
 =\frac{\lambda}{E_p+E_q}\frac{1}{N}\sum_{n=1}^N&
 \frac{T^{(g)}_{I\|nq|}-T^{(g)}_{I\|pq|}}{E_n-E_p}.
\end{align}
	
\item[b)] For $n=p$: There is one case where the correlation function
  is split into two components after deleting the first vertex. Depending
  on $k$, the three different subcases are drawn in
  Fig.~\ref{fig:pert2}.
	\begin{figure}[h]
		\includegraphics[width= 1.0\textwidth]{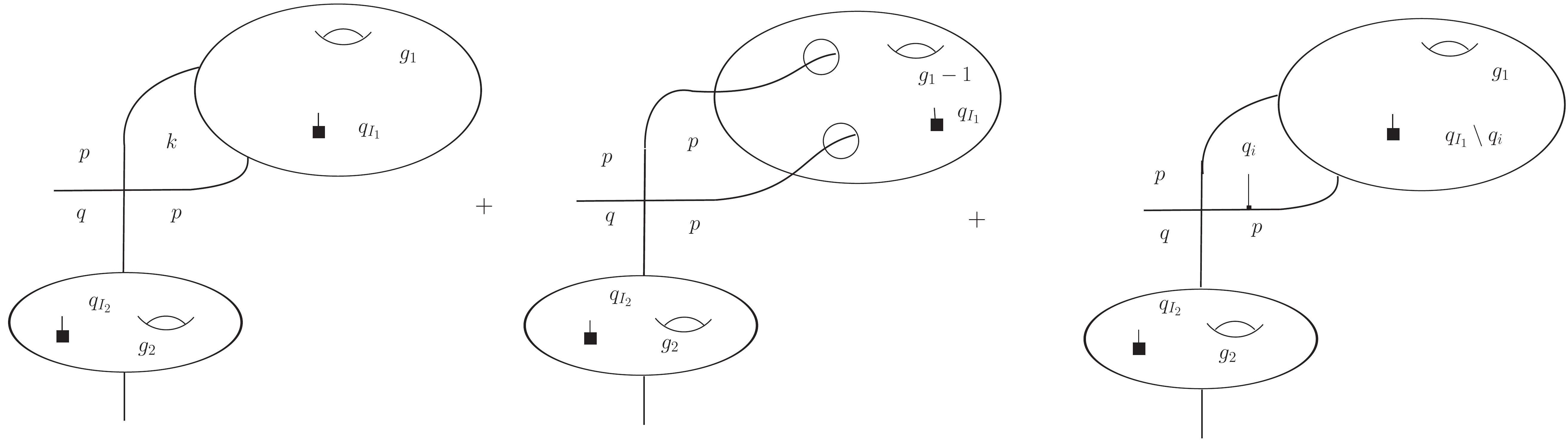}
		\caption{Three cases depending on $k$ for the generalised 2-point function with $n=p$. The first vertex is drawn in the middle seperately. 
			\label{fig:pert2}}
	\end{figure}
\begin{itemize}
\item For generic $k$: The sum over $k$ is taken and the two
  components have the form $T^{(g_1)}_{I_1\|pk|}T^{(g_2)}_{I_2\|pq|}$
  with $g_1+g_2=g$ and $I_1\uplus I_2=I$.
\item For $k=p$: A reduction of the genus by one can be generated with
  two boundaries for one of the correlation functions, we obtain
  $T^{(g_1-1)}_{I_1\|p|p|}T^{(g_2)}_{I_2\|pq|}$.
\item For $k=q_i\in I$: One correlation function gets an additional
  derivative wrt to the external face $q_i$ that is
  $(-\frac{\partial^{ext}}{\partial E_{q_i}}T^{(g_1)}_{I_1\setminus
    q_i\|pq_i|})T^{(g_2)}_{I_2\|pq|}$.
\end{itemize}
Including the deleted free propagator and applying the DSE
\eqref{DSE-Om} with $g_1$ and $I_1$, we finally obtain
	\begin{align}\nonumber
		\frac{-\lambda}{E_p+E_q}\sum_{\substack{g_1+g_2=g\\
				I_1 \uplus I_2=I}}T^{(g_2)}_{I_2\|pq|}\bigg[&\frac{1}{N}\sum_{k=1}^NT^{(g_1)}_{I_1\|pk|}+T^{(g_1-1)}_{I_1\|p|p|}-\sum_{q_i\in I}\frac{\partial^{ext}}{\partial E_{q_i}}T^{(g_1)}_{I_1\setminus q_i\|pq_i|}\bigg]\\\label{pert2}
		=\frac{-\lambda}{E_p+E_q}\sum_{\substack{g_1+g_2=g\\
				I_1 \uplus I_2=I}}T^{(g_2)}_{I_2\|pq|}&\Omega^{(g_1)}_{I_1,p}.
	\end{align}
	
      \item[c)] For $n=p$: There is a second case where the correlation
        function is not split into two components, see Fig.~\ref{fig:pert3}.
        The genus is reduced by one and we can again
        distinguish between different $k$.
\begin{figure}[h]
\includegraphics[width= 0.6\textwidth]{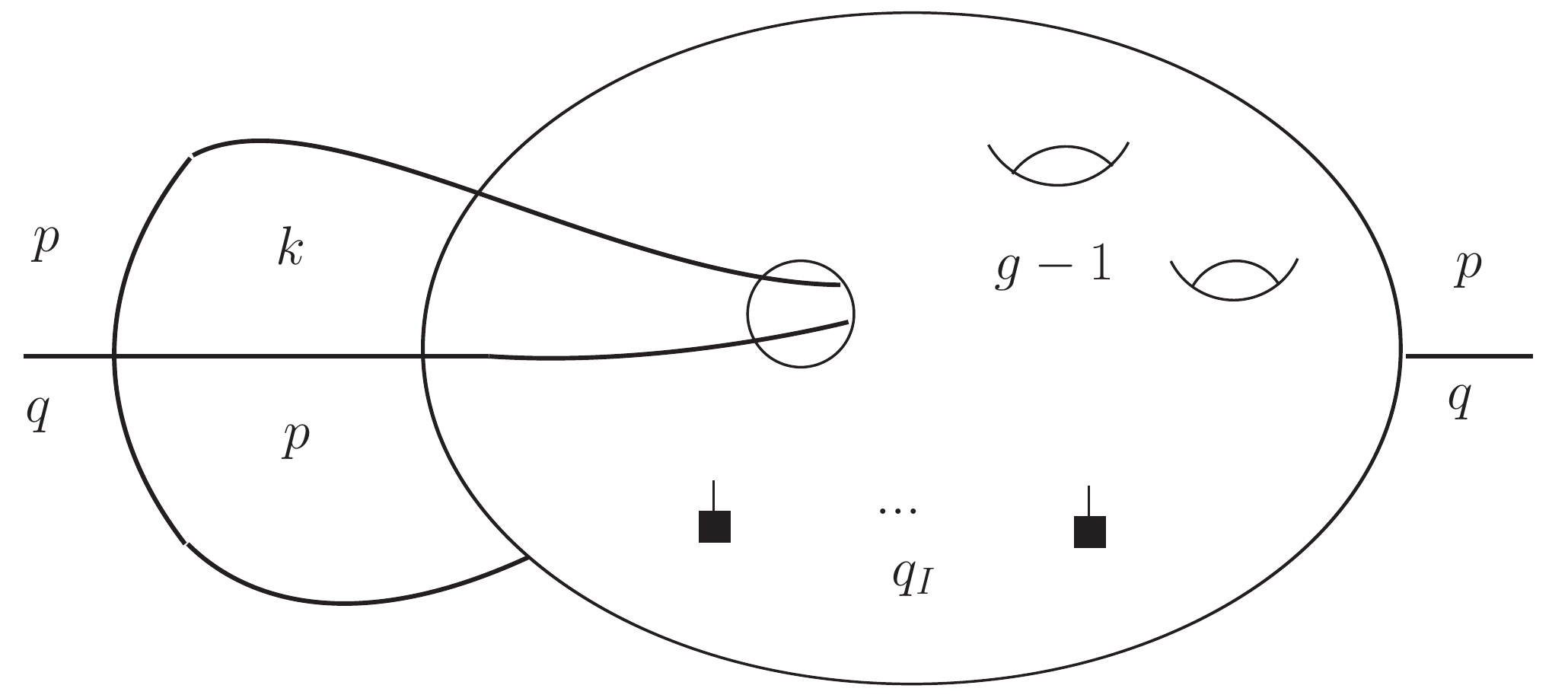}
\caption{The generalised 2-point function with $n=p$ but no seperation
  into two components. The first vertex is drawn in the middle seperately. 
			\label{fig:pert3}}
\end{figure}
\begin{itemize}
\item For generic $k$: There is an internal face labelled by $k$, we
  take the sum and have
  $\frac{1}{N}\sum_{k=1}^NT^{(g-1)}_{I\|pk|pq|}$.
\item For $k=q_i$: We have an additional derivative wrt to the
  external face labelled by $q_i$, that is
  $-\frac{\partial^{ext}}{\partial E_{q_i}}T^{(g-1)}_{I\|pq_i|pq|}$.
\item For $k=p$: The genus can be reduced by one more. In total three
  boundaries are generated two with length 1 and one with length
  2. The correlation function is of the form $T^{(g-2)}_{I\|p|p|pq|}$.
\item For $k=p$: Two boundaries can merge such that only one boundary
  of length 4 remains, $T^{(g-1)}_{I\|pppq|}$.
\item For $k=q$: Two boundaries can merge such that only one boundary
  of length 4 remains, $T^{(g-1)}_{I\|pqpq|}$.
\end{itemize}
These cases sum again together to the following
\begin{align}\nonumber
  \frac{-\lambda}{E_p+E_q}\bigg[&
  \frac{1}{N}\sum_{k=1}^NT^{(g-1)}_{I\|pk|pq|}
  -\sum_{q_i\in I}\frac{\partial^{ext}}{\partial E_{q_i}}T^{(g-1)}_{I\|pq_i|pq|}
  \\\nonumber
  &+T^{(g-2)}_{I\|p|p|pq|}+T^{(g-1)}_{I\|pppq|}+T^{(g-1)}_{I\|pqpq|}\bigg]
  \\\label{pert3}
=\frac{-\lambda}{E_p+E_q}&\,T^{(g-1)}_{I,p\|pq|}.
\end{align}
This can be seen by deriving
$-N\frac{\,\partial}{\partial E_{p'}}T^{(g)}_{I\|pq|}$, where we
consider the same action as described in Sec.~\ref{Sec.App.Om} plus
two additional cases: If the other adjacent face is an external face
either labelled by $p$ or $q$, then a boundary of length 4 is
generated, that is $T^{(g)}_{I\|pp'pq|}$ or $T^{(g)}_{I\|qp'qp|}$:
\begin{align*}
	&\frac{1}{N}\sum_{k=1}^NT^{(g)}_{I\|p'k|pq|}
	-\sum_{q_i\in I}\frac{\partial^{ext}}{\partial E_{q_i}}T^{(g)}_{I\|p'q_i|pq|}
	+T^{(g-1)}_{I\|p'|p'|pq|}+T^{(g)}_{I\|pp'pq|}+T^{(g)}_{I\|qp'qp|}\\
	&=T^{(g)}_{I,p'\|pq|}.
\end{align*}
This identity is then applied with $p'=p$ and $g\mapsto g-1$.

\item[d)] For $n=q_i\in I$: There is an overall derivative
  $-\frac{\partial^{ext}}{\partial E_{q_i}}$. Combinatorially, all the
  cases are exactly the same as for generic $n$, but with
  $I\setminus q_i$ instead of $I$, see Fig.~\ref{fig:pert4}.
\begin{figure}[h]
\includegraphics[width= 0.6\textwidth]{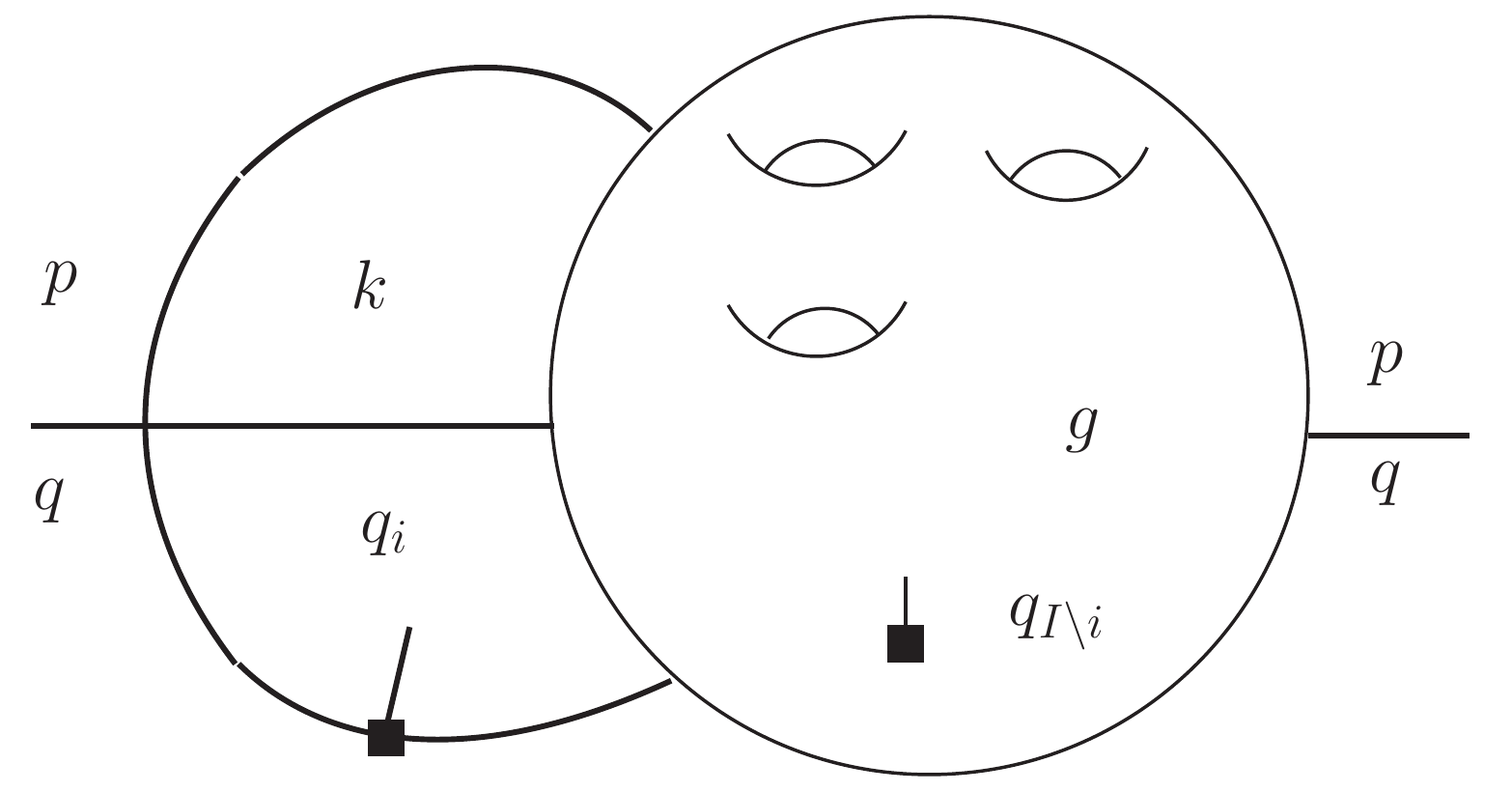}
\caption{The generalised 2-point function with $n=q_i$, which is
  similar to the generic $n$ case.
	\label{fig:pert4}}
\end{figure}
So we conclude equivalently to case a)
\begin{align}\label{pert4}
  -\frac{\lambda}{E_p+E_q}\sum_{q_i\in I}
  \frac{\partial^{ext}}{\partial E_{q_i}}\frac{T^{(g)}_{I\setminus q_i\|q_iq|}
    -T^{(g)}_{I\setminus q_i \|pq|}}{E_{q_i}-E_p}
\end{align}

\item[e)] For $n=q$: Fig.~\ref{fig:pert5} shows that for generic $k$
  the genus is reduced by one. Here again, we have several subcases:
\begin{figure}[h]
\includegraphics[width= 0.7\textwidth]{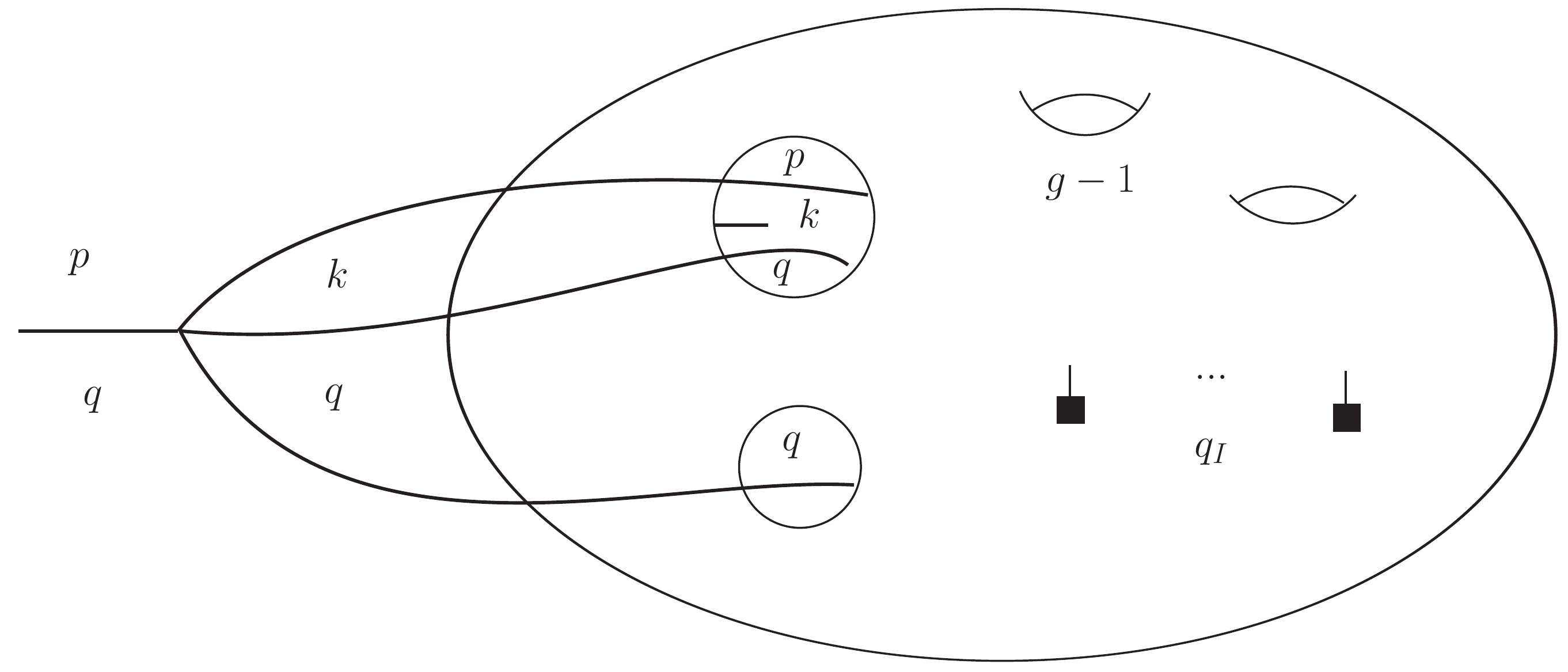}
\caption{The generalised 2-point function with $n=q$ is reduced by
  one genus, but has at least two boundaries.
\label{fig:pert5}}
\end{figure}
\begin{itemize}
\item For generic $k$: The sum over $k$ remains since $k$ is an
  interal face, that is
  $\frac{1}{N}\sum_{k=1}^NT^{(g-1)}_{I\|pkq|q|}$.
\item For generic $k=q_i\in I$: One overall derivative appears such
  that we have
  $-\frac{\partial^{ext}}{\partial E_{q_i}}T^{(g-1)}_{I\setminus
    q_i\|pq_iq|q|}$.
\item For $k=q$: Both boundaries can merge such that one boundary
  remains of length 4, $T^{(g-1)}_{I\|pqqq|}$.
\item For $k=q$: Also a splitting in two components can occur
  $T^{(g_1)}_{I_1\|pq|}T^{(g_2-1)}_{I_2\|q|q|}$ with $g_1+g_2=g$ and
  $I_1\uplus I_2=I$.
\item For $k=p$: A similar splitting in two components can
  occur $T^{(g_1)}_{I_1\|pq|}T^{(g_2-1)}_{I_2\|q|p|}$.
\item For $k=q$: Also a reduction by one further genus can occur,
  which has three boundaries of the form $T^{(g-2)}_{I\|q|q|pq|}$.
\item For $k=p$: Also a reduction by one further genus can occur,
  which has three boundaries of the form $T^{(g-2)}_{I\|q|p|pq|}$.
\end{itemize}
Summing all subcases, the identity of Lemma \ref{lem:WTI1+3P} applies
perfectly, and we conclude
\begin{align}\nonumber
  -\frac{\lambda}{E_p+E_q}\bigg[&
  \frac{1}{N}\sum_{k=1}^NT^{(g-1)}_{I\|pkq|q|}
  -\sum_{q_i\in I}\frac{\partial^{ext}}{\partial E_{q_i}}
  T^{(g-1)}_{I\setminus q_i\|pq_iq|q|}+T^{(g-1)}_{I\|pqqq|}
  \\\nonumber
  &+\sum_{\substack{g_1+g_2=g\\ I_1 \uplus I_2=I}}
  T^{(g_1)}_{I_1\|pq|}\bigg(T^{(g_2-1)}_{I_2\|q|q|}+T^{(g_2-1)}_{I_2\|q|p|}\bigg)
  +T^{(g-2)}_{I\|q|q|pq|}+T^{(g-2)}_{I\|q|p|pq|}\bigg]\\\label{pert5}
=\frac{\lambda}{E_p+E_q}&\,\frac{T^{(g-1)}_{I\|p|q|}-T^{(g-1)}_{I\|q|q|}}{E_p-E_q}.
\end{align}
\end{itemize}
Finally, including the free propagator for the genus $g=0$ and
$I=\emptyset$ we obtain the DSE
\begin{align*}
  T^{(g)}_{I\|pq|}=\frac{\delta_{0,g}\delta_{0,|I|}}{E_p+E_q}
  +\eqref{pert1}+\eqref{pert2}+\eqref{pert3}+\eqref{pert4}+\eqref{pert5}
\end{align*}
which coincides with \eqref{DSE-T2} after considering \eqref{delext} and its definition
\eqref{del}. 

\subsection{Graphical Derivation of \eqref{DSE-T1}}

The DSE 2) of Prop.~\ref{prop:TOm}  is also achieved by a bijection
between $T^{(g)}_{I\|p|q|}$, a generating series of ribbon graphs with
two boundaries of length 1 and other generating series via deletion of
the first vertex. For the generalised $1+1$-point function, the
prefactor after deletion becomes
\begin{align*}
	\frac{-\lambda}{E_p+E_p}
\end{align*}
instead of $\frac{-\lambda}{E_p+E_q}$ as it was for the generalised
2-point function. The first four cases are (up to some small
subtleties) the same as the cases a)-d) of the $T^{(g)}_{I\|pq|}$. The
fifth case, however, is different:
\begin{itemize}[e')]
\item[e')] We have two separated boundaries each of length 1
  labelled by $p$ and $q$. The deletion of the first vertex is
  divided in several subcases:
\begin{figure}[h]
\includegraphics[width= 0.5\textwidth]{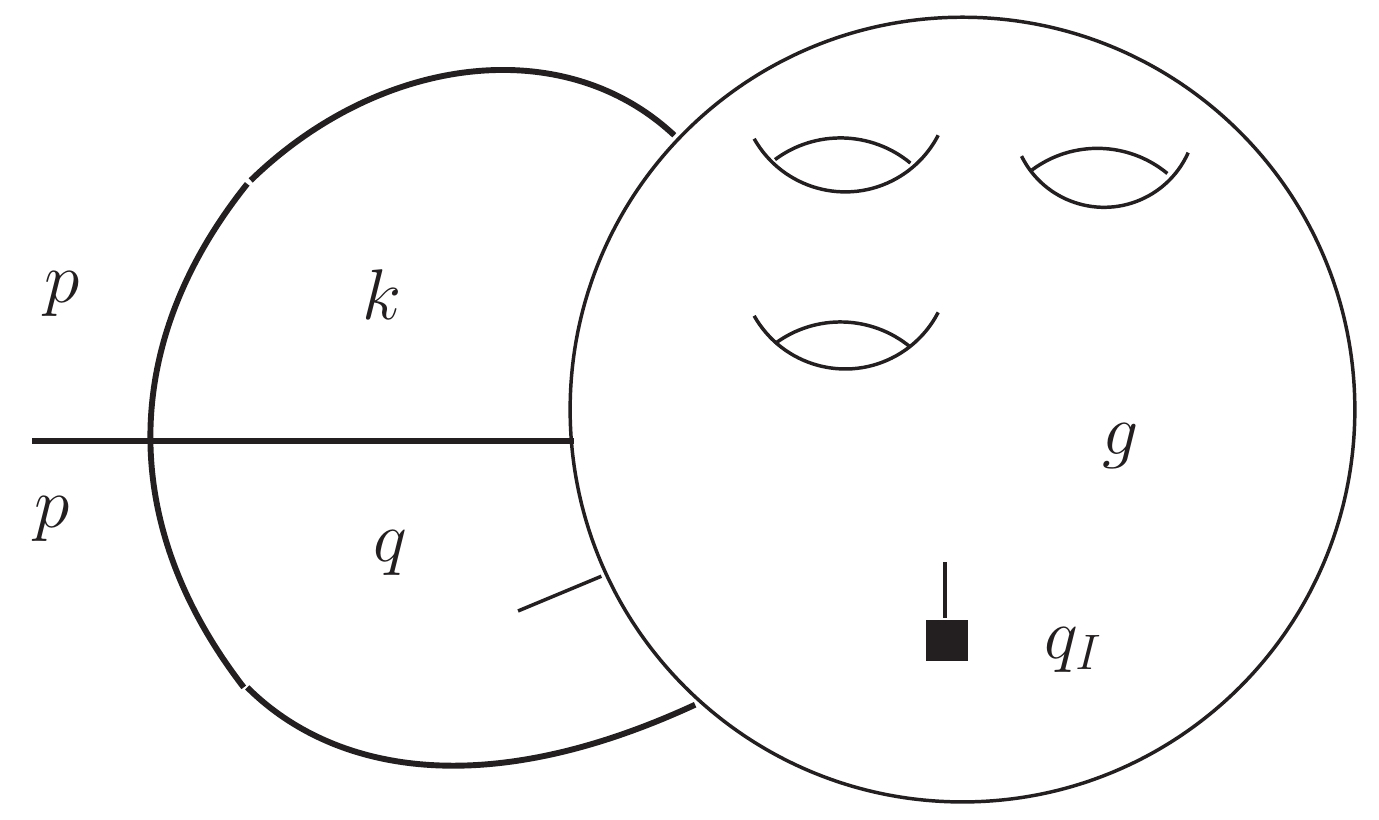}
\caption{The generalised $1+1$-point function with $n=q$ does not
  reduce the genus, but merges the two boundaries.
\label{fig:pert5b}}
\end{figure}
\begin{itemize}
\item For generic $k$: If $k$ is an internal face, the sum is taken
  and one boundary of length 4 remains, 
  $\frac{1}{N}\sum_{k=1}^NT^{(g)}_{I\|pkqq|}$.
\item For $k=q_i\in I$: The overall derivative is taken with $k=q_i$
  and $I\setminus q_i$, that is
  $-\frac{\partial^{ext}}{\partial E_{q_i}}T^{(g)}_{I\setminus
    q_i\|pq_iqq|}$.
\item For $k=q$: The graph can be split into two after deleting the
  first vertex. Two correlation functions occur each with boundary of
  length 2; we obtain $T^{(g_1)}_{I_1\|qq|}T^{(g_2)}_{I_2\|pq|}$ with
  $g_1+g_2=g$ and $I_1\uplus I_2=I$.
\item For $k=q$: There is also the case that two separated boundaries
  remain, but with one genus less,
  $T^{(g-1)}_{I\|q|qqp|}+T^{(g-1)}_{I\|qq|qp|}$.
\item For $k=p$: Two boundaries remain but genus is reduced by one,
  $T^{(g-1)}_{I\|p|qqp|}$.
\end{itemize}
In summary, we find for the case e') together with Lemma \ref{lem:WTI4P}
\begin{align*}
  \frac{-\lambda}{E_p+E_p}\bigg[&
  \frac{1}{N}\sum_{k=1}^NT^{(g)}_{I\|pkqq|}
  -\sum_{q_i\in I}\frac{\partial^{ext}}{\partial E_{q_i}}
  T^{(g)}_{I\setminus q_i\|pq_iqq|}
  \\
  &+\sum_{\substack{g_1+g_2=g\\ I_1 \uplus I_2=I}}
  T^{(g_1)}_{I_1\|qq|}T^{(g_2)}_{I_2\|pq|}+T^{(g-1)}_{I\|q|qqp|}
  +T^{(g-1)}_{I\|qq|qp|}+T^{(g-1)}_{I\|p|qqp|}\bigg]
  \\
=\frac{\lambda}{E_p+E_p}&\frac{T^{(g)}_{I\|qq|}-T^{(g)}_{I\|pq|}}{E_q-E_p}.
\end{align*}
\end{itemize} 
Including the first four cases similar to the graphical derivation of
the generalised 2-point function, we finally confirm the DSE
\eqref{DSE-T1} of the generalised $1+1$-point function
$T^{(g)}_{I\|p|q|}$ with the same considerations as for the
generalised 2-point function.

\section{Proofs for Section \ref{Sec:RecursT}}\label{App:RecT}

We will use the following well-known interpolation formula:
\begin{lemma}\label{lem:interpol}
	Let $f$ be a polynomial of degree $d-1\geq 0$ and 
	$x_1,...,x_d$ be pairwise distinct complex numbers.
	Then, for all $x\in \mathbb{C}$,
	\begin{align*}
		f(x)=L(x)\sum_{j=1}^d\frac{f_j}{(x-x_j)L'(x_j)},\qquad 
		\text{where } L(x)=\prod_{j=1}^d(x-x_j) \text{ and } f_j=f(x_j).
	\end{align*}
\end{lemma}
\noindent
We recall \cite{Grosse:2019jnv} that \eqref{eq:Gzw0} gives rise to 
the product representation
\begin{align}\label{2Pprod}
	\mathcal{G}^{(0)}(z,w)=\frac{1}{R(w)-R(-z)}
	\prod_{k=1}^{d}\frac{R(z)-R(-\hat{w}^k)}{R(z)-R(\varepsilon_k)}\;.
\end{align}

\begin{proposition} \label{prop2}
  Let $I=\{u_1,...,u_m\}$.  The DSE  \eqref{DSE-cT2} is solved by
\begin{align}
\mathcal{T}^{(g)}&(I\|z,w|)
=\lambda\mathcal{G}^{(0)}(z,w)\Res\displaylimits_{t\to z,-\hat{w}^j}
\frac{R'(t)\, dt}{(R(z)-R(t))(R(w)-R(-t))\mathcal{G}^{(0)}(t,w)}
\nonumber
\\
&\times\bigg[\sum_{\substack{I_1 \uplus I_2=I\\g_1+g_2=g\\
		(I_1,g_1)\neq (\emptyset,0)}}
\Omega^{(g_1)}_{|I_1|+1}(I_1,t) \mathcal{T}^{(g_2)}(I_2\|t,w|)
+\sum_{i=1}^m\frac{\partial}{\partial R(u_i)}
\frac{\mathcal{T}^{(g)}(I\backslash u_i\|u_i,w|)}{R(u_i)-R(t)}
\nonumber
\\
&+\frac{\mathcal{T}^{(g-1)}(I\|t|w|)
-\mathcal{T}^{(g-1)}(I\|w|w|)}{R(w)-R(t)}
+ \mathcal{T}^{(g-1)}(I,t\|t,w|)\bigg],
\label{res-tgm2}
\end{align}
where $v\in\{w,\hat{w}^1,...,\hat{w}^d\}$ are the solutions of
$R(v)=R(w)$. We employed the short-hand notation
$\Res\displaylimits_{t\to z,-\hat{w}^j} \equiv
\Res\displaylimits_{t\to z} + \sum_{j=1}^d\Res\displaylimits_{t\to
  -\hat{w}^j}$.
\begin{proof}
  The second term of the lhs of the DSE \eqref{DSE-cT2} is
  conveniently written as
\begin{align*}
-\frac{\lambda}{N}\sum_{k=1}^d r_k
\frac{\mathcal{T}^{(g)}(I\|\varepsilon_k,w|)}{R(\varepsilon_k)-R(z)}
&=\frac{\frac{\lambda}{N}\sum_{k=1}^d r_k
\mathcal{T}^{(g)}(I\|\varepsilon_k,w|) \prod_{i\neq k}^d
(R(z)-R(\varepsilon_i))}{\prod_{j=1}^d(R(z)-R(\varepsilon_j))}
\\
&=:\frac{f(R(z);w|I)}{\prod_{j=1}^d (R(z)-R(\varepsilon_j))}\;,
\end{align*}
where $f(\,.\,;w|I)$ is now a polynomial of degree $d-1$. Applying
Lemma \ref{lem:interpol} with
$L_w(t):=\prod_{j=1}^d (t-R(-\hat{w}^j))$, the interpolation formula
yields
\begin{align*}
  f(R(z);w|I)&= L_w(R(z))\sum_{j=1}^d\frac{f(R(-\hat{w}^j);w|I)}{
    (R(z)-R(-\hat{w}^j)) L_w'(R(-\hat{w}^j))}
\\
&=L_w(R(z))\sum_{j=1}^d\Res\displaylimits_{t\to -\hat{w}^j}
\frac{f(R(t);w|I)R'(t)dt}{(R(z)-R(t)) L_w(R(t))}\;,
\end{align*}
where the analyticity of $f(R(z);w|I)$ at $z=-\hat{w}^j$ was used.
Next, insert \eqref{DSE-cT2} again for $z\mapsto t$ near
$t=-\hat{w}^j$ at which the first term of the lhs vanishes (here it is
important that the integrand has only a simple pole at
$t=-\hat{w}^j$). Inserting it for $f(R(t);w|I)$ leads to
\begin{align}
&-\frac{\lambda}{N}\sum_{k=1}^dr_k 
\frac{\mathcal{T}^{(g)}(I\|\varepsilon_k,w|)}{R(\varepsilon_k)-R(z)}
\label{eq:Tsum}
\\
&=-\lambda(R(w){-}R({-}z))\mathcal{G}^{(0)}(z,w)
\sum_{j=1}^d \Res\displaylimits_{t\to -\hat{w}^j}
\frac{R'(t)dt}{(R(z){-}R(t))(R(w){-}R({-}t))\mathcal{G}^{(0)}(t,w)}  
\nonumber
\\
&\times\bigg[\sum_{\substack{I_1\uplus I_2=I\\g_1+g_2=g\\
    (I_1,g_1)\neq (\emptyset,0)}}
\Omega^{(g_1)}_{|I_1|+1}(I_1,t)\mathcal{T}^{(g_2)}(I_2\|t,w|)
+\sum_{i=1}^m\frac{\partial}{\partial R(u_i)}
\frac{\mathcal{T}^{(g)}(I\backslash u_i\|u_i,w|)}{R(u_i)-R(t)}
\nonumber
\\
&+\frac{\mathcal{T}^{(g-1)}(I\|t|w|)
-\mathcal{T}^{(g-1)}(I\|w|w|)}{R(w)-R(t)}
+ \mathcal{T}^{(g-1)}(I,t\|t,w|)\bigg]\;,
\nonumber
\end{align}
where the product representation \eqref{2Pprod} was inserted.
		
Next, compute for the same integrand the residue at $t= z$ (for
arbitrary $z$)
\begin{align*}
&\lambda(R(w)-R(-z))\mathcal{G}^{(0)}(z,w)
\Res\displaylimits_{t\to z}\frac{R'(t)dt}{(R(z)-R(t))(R(w)-R(-t))
	\mathcal{G}^{(0)}(t,w)}  \\
&\times\bigg[\sum_{\substack{I_1\uplus I_2=I\\g_1+g_2=g\\
		(I_1,g_1)\neq (\emptyset,0)}}
\Omega^{(g_1)}_{|I_1|+1}(I_1,t)\mathcal{T}^{(g_2)}(I_2\|t,w|)
+\sum_{i=1}^m\frac{\partial}{\partial R(u_i)}
\frac{\mathcal{T}^{(g)}(I\backslash u_i\|u_i,w|)}{R(u_i)-R(t)}\\
&+\frac{\mathcal{T}^{(g-1)}(I\|t|w|)
	-\mathcal{T}^{(g-1)}(I\|w|w|)}{R(w)-R(t)}
+ \mathcal{T}^{(g-1)}(I,t\|t,w|)\bigg]\\
&=-\lambda\bigg[\sum_{\substack{I_1\uplus I_2=I\\g_1+g_2=g\\
		(I_1,g_1)\neq (\emptyset,0)}} \!\!\!\!\!\!
\Omega^{(g_1)}_{|I_1|+1}(I_1,z)\mathcal{T}^{(g_2)}(I_2\|z,w|)
+\sum_{i=1}^m\frac{\partial}{\partial R(u_i)}
\frac{\mathcal{T}^{(g)}(I\backslash u_i\|u_i,w|)}{R(u_i)-R(z)}
\\[-3ex]
&\qquad\qquad\qquad\qquad +\frac{\mathcal{T}^{(g-1)}(I\|z|w|)
	-\mathcal{T}^{(g-1)}(I\|w|w|)}{R(w)-R(z)}
+ \mathcal{T}^{(g-1)}(I,z\|z,w|)\bigg]
\\
&=(R(w)-R(-z))\mathcal{T}^{(g)}(I\|z,w|)
-\frac{\lambda}{N}\sum_{k=1}^d r_k 
\frac{\mathcal{T}^{(g)}(I\|\varepsilon_k,w|)}{R(\varepsilon_k)-R(z)}\;.
\end{align*}
Summing both expressions gives the assertion. 
\end{proof} 
\end{proposition}

\begin{remark}
  The residue formula of Proposition \ref{prop2} is equivalent to the
  formula found in \cite{Schurmann:2019mzu-v3} via inversion of Cauchy
  matrices
  $\big(\frac{1}{R(\varepsilon_j)-R(-\hat{w}^k)}\big)_{j,k}$. This is
  not surprising as the derivation of the inverse Cauchy matrix in
  \cite{Schechter:1959??} is mainly based on the interpolation
  formula.
\end{remark}

The proof of Corollary \ref{cor2+} is:
\begin{proof}
We rewrite one of the terms in Proposition \ref{prop2} as 
\begin{align*}
&\frac{\partial}{\partial R(u_i)}\Res\displaylimits_{t\to z,-\hat{w}^j}
\frac{R'(t)dt\prod_{k=1}^d(R(t)-R(\varepsilon_k)) }{(R(z)-R(t)) L_w(R(t))}
\frac{\mathcal{T}^{(g)}(I\backslash u_i\|u_i,w|)}{R(u_i)-R(t)}
\\
&=\frac{\partial}{\partial R(u_i)}\Res\displaylimits_{x\to R(z),R(-\hat{w}^j)}
\frac{dx\prod_{k=1}^d(x-R(\varepsilon_k)) }{(R(z)-x) L_w(x)}
\frac{\mathcal{T}^{(g)}(I\backslash u_i\|u_i,w|)}{R(u_i)-x}
\\
&=-\frac{\partial}{\partial R(u_i)}\Res\displaylimits_{x\to R(u_i)}
\frac{\prod_{k=1}^d(x-R(\varepsilon_k)) }{(R(z)-x) L_w(x)}
\frac{\mathcal{T}^{(g)}(I\backslash u_i\|u_i,w|)}{R(u_i)-x}
\\
&=\frac{1}{R'(u_i)}\frac{\partial}{\partial u_i}
\frac{\prod_{k=1}^d(R(u_i)-R(\varepsilon_k)) }{(R(z)-R(u_i)) 
	L_w(R(u_i))}\mathcal{T}^{(g)}(I\backslash u_i\|u_i,w|)
\\
&=\Res\displaylimits_{t\to u_i}\frac{R'(t)dt
	\prod_{k=1}^d(R(t)-R(\varepsilon_k)) }{(R(z)-R(t))
	L_w(R(t))}\Big\{\frac{1}{R'(u_i)R'(t)(t-u_i)^2 }
\mathcal{T}^{(g)}(I\backslash u_i\|t,w|)\Big\}\;,
\end{align*}
where we substituted $t\mapsto x=R(t)$, then moved the integration
contour and finally represented the result in form of a residue
formula.  Proposition \ref{prop:Om02} implies that
$\frac{1}{R'(u_i)R'(t)(t-u_i)^2 }$ is partially given in
$\Omega^{(0)}_{2}(u_i,t)$.  According to Lemma \ref{lem:analy},
$\Omega^{(0)}_2(z,w)$ is the only correlation function divergent on
the diagonal so that the terms in $\{~ \}$ extend to
$\sum_{I_1,I_2,g_1,g_2}
\Omega^{(g_1)}_{|I_1|+1}(I_1;t)\mathcal{T}^{(g_2)}(I_2\|t,w|)$ and
finally to
$\sum_{I_1,I_2,g_1,g_2}
\Omega^{(g_1)}_{|I_1|+1}(I_1;t)\mathcal{T}^{(g_2)}(I_2\|t,w|)+
\mathcal{T}^{(g-1)}(I,t\|t,w|)
+\frac{\mathcal{T}^{(g-1)}(I\|t|w|)}{R(w)-R(t)}$.
	
Analogously, the term 
\begin{align*}
&\Res\displaylimits_{t\to z,-\hat{w}^j}
\frac{R'(t)dt\prod_{k=1}^d(R(t)-R(\varepsilon_k)) }{(R(z)-R(t)) L_w(R(t))}
\frac{\mathcal{T}^{(g-1)}(I\|w|w|)}{R(w)-R(t)}
\end{align*}
is represented by the $w$-residue in the Corollary, where again
vanishing terms are added after substitution and moving the
integration contour.
\end{proof}
\noindent
As argued in the proof of Corollary \ref{cor2+}, Lemma
\ref{lem:analy} implies that the residue at
$t=u_i$ contributes only via $\Omega^{(0)}_2(u_i,t)$ and
the residue at $t=w$ only via
$\frac{\mathcal{T}^{(g-1)}(I\|t|w|)}{R(w)-R(t)}$.

\begin{proposition}\label{prop1+1}
Let $I=\{u_1,...,u_m\}$. The DSE \eqref{DSE-cT1} is solved by
\begin{align}
\mathcal{T}^{(g)}(I\|z|w|)
&=\frac{\lambda\prod_{j=1}^d\frac{R(z)-R(\alpha_j)}{
		R(z)-R(\varepsilon_j)}}{(R(z)-R(-z))}
\Res\displaylimits_{t\to z ,\alpha_j}\frac{R'(t)\, dt
  \prod_{k=1}^d(R(t)-R(\varepsilon_k))}{(R(z)-R(t))
  \prod_{k=1}^d (R(t)-R(\alpha_k))}  
\nonumber
\\
&\times\bigg[\sum_{\substack{I_1\uplus I_2=I\\g_1+g_2=g\\ 
    (I_1,g_1)\neq (\emptyset,0)	}} \!\!\!\!\!\!
\Omega^{(g_1)}_{|I_1|+1}(I_1,t) \mathcal{T}^{(g_2)}(I_2\|t|w|) 
+ \mathcal{T}^{(g-1)}(I,t\|t|w|)
\nonumber
\\
&  +\sum_{i=1}^m\frac{\partial}{\partial R(u_i)}
\frac{\mathcal{T}^{(g)}(I\backslash u_i\|u_i|w|)}{R(u_i)-R(t)}  
+\frac{\mathcal{T}^{(g)}(I\|t,w|)
	-\mathcal{T}^{(g)}(I\|w,w|)}{R(w)-R(t)}\bigg],
\label{res-tgm1}
\end{align}
where $v\in\{0,\pm\alpha_j\}$ are the $2d+1$ solutions of
$R(v)-R(-v)=0$.
\begin{proof}
  Similar to the proof of Proposition \ref{prop2}, but with $d$
  distinct points $x_k=R(\alpha_k)$ for the interpolation formula of
  Lemma \ref{lem:interpol}.
\end{proof}
\end{proposition}
The proof of Corollary \ref{cor1+} works in a completely analogous way.

\section{Proof of 
Conjecture~\ref{conj:Omega-poles} for 
\texorpdfstring{$g=0$}{g=0}}
\label{app:poles-g0} 

It is convenient to introduce
\begin{align}
\mathcal{T}^{(g)}(u_1,..,u_m\|z,w|)
&=:\frac{\partial^m \mathcal{U}^{(g)} (u_1,..,u_m\|z,w|)
}{\partial R(u_1)\cdots \partial R(u_m)} \:,
\nonumber
\\
\mathcal{T}^{(g)}(u_1,..,u_m\|z|w|)
&=:\frac{\partial^m\mathcal{U}^{(g)}(u_1,..,u_m\|z|w|)
}{\partial R(u_1)\cdots \partial R(u_m)} \:,
\nonumber
\\
\Omega^{(g)}_{m+1}(u_1,..,u_m,z)&=:\frac{\partial^m
\mathcal{W}^{(g)}_{m+1}(u_1,..,u_m;z)}{\partial R(u_1)\cdots 
\partial R(u_m)} 
\;.
\label{TUOmW}
\end{align}
In these variables the DSE (\ref{DSE-cT2}) reads for $g=0$ and $m\geq 1$
\begin{align}
&(R(w)-R(-z))\mathcal{U}^{(0)}(u_1,...,u_m\|z,w|)
-\frac{\lambda}{N}\sum_{k=1}^d \frac{r_k
\mathcal{U}^{(0)}(u_1,...,u_m\|\varepsilon_k,w|)}{R(\varepsilon_k)-R(z)}
\nonumber
\\
&=-\lambda\!\!\!\!\!\!
\sum_{\substack{ I_1\uplus I_2=\{u_1,...,u_m\}\\ 
I_1\neq\emptyset}} \hspace*{-2em}
\mathcal{W}^{(0)}_{|I_1|+1}(I_1;z)
\mathcal{U}^{(0)}(I_2\|z,w|) 
-\lambda   \sum_{j=1}^m
\frac{\mathcal{U}^{(0)}(u_1,..\check{u_j}..,u_m\|u_j,w|)}{
R(u_j)-R(z)}\;,
\label{DSE-U0}
\end{align}
where $\mathcal{U}^{(0)}(\emptyset\|z,w|)=\mathcal{G}^{(0)}(z,w)$.
The DSE (\ref{DSE-Omega-complex}) becomes for $m\geq 2$
\begin{align}
&R'(z)\mathfrak{G}_0(z)\mathcal{W}^{(0)}_{m+1}(u_1,...,u_m;z)
-\frac{\lambda}{N^2}\sum_{n,k=1}^d 
\frac{r_nr_k \mathcal{U}^{(0)}(u_1,...,u_m\|\varepsilon_k,\varepsilon_n|)}{
(R(\varepsilon_k)-R(z))(R(\varepsilon_n)-R(-z))}
\nonumber
\\
&=
-\sum_{\substack{I_1\uplus I_2=\{u_1,...,u_m\}\\ I_1\neq \emptyset \neq I_2}}
\hspace*{-2em}
\mathcal{W}^{(0)}_{|I_1+1|}(I_1;z)\frac{\lambda}{N}\sum_{n=1}^d 
\frac{r_n  \mathcal{U}^{(0)}(I_2\|z,\varepsilon_n|)}{
R(\varepsilon_n)-R(-z)} 
\nonumber
\\
&-\sum_{j=1}^m
\frac{\frac{\lambda}{N}\sum_{n=1}^d r_n
\frac{\mathcal{U}^{(0)}(u_1,..\check{u_j}..,u_m\|u_j,\varepsilon_n|)}{
R(\varepsilon_n)-R(-z)} }{R(u_j)-R(z)}
-\sum_{j=1}^m\mathcal{U}^{(0)}(u_1,..\check{u_j}..,u_m\|u_j,z|)\;.
\label{DSE-W0}
\end{align}
\begin{lemma}
  For all $m\geq 1$,  the function $\mathcal{W}^{(0)}_{m+1}(u_1,\dots,u_m;z)$ is
  holomorphic in every $z=\widehat{u_j}^k$, whereas 
  $\mathcal{U}^{(0)}(u_1,\dots,u_m\|z,w|)$ has simple poles there 
  with residue
  \begin{align}
&    \Res\displaylimits_{z\to \widehat{u_j}^k}   \mathcal{U}^{(0)}(u_1,...,u_m\|z,w|)dz
\label{ResU0}
\\
&=-\sum_{l=1}^{m} \sum_{\substack{I_1\uplus...\uplus I_{l}=\{u_1,...\check{u_j}...,u_m\}
   \\ I_2,...,I_l\neq \emptyset \text{ for }l>1}}
\hspace*{-1cm}   \frac{(-\lambda)^l\mathcal{U}^{(0)}(I_1\|u_j,w|)
\prod_{i=2}^l  \mathcal{W}^{(0)}_{|I_i|+1}(I_i;z)
}{R'(z)(R(w)-R(-z))^l}\Big|_{z=\widehat{u_j}^k}\;.
\nonumber
\end{align}
\begin{proof} 
  By induction in $m$, starting with Proposition \ref{prop:Om02} for
  $m=1$.  Assume that the assertion concerning
  $\mathcal{W}_{k+1}^{(0)}$ is true for all $k\leq m-1$. Then
  (\ref{ResU0}) is recursively obtained when taking the residue in
  (\ref{DSE-U0}) and inserting it repeatedly into itself. Next, taking
  (\ref{ResU0}) into account, the residue of (\ref{DSE-W0}) at
  $z=\widehat{u_j}^k$ collapses to
\begin{align*}
  &\Res\displaylimits_{z\to \widehat{u_j}^k}
  R'(z)\mathfrak{G}_0(z)\mathcal{W}^{(0)}_{m+1}(u_1,...,u_m;z)dz
  \\
  &=
  \Res\displaylimits_{z\to \widehat{u_j}^k}\Big[
\frac{1}{N}  \sum_{n=1}^d r_n \mathcal{U}^{(0)}(u_1,...,u_m\|z,\varepsilon_n|)
-\sum_{i=1}^m\mathcal{U}^{(0)}(u_1,..\check{u_i}..,u_m\|u_i,z|)\Big]dz\;.
\nonumber
\end{align*}
But the rhs is $\Res\displaylimits_{z\to \widehat{u_j}^k}
\mathcal{W}^{(0)}_{m+1}(u_1,...,u_m;z)dz$ when expressing (\ref{DSE-cOm}) in terms of
$\mathcal{W}$ and $\mathcal{U}$. With
$ R'(\widehat{u_j}^k)\mathfrak{G}_0(\widehat{u_j}^k)\neq 1$ we finish the proof.
\end{proof}  
\end{lemma}

\begin{lemma}  
 For all $m\geq 1$,  the function $\mathcal{W}^{(0)}_{m+1}(u_1,\dots,u_m;z)$ is
  holomorphic in every $z=\pm \widehat{\varepsilon_k}^j$, whereas 
  $\mathcal{U}^{(0)}(u_1,\dots,u_m\|z,w|)$ has simple poles 
at $z=\widehat{\varepsilon_k}^j$ and is regular at 
$z=-\widehat{\varepsilon_k}^j$.
\begin{proof}
By induction in $m$, starting from the true statement 
for $\mathcal{G}^{(0)}(z,w)$ and $\mathcal{W}^{(0)}_{2}(u;z)$. 
If the statement is true for all $\mathcal{W}^{(0)}_{|I|+1}(I;z)$ with
$|I|+1\leq m$ and $\mathcal{U}^{(0)}(I\|..|)$ with $|I|\leq
m-1$, then the rhs of (\ref{DSE-U0}) has at most simple poles at 
$z=\widehat{\varepsilon_k}^j$ and no poles at
$z=-\widehat{\varepsilon_k}^j$. The same is true for the second 
term of the lhs of  (\ref{DSE-U0}) so that the statement extends to 
$\mathcal{U}^{(0)}(u_1,\dots,u_m\|z,w|)$. 
This means that the rhs and the second term of the lhs of 
(\ref{DSE-W0}) have at most simple poles at 
$z=\pm \widehat{\varepsilon_k}^j$. Since the prefactor
$\mathfrak{G}_0(z)$ has simple poles at every 
$z=\pm \widehat{\varepsilon_k}^j$ and 
$R'(\pm \widehat{\varepsilon_k}^j)$ is regular, 
the function $\mathcal{W}^{(0)}_{m+1}(u_1,\dots,u_m;z)$ must be
regular at $z=\pm \widehat{\varepsilon_k}^j$.
\end{proof}
\end{lemma}

\begin{lemma}
The functions $\mathcal{W}^{(0)}_{m+1}(u_1,..,u_m;z)$  and 
$\mathcal{U}^{(0)}(u_1,..,u_m\|z,w|)$  are regular at $z=-\varepsilon_n$.
\begin{proof}
No term in (\ref{DSE-U0}) is singular for $z=-\varepsilon_n$, some of
them even vanish because of $R(-\varepsilon_n)=\infty$. 
The singular denominators $\frac{1}{R(\varepsilon_n)-R(-z)}$ in 
(\ref{DSE-W0}) are protected by $\frac{1}{R'(z)}\to 0$ for 
$z\to -\varepsilon_n$.
\end{proof}
\end{lemma}
By construction all functions are holomorphic at
$z=\varepsilon_n$. This leaves
the opposite diagonals $z=-u_k$ and the ramification points 
$z=\beta_i$ (from the prefactor $R'(z)$ in (\ref{DSE-W0})) as the only
possible location of poles in
$\mathcal{W}^{(0)}_{m+1}(u_1,...,u_m;z)$. These are preserved by
differentiation to $\Omega^{(0)}_{m+1}(u_1,...,u_m,z)$, so that the 
proof of Conjecture~\ref{conj:Omega-poles} for 
$g=0$ is complete. 

For $g\geq 1$ we also expect poles at $z=0$ inherited from 
the initial value $\mathcal{G}^{(g-1)}(z|z)$ and from the poles at 
$z=-z$ in $\mathcal{T}^{(g-1)}(u_1,...,u_m,z\|z,\varepsilon_n|)$.  
Also absence of poles at $z=\pm \alpha_k$ is only relevant for $g\geq 1$.
We also note
\begin{lemma}
We have
\begin{align}
&    \Res\displaylimits_{z\to \widehat{u_j}^k}  
\mathcal{U}^{(0)}(u_1,...,u_m\|z|w|)dz
\label{ResU011} 
\\
&=-\sum_{l=1}^{m} \sum_{\substack{I_1\uplus...\uplus I_{l}
=\{u_1,...\check{u_j}...,u_m\} \\ I_2,...,I_l\neq \emptyset \text{ for }l>1}}
\hspace*{-1cm}   
\frac{(-\lambda)^l\mathcal{U}^{(0)}(I_1\|u_j|w|)
\prod_{i=2}^l  \mathcal{W}^{(0)}_{|I_i|+1}(I_i;z)
}{R'(z)(R(z)-R(-z))^l}\Big|_{z=\widehat{u_j}^k}\;
\nonumber
\\
&-\lambda \sum_{n=1}^m \frac{1}{R'(z)(R(z)-R(-z))^n}
\nonumber
\\
&\quad\times 
  \sum_{l=n}^{m} 
\sum_{\substack{I_1\uplus...\uplus I_{l}=\{u_1,...\check{u_j}...,u_m\}
 \\ I_2,...,I_l\neq \emptyset \text{ for }l>1}}
\frac{(-\lambda)^l\mathcal{U}^{(0)}(I_1\|u_j,w|)
\prod_{i=2}^l  \mathcal{W}^{(0)}_{|I_i|+1}(I_i;z)
 }{(R(w)-R(z)) (R(w)-R(-z))^{l+1-n}}
\Big|_{z=\widehat{u_j}^k}.\nonumber
\end{align}
\end{lemma}

\section{Solution of the Recursion for Small Degree}

\label{app:solution}

\subsection{Preparations for $g=0$}

We formulate the proof in terms of $\mathcal{U}^{(0)}$ 
and $\mathcal{W}^{(0)}_m$ introduced in (\ref{TUOmW}). 
Equation (\ref{eq:Tsum}) then translates for $g=0$ to
\begin{align}
&\frac{\lambda}{N}\sum_{k=1}^dr_k 
\frac{\mathcal{U}^{(0)}(I\|\varepsilon_k,w|)}{R(\varepsilon_k)-R(z)}
\label{eq:Tsum1}
\\
 &=\lambda(R(w){-}R({-}z))\mathcal{G}^{(0)}(z,w)
\sum_{j=1}^d \Res\displaylimits_{t\to -\hat{w}^j}
\frac{R'(t)dt}{(R(z){-}R(t))(R(w){-}R({-}t))\mathcal{G}^{(0)}(t,w)}  
\nonumber
\\
&\times\bigg[\sum_{\substack{I_1\uplus I_2=I\\
I_1\neq \emptyset}} \mathcal{W}^{(0)}_{|I_1|+1}(I_1;t)\mathcal{U}^{(0)}(I_2\|t,w|)
+\sum_{i=1}^m
\frac{\mathcal{U}^{(0)}(I\backslash u_i\|u_i,w|)}{R(u_i)-R(t)}
\bigg],
\nonumber
\end{align}
where $I=\{u_1,...,u_m\}$ and $m\geq 1$. Using the
product representation (\ref{2Pprod}) and the interpolation formula
of Lemma~\ref{lem:interpol} it is straightforward to establish
\begin{align*}
&\sum_{j=1}^d \Res\displaylimits_{t\to -\hat{w}^j}
\frac{R'(t)dt}{(R(z){-}R(t))(R(u_i)-R(t))(R(w){-}R({-}t))\mathcal{G}^{(0)}(t,w)}  
\\
&= \frac{1}{R(z){-}R(u_i)} \Big(\frac{1}{(R(w)-R({-}u_i))\mathcal{G}^{(0)}(u_i,w)}
-\frac{1}{(R(w)-R({-}z))\mathcal{G}^{(0)}(z,w)}\Big).
\end{align*}
Defining $\mathcal{U}^{(g)}(I\|z,w|)
=:\mathcal{G}^{(0)}(z,w)\tilde{\mathcal{U}}^{(g)}(I\|z,w|)$
(with $\tilde{\mathcal{U}}^{(g)}(\emptyset\|z,w|)\equiv 1$),
equation
(\ref{eq:Tsum1}) becomes
\begin{align}
&\frac{\lambda}{N}\sum_{k=1}^dr_k 
\frac{\mathcal{U}^{(0)}(I\|\varepsilon_k,w|)}{R(\varepsilon_k)-R(z)}
\label{eq:Tsum2}
\\
&=\lambda(R(w){-}R({-}z))\mathcal{G}^{(0)}(z,w)
\Big(\sum_{i=1}^m
\frac{\tilde{\mathcal{U}}^{(0)}(I\backslash u_i\|u_i,w|)}{(R(z)-R(u_i))(R(w)-R(-u_i))}
\nonumber
\\
&+\sum_{\substack{I_1\uplus I_2=I\\
    I_1\neq \emptyset}}
\sum_{j=1}^d \frac{R'({-}\hat{w}^j)\mathcal{W}^{(0)}_{|I_1|+1}(I_1;{-}\hat{w}^j)
\tilde{\mathcal{U}}^{(0)}(I_2\|{-}\hat{w}^j,w|)}{R'(\hat{w}^j) (R(z)-R(-\hat{w}^j))}
\Big) 
\nonumber
\\
&-\sum_{i=1}^m
\frac{\lambda \mathcal{U}^{(0)}(I\backslash u_i\|u_i,w|)}{R(z)-R(u_i)}\;.
\nonumber
\end{align}
The limit $w=q$, $z\to -q$ of (\ref{eq:Tsum2}) is
\begin{align}
\label{eq:Tsum3}
\frac{\lambda}{N}\sum_{k=1}^dr_k 
\frac{\mathcal{U}^{(0)}(I\|\varepsilon_k,q|)}{R(\varepsilon_k)-R(-q)}
&=\lambda R'(q)\mathfrak{G}_0(q) \mathfrak{U}^{(0)}(I\|q)
-\lambda \sum_{i=1}^m
\frac{\mathcal{U}^{(0)}(I\backslash u_i\|u_i,q|)}{R(-q)-R(u_i)}
\end{align}
where
\begin{align}
  \mathfrak{U}^{(0)}(I\|q)
&=\sum_{\substack{I_1\uplus I_2=I\\
    I_1\neq \emptyset}}
\sum_{j=1}^d \frac{R'(-\hat{q}^j)\mathcal{W}^{(0)}_{|I_1|+1}(I_1;-\hat{q}^j)
\tilde{\mathcal{U}}^{(0)}(I_2\|{-}\hat{q}^j,q|)}{R'(\hat{q}^j) (R(-q)-R(-\hat{q}^j))}
\nonumber
\\
&-\sum_{i=1}^m
\frac{\tilde{\mathcal{U}}^{(0)}(I\backslash u_i\|u_i,q|)}{(R(u_i)-R(-q))(R(q)-R(-u_i))}
\;.
\label{frakUIq}
\end{align}
On the other hand we insert (\ref{eq:Tsum2}) into the lhs of
the Dyson-Schwinger equation (\ref{DSE-cT2}), restricted
to $g=0$ and translated
to $\mathcal{U}^{(0)}$ and $\mathcal{W}^{(0)}_m$:
\begin{align}
\tilde{\mathcal{U}}^{(0)}(I\|z,w|)
&=
\sum_{\substack{I_1\uplus I_2=I\\
    I_1\neq \emptyset}}
\sum_{j=1}^d \frac{\lambda R'(-\hat{w}^j)\mathcal{W}^{(0)}_{|I_1|+1}(I_1;-\hat{w}^j)
  \tilde{\mathcal{U}}^{(0)}(I_2\|{-}\hat{w}^j,w|)}{R'(\hat{w}^j) (R(z)-R(-\hat{w}^j))}
\nonumber
\\
&+\sum_{i=1}^m
\frac{\lambda \tilde{\mathcal{U}}^{(0)}(I\backslash u_i\|u_i,w|)}{(R(z)-R(u_i))(R(w)-R(-u_i))}
\nonumber
\\
&-
\sum_{\substack{ I_1\uplus I_2=I\\ 
I_1\neq\emptyset}} 
\frac{\lambda\mathcal{W}^{(0)}_{|I_1|+1}(I_1;z)
\tilde{\mathcal{U}}^{(0)}(I_2\|z,w|) }{R(w)-R(-z)}\;.
\label{UIzw}
\end{align}

With these preparations we can eliminate $\mathfrak{G}_0$ in the residue formula
of Corollary~\ref{corr:Omega-residue}:
\begin{proposition}
For $I=\{u_1,...,u_m\}$ and $m\geq 2$ one has
\begin{align}
&R'(z)\mathcal{W}^{(0)}_{|I|+1}(I;z)
\label{W0-noG}
\\
&=\Res\displaylimits_{q\to -u_{1,...,m},\beta_{1,...2d}}
\frac{\lambda\,dq}{(q-z)}
\bigg[
\sum_{\substack{I_1\uplus I_2=I\\ I_1,I_2\neq \emptyset}}
R'(q)\mathcal{W}^{(0)}_{|I_1|+1}(I_1;q) \mathfrak{U}^{(0)}(I_2\|q)\bigg]
\nonumber
\\
&-
\lambda \sum_{k=1}^m \frac{\mathfrak{U}^{(0)}(I{\setminus}u_k \|u_k)}{z+u_k}\;.
\nonumber
\end{align}
\begin{proof}
We insert (\ref{eq:Tsum3}) into Corollary~\ref{corr:Omega-residue},
restricted to $g=0$ and translated to $\mathcal{W}^{(0)}_m$.
The second term on the rhs (sum over $i$) then cancels
the last line of (\ref{UIzw}):
\begin{align*}
&R'(z)\mathcal{W}^{(0)}_{|I|+1}(I;z)
\\
&=\Res\displaylimits_{q\to -u_{1,...,m},\beta_{1,...2d}}
\frac{\lambda\,dq}{(q-z)}
\bigg[
\sum_{\substack{I_1\uplus I_2=I\\ I_1,I_2\neq \emptyset}}
R'(q)\mathcal{W}^{(0)}_{|I_1|+1}(I_1;q) \mathfrak{U}^{(0)}(I_2\|q)
\\
&+
\sum_{k=1}^m \frac{\mathcal{G}^{(0)}(u_k,q)}{\mathfrak{G}_0(q)}
\Big\{
\sum_{\substack{I_1\uplus I_2=I\setminus u_k\\
    I_1\neq \emptyset}}
\sum_{j=1}^d \frac{R'(-\widehat{u_k}^j)\mathcal{W}^{(0)}_{|I_1|+1}(I_1;-\widehat{u_k}^j)
\tilde{\mathcal{U}}^{(0)}(I_2\|{-}\widehat{u_k}^j,u_k|)}{R'(\widehat{u_k}^j) (R(q)-R(-\widehat{u_k}^j))}
\\
&\qquad\qquad\qquad +\sum_{\substack{l=1 \\ l\neq k}}^m
\frac{\tilde{\mathcal{U}}^{(0)}(I\backslash \{u_l,u_k\}\|u_l,u_k|)}{(R(q)-R(u_l))(R(u_k)-R(-u_l))}
\Big\}\bigg]\;.
\nonumber
\end{align*}
The last two lines only contribute to the residue at $q=-u_k$ via the
first-order pole of $\mathcal{G}^{(0)}(u_k,q)$. The residue cancels
$\mathfrak{G}_0$ and otherwise amounts to replace $q\mapsto
-u_k$. This produces $\mathfrak{U}^{(0)}(I{\setminus}u_k \|u_k)$
according to (\ref{frakUIq}).
\end{proof}
\end{proposition}
We prove in the next subsections that the solution of (\ref{W0-noG})
confirms the following conjecture for $m=2$ and $m=3$:
\begin{conjecture}
\label{conj-Wplanar}
  For any $I=\{u_1,..,u_m\}$ with $m\geq 2$ one has
\begin{align}
&R'(z)  \mathcal{W}^{(0)}_{m+1}(I;z)
\\
&=
\lambda \sum_{i=1}^{2d}\Res\displaylimits_{q\to \beta_i} \tilde{K}_i(z,q)
\sum_{\substack{I_1\uplus I_2=I\\  I_1,I_2\neq \emptyset}}
R'(q)\mathcal{W}^{(0)}_{|I_1|+1}(I_1;q) R'(\sigma_i(q))\mathcal{W}^{(0)}_{|I_2|+1}(I_2;\sigma_i(q))
\nonumber
\\
&+
\sum_{k=1}^{m} 
\Big(\Res\displaylimits_{q\to u_k} \tilde{K}_{u_k}(z,q) \!\!\!\!
\sum_{\substack{I_1\uplus ... \uplus I_s=I{\setminus} u_k\\ I_1,...,I_s \neq  \emptyset}}
\!\!\!\!\!
\frac{\lambda R'({-}q)\mathcal{W}^{(0)}_{|I_1|+1}(I_1;{-}q)}{R({-}u_k)-R({-}q)}
\prod_{r=2}^s \frac{\lambda \mathcal{W}^{(0)}_{|I_r|+1}(I_r;u_k)}{
  R(-u_k){-}R({-}q)}\Big)\,,
\nonumber
\end{align}
where
\begin{align*}
  \tilde{K}_i(z,q)&:= \frac{\frac{1}{2}(\frac{1}{z-q}  -\frac{1}{z-\sigma_i(q)})dq}{
R'(\sigma_i(q))(-R(-q)+R(-\sigma_i(q)))}\;,
&\tilde{K}_u(z,q) &:= \frac{(\frac{1}{z+q}  -\frac{1}{z+u})dq}{
    R(u)-R(q)}\;.
\end{align*}
\end{conjecture}
\noindent
Conjecture (\ref{conj-Wplanar}) is now a Theorem proved in
\cite{Hock:2021tbl}. The statement translates easily to
Conjecture~\ref{conj-omplanar} about $\omega_{0,m}$.

In intermediate steps of the proof the following residues become important:
\begin{align}
\nabla^n_z f(z):=
\Res\displaylimits_{q\to z} \frac{f(q)dq}{(R(q)-R(z))^n (R(-z)-R(-q))}\;.
\label{nabla}
\end{align}
Note that the upper index $n$ is merely an index, not an exponent.
The first such residues read
\begin{align*}
 \nabla^1_zf(z)&=\frac{f'(z)+ f(z)(\frac{R''(-z)}{2R'(-z)}
    -\frac{R''(z)}{2R'(z)})}{R'(z)R'(-z)}\;,
  \\
 \nabla^2_zf(z)&=\frac{\frac{1}{2}f''(z)+ f'(z)(\frac{R''(-z)}{2R'(-z)}
   -\frac{R''(z)}{R'(z)})
 }{(R'(z))^2R'(-z)}
 \\*
 &+\frac{f(z)
(\frac{(R''(-z))^2}{4(R'(-z))^2}
+\frac{3(R''(z))^2}{4(R'(z))^2}
-\frac{R''(z) R''(-z)}{2R'(z) R'(-z)}
-\frac{R'''(-z)}{6R'(-z)}
-\frac{R'''(z)}{3R'(z)})
}{(R'(z))^2R'(-z)}\;.
\end{align*}  
These operations arise in the limit $w=q,z\to -\hat{q}^j$ of (\ref{UIzw}):
\begin{align}
\tilde{\mathcal{U}}^{(0)}(I\|{-}\hat{q}^j,q|)
&=
\sum_{\substack{I_1\uplus I_2=I\\
    I_1\neq \emptyset}}
\sum_{\substack{l=1 \\l\neq j}}^d
\frac{\lambda R'(-\hat{q}^l)\mathcal{W}^{(0)}_{|I_1|+1}(I_1;-\hat{q}^l)
  \tilde{\mathcal{U}}^{(0)}(I_2\|{-}\hat{q}^l,q|)}{R'(\hat{q}^l)
  (R(-\hat{q}^j)-R(-\hat{q}^l))}
\label{calUIqq}
\\
&+\sum_{i=1}^m
\frac{\lambda \tilde{\mathcal{U}}^{(0)}(I\backslash u_i\|u_i,q|)}{
  (R(-\hat{q}^j)-R(u_i))(R(q)-R(-u_i))}
\nonumber
\\
&-
\sum_{\substack{ I_1\uplus I_2=I\\ I_1\neq\emptyset}} 
\lambda \nabla_z^1\big(R'(z)\mathcal{W}^{(0)}_{|I_1|+1}(I_1;z)
\tilde{\mathcal{U}}^{(0)}(I_2\|z,q|) \big)\big|_{z=-\hat{q}^j}\;.
\nonumber
\end{align}

\subsection{Proof of Conjectures~\ref{conj-omplanar} and
\ref{conj-Wplanar} for $m=2$}

The case $m=1$ of \eqref{frakUIq} reduces to 
\begin{align}
  \mathfrak{U}^{(0)}(u\|q)
&=\sum_{j=1}^d \frac{\mathcal{W}^{(0)}_{2}(u;\hat{q}^j)}{(R(-q)-R(-\hat{q}^j))}
-\frac{1}{(R(u)-R(-q))(R(q)-R(-u))}\;,
\label{frakU03}
\end{align}
where the symmetry
$\frac{R'(-z)}{R'(z)}
\mathcal{W}^{(0)}_{2}(u_1;-z)=\mathcal{W}^{(0)}_{2}(u_1;z)$ has been
used.  This is inserted into (\ref{W0-noG}) for $m=2$. The first term
of the rhs of (\ref{frakU03}) contributes to the residue at
$q=\beta_i$, the second term to the residue at $q=-u$.  Moreover,
$R'(q)\mathcal{W}^{(0)}_{2}(u;q)=-\big(\frac{1}{u+q}+\frac{1}{u-q}\big)$
has a simple pole at $q=-u$. We thus arrive at
\begin{align*}
& R'(z)  \mathcal{W}^{(0)}_3(u_1,u_2,z)
\nonumber
\\
&=
\sum_{i=1}^{2d} \Res\displaylimits_{q\to \beta_i}
\frac{\lambda\,dq}{q-z}
\sum_{j=1}^d \frac{\big[R'(q)\mathcal{W}^{(0)}_2(u_1;q)\mathcal{W}^{(0)}_2(u_2;\hat{q}^j)
+u_1{\leftrightarrow} u_2\big]}{
R( -q)-R(-\hat{q}^j)}
\\
&
+\Big[
\frac{\lambda \mathfrak{U}^{(0)}(u_2\|{-}u_1)}{z+u_1}
-\frac{\lambda \mathfrak{U}^{(0)}(u_2\|u_1)}{z+u_1}
\\
&- \Res\displaylimits_{q\to -u_{1}}
\frac{\lambda\,dq}{q-z}
\frac{R'(q)\mathcal{W}^{(0)}_2(u_2;q)}{
(R(u_1)-R(-q))(R(q)-R(-u_1))}
+ u_1{\leftrightarrow} u_2\Big]\,.
\end{align*}
In the first line there is, under the assumption of simple
ramification points, a unique preimage $\hat{q}^{j_i}=\sigma_i(q)$
with $\lim_{q\to \beta_i}\hat{q}^{j_i} =\beta_i$. Only this one
contributes to the pole at $q=\beta_i$. In the last line we
change variables $q\to -q$,
arrange
$\frac{\lambda\,dq}{z+q}
=(\frac{\lambda\,dq}{z+q}-\frac{\lambda\,dq}{z+u_1})+\frac{\lambda\,dq}{z+u_1}$
and note that $\frac{\lambda\,dq}{z+u_1}$ produces the residue
(\ref{nabla}):
\begin{align*}
& R'(z)  \mathcal{W}^{(0)}_3(u_1,u_2;z)
\\
&=
\sum_{i=1}^{2d} \Res\displaylimits_{q\to \beta_i}
\frac{\lambda\,dq}{z-q}
\frac{\big[R'(q)\mathcal{W}^{(0)}_2(u_1;q)\,R'(\sigma_i(q))\mathcal{W}^{(0)}_2(u_2;\sigma_i(q))
+u_1{\leftrightarrow} u_2\big]}{
R'(\sigma_i(q))(-R( -q)+R(-\sigma_i(q)))}
\\
&
+\Big[
\frac{\lambda}{z+u_1} \Big\{\mathfrak{U}^{(0)}(u_2\|{-}u_1)
-\mathfrak{U}^{(0)}(u_2\|u_1) +\nabla^1_{-u_1} \big(R'(-u_1)
\mathcal{W}^{(0)}_2(u_2;-u_1)\big)\Big\}
\\
&+ \Res\displaylimits_{q\to u_{1}}
\Big(\frac{\lambda\,dq}{z+q}-\frac{\lambda\,dq}{z+u_1}\Big)
\frac{R'(-q)\mathcal{W}^{(0)}_2(u_2;-q)}{
(R(u_1)-R(q))(R(-u_1)-R(-q))}
+ u_1{\leftrightarrow} u_2\Big]\,.
\end{align*}
One checks that the terms in braces $\{~~\}$ sum up to zero. The first term on the rhs can
be rearranged using the symmetry
\[
  \frac{dq}{(z-q)x'(\sigma_i(q))}
  =  \frac{dq d\sigma_i(q)}{(z-q)x'(\sigma_i(q))d\sigma_i(q)}
  =  \frac{dq d\sigma_i(q)}{(z-q) x'(q)dq}
  =  \frac{d\sigma_i(q)}{(z-q) x'(q)}\;.
\]
Since an odd function under the involution
$q\leftrightarrow \sigma_i(q)$ is integrated, we may replace
$\frac{dq}{(z-q)}\mapsto \frac{1}{2}
(\frac{dq}{(z-q)}-\frac{dq}{(z-\sigma_i(q))})$ and thus establish
Conjecture~\ref{conj-Wplanar} for $m=2$.  The result immediately
translates into Conjecture~\ref{conj-omplanar} for $m=2$. The residue
at $q=\beta_i$ is evaluated with formulae of
Appendix~\ref{app:Galois} and translates to (\ref{Pom03}).
It is straightforward to evaluate the residue at $q=u_1$ and $q=u_2$ to 
(\ref{Hom03}).

\subsection{Proof of Conjectures~\ref{conj-omplanar} and
  \ref{conj-Wplanar} for $m=3$}

\begin{lemma}
For $I=\{u_1,u_2\}$ one has   
\begin{align}
\mathfrak{U}^{(0)}(I\|q)
\label{frakUuvq}
&=  
\sum_{j=1}^d \frac{\mathcal{W}^{(0)}_{3}(I;\hat{q}^j)
  +\lambda \sum_{k=1}^2 \mathcal{W}^{(0)}_{2}(I{\setminus}u_k; \hat{q}^j)
  \check{\mathfrak{U}}_j^{(0)}(u_k\|q)
}{(R(-q)-R(-\hat{q}^j))}
\\
&
+\sum_{k=1}^2 \frac{\lambda \mathcal{W}^{(0)}_2(I{\setminus} u_k;u_k) }{
  (R(q)-R(-u_k))^2(R(u_k)-R(-q))}
\nonumber
\\
&+\lambda \prod_{k=1}^2 \Big(\frac{1}{ (R(q)-R(-u_k))(R(u_k)-R(-q))}\Big)
\;,
\nonumber
\end{align}
where
\begin{align}
\check{\mathfrak{U}}^{(0)}_j(u\|q):=  
\sum_{\substack{l=1 \\l\neq j}}^d
  \frac{\mathcal{W}^{(0)}_2(u;\hat{q}^l)  }{R(-\hat{q}^j)-R(-\hat{q}^l)}
  -\frac{1}{(R(u)-R(-q))(R(q)-R(-u))}\;.
  \label{cfrakU}
\end{align}
\begin{proof}
  Inserting (\ref{calUIqq}) for $I=u_k$ into the first line of
(\ref{frakUIq}) for $I=\{u_1,u_2\}$ and 
(\ref{UIzw}) into the second line of (\ref{frakUIq})
leads after simplifications to 
\begin{align}
&\mathfrak{U}^{(0)}(u_1,u_2\|q)
\label{Uuvq}
\\
&=  
\sum_{j=1}^d \frac{R'(-\hat{q}^j)
  \mathcal{W}^{(0)}_{3}(u_1,u_2;-\hat{q}^j)}{R'(\hat{q}^j) (R(-q)-R(-\hat{q}^j))}
+ \lambda \sum_{j=1}^d \frac{
\big[\mathcal{W}^{(0)}_{2}(u_1;\hat{q}^j)
\check{\mathfrak{U}}_j^{(0)}(u_2\|q)
+ u_1\leftrightarrow u_2\big]
}{(R(-q)-R(-\hat{q}^j))}
\nonumber
\\
&- 
\sum_{j=1}^d \frac{
\big[\lambda R'(\hat{q}^j)  \mathcal{W}^{(0)}_{2}(u_1;\hat{q}^j)
\nabla_{-\hat{q}^j} \big(R'(-\hat{q}^j)\mathcal{W}^{(0)}(u_2;-\hat{q}^j)\big)
+ u_1\leftrightarrow u_2\big]
}{R'(\hat{q}^j) (R(-q)-R(-\hat{q}^j))}
\nonumber
\\
&+ \frac{\lambda}{(R(q)-R(-u_1))(R(u_1)-R(-q))
  (R(q)-R(-u_2))(R(u_2)-R(-q)) }
\nonumber
\\
& +
\Big[\frac{\lambda\mathcal{W}^{(0)}_2(u_2;u_1)}{(R(q)-R(-u_1))^2(R(u_1)-R(-q))}
+u_1\leftrightarrow u_2\Big]\;.
\nonumber
\end{align} 
The explicit formulae for $\mathcal{W}^{(0)}_3$ allow to prove the
following identity\footnote{The generalisation of this identity
  to any $\mathcal{W}^{(0)}_{|I|+1}(I,\pm z)$, and its proof,
  is the key step of the proof of
  Conjecture~\ref{mconj} for $g=0$ in \cite{Hock:2021tbl}.
  The identity seems analogous
  to \cite[Appendix A, eq.\
(1--5)]{Chekhov:2006vd} in the Hermitian 2-matrix model.}:
\begin{align}
  &R'(z)\mathcal{W}^{(0)}_3(u_1,u_2;z)-R'(-z)\mathcal{W}^{(0)}_3(u_1,u_2;-z)
  \nonumber
  \\
  &=\big[\lambda R'(-z)\mathcal{W}^{(0)}_{2}(u_1;-z)
\nabla_z^1 (R'(z)\mathcal{W}^{(0)}(u_2;z))
+u_1\leftrightarrow u_2\big]\;.
\label{flip-W03}
\end{align}
Inserted back into (\ref{Uuvq}) for $z\mapsto -\hat{q}^j$ gives 
the assertion.
\end{proof}
\end{lemma}

We insert (\ref{frakUuvq}) and (\ref{frakU03}) into (\ref{W0-noG}) for
$m=3$.  Let again $\hat{q}^{j_i}=\sigma_i(q)$ be the unique (for
simple ramification points) preimage with
$\lim_{q\to\beta_i}\sigma_i(q)=\beta_i$. Only the first line of
(\ref{frakUuvq}) contributes to the residue at $q=\beta_i$.  More
precisely, in the sum in this first line we have contributions from
all terms with $j=j_i$, whereas for $j\neq j_i$ we only have poles in
$\check{\mathfrak{U}}_{j}(u\|q)$, namely in the single term
$\frac{\lambda
  \mathcal{W}^{(0)}_2(u;\hat{q}^{j_i})}{R(-\hat{q}^{j})-R(-\hat{q}^{j_i})}$
of the sum in (\ref{cfrakU}).  To the residue at $q=-u_3$ we have
contributions from
$R'(q)\mathcal{W}^{(0)}_3(u_i,u_3;q)\mapsto \frac{\lambda
  \mathcal{W}^{(0)}_2(u_i,-u_3)}{(q+u_3)^2R'(u_3)}$, from
$R'(q)\mathcal{W}^{(0)}_2(u_3;q)\mapsto -\frac{1}{(q+u_3)}$, from
$\mathfrak{U}^{(0)}(u_3\|q)\mapsto
-\frac{1}{(R(u_3)-R(-q))(R(q)-R(-u_3))}$ according to (\ref{frakU03})
and from $\mathfrak{U}^{(0)}(u_i,u_3\|q)$ given by
(\ref{frakUuvq}). Here the first line of (\ref{frakUuvq}) contributes
via
$\check{\mathfrak{U}}_j^{(0)}(u_3\|q) \mapsto
-\frac{1}{(R(u_3)-R(-q))(R(q)-R(-u_3))}$, which together with the
other lines amount to
\begin{align*}
\mathfrak{U}^{(0)}(u_i,u_3\|q)&\mapsto 
\frac{\lambda \mathcal{W}^{(0)}_2(u_i;u_3)}{(R(u_3)-R(-q))(R(q)-R(-u_3))^2}
\\
&-\frac{\lambda\mathfrak{U}^{(0)}(u_i\|q)}{(R(u_3)-R(-q))(R(q)-R(-u_3))}\;.
\end{align*}
After partial rearrangement of permutations and
change of variables $q\to -q$ to achieve the residue at $q=u_3$
we arrive at
\begin{align}
&R'(z)  \mathcal{W}^{(0)}_4(u_1,u_2,u_3;z)
\nonumber
\\
&=\bigg\{
\sum_{i=1}^{2d} \Res\displaylimits_{q\to \beta_i} \frac{\lambda\,dq}{q-z}\Big[
\frac{R'(q)\mathcal{W}^{(0)}_3(u_1,u_2;q)  \mathcal{W}^{(0)}_2(u_3;\sigma_i(q))
}{R(-q)-R(-\sigma_i(q))}
\nonumber
\\
&
+\frac{R'(q)\mathcal{W}^{(0)}_2(u_3;q)  \mathcal{W}^{(0)}_3(u_1,u_2;\sigma_i(q))  
}{R(-q)-R(-\sigma_i(q))}
\nonumber
\\
&+
\Big(R'(q)\mathcal{W}^{(0)}_3(u_1,u_2;q)
{+} \frac{\big[\lambda R'(q)\mathcal{W}^{(0)}_2(u_2;q)
  \mathcal{W}^{(0)}_2(u_1;\sigma_i(q))
  {+} u_1{\leftrightarrow}u_2\big]}{  R(-q)-R(-\sigma_i(q))}\Big)
\tag{*}
\\
&\qquad \qquad \times \Big(\check{\mathfrak{U}}_{j_i}^{(0)}(u_3\|q)
+ \sum_{\substack{l=1 \\ l\neq j_i}}^d
\frac{(R(-q)-R(-\sigma_i(q)))\mathcal{W}^{(0)}_2(u_3;\hat{q}^l)}{R(-q)-R(-\hat{q}^l)
  (R(-\hat{q}^l)-R(-\sigma_i(q)))}
\Big)\Big]
\nonumber
\\
&+
\Res\displaylimits_{q\to u_3}
\frac{\lambda\,dq}{z+q} 
\Big[-
\frac{R'(-q)\mathcal{W}^{(0)}_3(u_1,u_2;-q)}{(R(u_3)-R(q))(R(-q)-R(-u_3))}
\tag{**}
\\
&
+ \frac{\lambda R'(-q)(\mathcal{W}^{(0)}_2(u_1;-q) \mathcal{W}^{(0)}_2(u_2;u_3)+
\mathcal{W}^{(0)}_2(u_2;-q) \mathcal{W}^{(0)}_2(u_1;u_3))}{
(R(u_3)-R(q))(R(-q)-R(-u_3))^2}
\tag{**}
\\
&
+\lambda \mathfrak{U}^{(0)}(u_1\|{-}q)
\Big(\frac{\mathcal{W}^{(0)}(u_2;-u_3)}{(q-u_3)^2R'(u_3)}
- \frac{R'(-q)\mathcal{W}^{(0)}(u_2;{-}q)}{(R(u_3)-R(q))(R(-q)-R(-u_3))}\Big)
\tag{$\dag$}
\\
&+\lambda \mathfrak{U}^{(0)}(u_2\|{-}q)
\Big(\frac{\mathcal{W}^{(0)}(u_1;-u_3)}{(q-u_3)^2R'(u_3)}
- \frac{R'(-q)\mathcal{W}^{(0)}(u_1;-q)}{(R(u_3){-}R(q))(R({-}q)-R({-}u_3))}\Big)
\Big]
\tag{$\dag$}
\\
&
+ \frac{\lambda}{z+u_3} \Big(\mathfrak{U}^{(0)}(u_1,u_2\|{-}u_3)
-\mathfrak{U}^{(0)}(u_1,u_2\|u_3)\Big)
+ [u_3\leftrightarrow u_1]+ [u_3\leftrightarrow u_2]\bigg\}\;.
\nonumber
\end{align}
As shown in the previous subsection, the line labelled by (*) is
regular for $q\to \beta_i$ so that (*) and the line after can be
discarded. In the two lines (**) we arrange
$\frac{\lambda\,dq}{z+q}
=(\frac{\lambda\,dq}{z+q}-\frac{\lambda\,dq}{z+u_3})+\frac{\lambda\,dq}{z+u_3}$
as before, where the second $\frac{\lambda\,dq}{z+u_3}$ produces the
residue (\ref{nabla}).  The lines ($\dag$) have only a simple pole
whose residue can also be written in terms of (\ref{nabla}). We thus
find
\begin{align}
&R'(z)  \mathcal{W}^{(0)}_4(u_1,u_2,u_3;z)
\label{W04-final}
\\
&=
\sum_{i=1}^{2d} \Res\displaylimits_{q\to \beta_i} \frac{\lambda\,dq}{q-z}
\sum_{\substack{I_1\uplus I_2=\{u_1,u_2,u_3\} \\ I_1,I_2\neq \emptyset}}
\frac{R'(q)\mathcal{W}^{(0)}_{|I_1|+1}(I_1;q)  \mathcal{W}^{(0)}_{|I_2|+1}(I_2;\sigma_i(q))
}{R(-q)-R(-\sigma_i(q))}
\nonumber
\\
&
+\bigg\{ \Res\displaylimits_{q\to u_3}
\Big(\frac{\lambda\, dq}{z+q} -\frac{\lambda\,dq}{z+u_3}\Big) 
\Big[
\frac{R'(-q)\mathcal{W}^{(0)}_3(u_1,u_2;-q)}{(R(u_3)-R(q))(R(-u_3)-R(-q))}
\nonumber
\\
&
+ \frac{\lambda R'(q)(\mathcal{W}^{(0)}_2(u_1;q) \mathcal{W}^{(0)}_2(u_2;u_3)+
\mathcal{W}^{(0)}_2(u_2;q) \mathcal{W}^{(0)}_2(u_1;u_3))}{
(R(u_3)-R(-q))(R(-u_3)-R(-q))^2} \Big]
\nonumber
\\
&+\frac{\lambda X(u_1,u_2;-u_3)}{z+u_3}  + [u_3\leftrightarrow u_1]+ [u_3\leftrightarrow u_2]\bigg\}\;,
\nonumber
\end{align}
where the remaining collection of terms is shown to
vanish identically\footnote{This is a consequence of sophisticated
  combinatorial structures, see \cite{Hock:2021tbl}.}:
\begin{align*}
X(u_1,u_2;q)
&:=
\mathfrak{U}^{(0)}(u_1,u_2\|q)
-\mathfrak{U}^{(0)}(u_1,u_2\|-q)
+ \nabla_q^1 \big(R'(q)\mathcal{W}^{(0)}_3(u_1,u_2;q)\big)
\\
&
+ \lambda\big[\mathfrak{U}^{(0)}(u_2\|q)
\nabla^1_q\big(R'(q)\mathcal{W}^{(0)}_2(u_1;q)\big)
+u_1\leftrightarrow u_2\big]
\\
&
-\lambda \big[\mathcal{W}^{(0)}_2(u_2;-q) \nabla^2_q\big(
R'(q)\mathcal{W}^{(0)}_2(u_1;q)\big)
+u_1\leftrightarrow u_2\big]
\equiv 0\;.
\end{align*}
After symmetrisation $q\leftrightarrow \sigma_i(q)$ 
we confirm Conjecture~\ref{conj-Wplanar} for $m=3$.
The result immediately
translates into Conjecture~\ref{conj-omplanar} for $m=3$. The residue
at $q=\beta_i$ is evaluated with the formulae given in
Appendix~\ref{app:Galois} and translates to (\ref{Pom04}).
The evaluation of the residue at $q=-u_3$ gives in a first step rise to
$\mathcal{W}^{(0)}_3(u_1,u_2;-u_3)$ and derivatives 
$\partial_{u_3}\mathcal{W}^{(0)}_3(u_1;u_3)$ and
$\partial_{u_3}\mathcal{W}^{(0)}_3(u_2;u_3)$. The reflection~(\ref{flip-W03})
simplifies this to an equation  which translates into (\ref{Hom04}).

\subsection{Proof of Proposition~\ref{prop:om11}}

According to Corollary~\ref{corr:Omega-residue} for $m=0$ and $g=1$
we have with (\ref{eq:Tsum3})
\begin{align}
&R'(z)\Omega^{(1)}_{1}(z)
\label{eq:resom11}
\\
&=\Res\displaylimits_{q\to 0,\beta_i}
\frac{dq}{(q-z)\mathfrak{G}_0(q)}
\Big[
\frac{\partial}{\partial R(u)}
\Big(\lambda R'(q)\mathfrak{G}_0(q)
\mathfrak{U}^{(0)}(u\|q)
-\frac{\lambda \mathcal{G}^{(0)}(u,q)}{R(-q)-R(u)}
\Big)
\Big|_{u=q}
\nonumber
\\
&
+\frac{\lambda}{N}\sum_{n=1}^d \frac{r_n 
\mathcal{G}^{(0)}(q|\varepsilon_n)}{
(R(\varepsilon_n)-R(q))(R(\varepsilon_n)-R(-q))}
-\mathcal{G}^{(0)}(q|q)\Big]\,,
\nonumber
\end{align}
provided that the part in $[\dots]$ has only simple poles at
$z=\pm \widehat{\varepsilon_k}^j$ (which we will confirm).
From (\ref{DSE-cT1}) at $m=0$ and $g=0$ we get 
\begin{align}
&\frac{\lambda}{N}\sum_{n=1}^d 
\frac{r_n  \mathcal{G}^{(0)}(q|\varepsilon_n)}{
(R(\varepsilon_n)-R(q))(R(\varepsilon_n)-R(-q))}
-\mathcal{G}^{(0)}(q|q)
\label{Omega11-G11}
\\
&=
\lim_{u\to q} \Big(\mathcal{G}^{(0)}(u|{-}q)
-\lambda \frac{\mathcal{G}^{(0)}(u,{-}q)}{(R(u)-R(-q))^2}\Big)
-\frac{\lambda R'(q)\mathfrak{G}_0(q)}{(R(q)-R(-q))^3}
\nonumber
\\
&
-\frac{\partial}{\partial R(u)}
\frac{\lambda 
 \mathcal{G}^{(0)}(u,q)}{
(R(u)-R(-q))}\Big|_{u=q}\;.
\nonumber
\end{align}
The limit in the second line follows with \cite[Prop.\ 17]{Schurmann:2019mzu-v3},
\begin{align}
&\lim_{u\to q}\Big(\mathcal{G}^{(0)}(u|{-}q)
-\lambda \frac{\mathcal{G}^{(0)}(u,{-}q)}{(R(u)-R(-q))^2}\Big)
\label{Omega11-G11l}
\\
&= \frac{\lambda(R(q)+R(-q)-2R(0))}{(R(q)-R(-q))^4}
\prod_{j=1}^d \frac{(R(q)-R(\alpha_j))(R(-q)-R(\alpha_j))}{
(R(q)-R(\varepsilon_j))(R(-q)-R(\varepsilon_j))}\;.
\nonumber
\end{align}
With these considerations and (\ref{frakU03}) we get
\begin{align}
&R'(z)\Omega^{(1)}_{1}(z)
\label{Om11}
\\
&=\Res\displaylimits_{q\to 0,\beta_i}
\frac{\lambda\,dq}{(q-z)}
\Big[
\sum_{j=1}^d
\frac{R'(q) \Omega_2^{(0)}(q,\hat{q}^j)}{
R(-q)-R(-\hat{q}^j)}
+\frac{R'(-q)}{(R(q)-R(-q))^3}
\nonumber
\\
&
+\frac{(R(q)+R(-q)-2R(0))}{\mathfrak{G}_0(q)(R(q)-R(-q))^4}
\prod_{j=1}^d \frac{(R(q)-R(\alpha_j))(R(-q)-R(\alpha_j))}{
(R(q)-R(\varepsilon_j))(R(-q)-R(\varepsilon_j))}
\Big]\,.
\nonumber
\end{align}
In particular, the terms in brackets are regular at
$q=\pm \widehat{\varepsilon_n}^j$ so that
Conjecture~\ref{conj:Omega-poles} is true for $g=1$, $m=0$.
Being an
even function of $q$, the expression (\ref{Omega11-G11l}) has a
second-order pole at $q=0$ without residue; it is regular at 
$q=\beta_i$:
\[
  \eqref{Omega11-G11l}
  =\frac{\lambda R''(0)}{16 q^2 (R'(0))^4}
\prod_{j=1}^d \frac{(R(0)-R(\alpha_j))^2}{(R(0)-R(\varepsilon_j))^2}
+\text{regular terms at $q\in \{0,\beta_i\}$}\;.
\]
One has $\prod_{j=1}^d
\frac{(R(0)-R(\alpha_j))^2}{(R(0)-R(\varepsilon_j))^2} =\lim_{q\to 0}
\frac{(R(q)-R(-q))^2}{2R(q)-2R(0)}\mathcal{G}^{(0)}(q,q)
=R'(0)\mathfrak{G}_0(0)$ as shown in \cite[Prop.\
15]{Schurmann:2019mzu-v3}.
The same discussion as for $\mathcal{W}^{(0)}_3$
shows that only one preimage $\hat{q}^{j_i}=\sigma_i(q)$
of the first term on the rhs of (\ref{Om11})
contributes
to the pole at $q=\beta_i$, and again the standard recursion kernel of
topological recursion arises:
\begin{align}
R'(z)\Omega^{(1)}_{1}(z)dz
&=\sum_{i=1}^{2d} \Res\displaylimits_{q\to \beta_i} \frac{\lambda dqdz}{z-q}
\frac{ R'(q)R'(\sigma_i(q))\Omega^{(0)}_2(q,\sigma_i(q))}{
R'(\sigma_i(q))(-R(-q)-(-R(-\sigma_i(q))))}
\nonumber
\\
&+\Res\displaylimits_{q\to 0}\frac{\lambda dqdz}{z-q}\Big[
-\frac{R'(-q)}{(R(q)-R(-q))^3}
-\frac{R''(0)}{16q^2 (R'(0))^3}\Big]
\nonumber
\\
&=\lambda \sum_{i=1}^{2d} \Res\displaylimits_{q\to \beta_i} \big[K_i(z,q)
\omega_{0,2}(q,\sigma_i(q))\big]
\nonumber
\\
&+
\lambda dz
\Big[-\frac{1}{8(R'(0))^2z^3}+
 \frac{R''(0)}{16(R'(0))^3z^2}\Big]\;.
\end{align}
The expansion of the recursion kernel given in Appendix~\ref{app:Galois}
evaluates the residue to the explicit formula given in
Proposition~\ref{prop:om11}. \hspace*{\fill} $\square$%

%\bibliography{omega-cmp}

\end{document}